\pdfoutput=1
\documentclass[pdflatex,sn-mathphys-num]{sn-jnl}

\usepackage{graphicx}%
\usepackage{multirow}%
\usepackage{amsmath,amssymb,amsfonts}%
\usepackage{amsthm}%
\usepackage{mathrsfs}%
\usepackage[title]{appendix}%
\usepackage{xcolor}%
\usepackage{textcomp}%
\usepackage{manyfoot}%
\usepackage{booktabs}%
\usepackage{algorithm}%
\usepackage{algorithmicx}%
\usepackage{algpseudocode}%
\usepackage{listings}%
\usepackage{geometry}
\usepackage{subfig}

\theoremstyle{thmstyleone}%
\newtheorem{theorem}{Theorem}
\theoremstyle{thmstyletwo}%

\theoremstyle{thmstylethree}%

\begin{document}



\title[ArticleTitle]{PGD-TO: A Scalable Alternative to MMA Using Projected Gradient Descent for Multi-Constraint Topology Optimization}

\author*[1]{\fnm{Amin} \sur{Heyrani Nobari}}\email{ahnobari@mit.edu}

\author[1]{\fnm{Faez} \sur{Ahmed}}

\affil[1]{\orgdiv{Department of Mechanical Engineering}, \orgname{Massachusetts Institute of Technology}, \orgaddress{\street{77 Massachusetts Ave.}, \city{Cambridge}, \state{MA}, \postcode{02139}, \country{USA}}}

\abstract{
Projected Gradient Descent (PGD) methods offer a simple and scalable approach to topology optimization (TO), yet they often struggle with nonlinear and multi-constraint problems due to the complexity of active-set detection. This paper introduces PGD-TO, a framework that reformulates the projection step into a regularized convex quadratic problem, eliminating the need for active-set search and ensuring well-posedness even when constraints are infeasible. The framework employs a semismooth Newton solver for general multi-constraint cases and a binary search projection for single or independent constraints, achieving fast and reliable convergence. It further integrates spectral step-size adaptation and nonlinear conjugate-gradient directions for improved stability and efficiency. We evaluate PGD-TO on four benchmark families representing the breadth of TO problems: (i) minimum compliance with a linear volume constraint, (ii) minimum volume under a nonlinear compliance constraint, (iii) multi-material minimum compliance with four independent volume constraints, and (iv) minimum compliance with coupled volume and center-of-mass constraints. 
Across these single- and multi-constraint, linear and nonlinear cases, PGD-TO achieves convergence and final compliance comparable to the Method of Moving Asymptotes (MMA) and Optimality Criteria (OC), while reducing per-iteration computation time by 10--43$\times$ on general problems and 115--312$\times$ when constraints are independent.
Overall, PGD-TO establishes a fast, robust, and scalable alternative to MMA, advancing topology optimization toward practical large-scale, multi-constraint, and nonlinear design problems.
Public code available at: \href{https://github.com/ahnobari/pyFANTOM}{https://github.com/ahnobari/pyFANTOM}
}

\keywords{Topology Optimization, Projected Gradient Descent, Minimum Compliance, Nonlinear Optimization}

\maketitle

\section{Introduction}
With the advent of large computational capacity and modern manufacturing capabilities, most notably additive manufacturing, the optimal geometric/topological design of parts and structures under different physics-based and geometric constraints and objectives has emerged as a prominent problem in computational design. Specifically, the problem of structural Topology Optimization (TO) aims to optimally distribute material in a physical domain to maximize performance with respect to some physical or geometric objective while adhering to constraints. In many TO approaches, the search problem is solved using gradient-based optimization methods that rely on Finite Element Analysis (FEA), which can be expensive in high-fidelity scenarios. 
Beyond the expensive simulations, the nonlinear and often nonconvex nature of the problem makes it challenging to solve efficiently; many nonlinear optimizers incur high computational costs in high‑fidelity settings.
Several efforts have been made to introduce different algorithms for TO. The most popular among these efforts are the Method of Moving Asymptotes (MMA)~\cite{svanberg1987method} and Optimality Criteria (OC)~\cite{hassani1998review}. Despite their popularity, these approaches have drawbacks: OC does not generalize to multiple constraints, and MMA involves a large number of optimizer parameters. These limitations have given rise to alternatives such as Sequential Linear Programming (SLP) \cite{gomes2011slp}, Sequential Quadratic Programming (SQP) \cite{rojas2016efficient}, or Interior Point Method (IPM) \cite{hoppe2002primal}. Recently, however, many have focused on Projected Gradient Descent (PGD) methods for topology optimization~\cite{tavakoli2014multimaterial, nishioka2023inertial, Barbeau_2025}. With a much simpler overall approach and computational efficiency, PGD has promising features that make it an attractive candidate for TO problems. 

However, PGD can become problematic in many situations. For example, nonlinear geometric constraints such as overhang angle constraints \cite{gaynor2014topology,qian2017undercut,lamarche2024additively} in TO problems can cause PGD to become unstable or converge poorly. While such geometric-constraint-induced instabilities motivate the need for robust frameworks, PGD-TO is not a dedicated method for resolving specific nonconvex/nonsmooth pathologies. Instead, PGD-TO improves stability more generally through projection regularization and practical safeguards (e.g., step-size adaptation), and is compatible with geometric constraints when provided in a suitable differentiable form. Moreover, projected gradient descent approaches often handle a single constraint relatively efficiently, yet they encounter a complexity issue in the presence of multiple constraints. This stems from the fact that a general convex Quadratic Program (QP) arises in the projection step, which can be costly, requiring computationally intensive algorithms for the projection alone. Some efforts have been made to solve this problem more efficiently using sophisticated active-set methods~\cite{gu2024random,dostal2005minimizing, Barbeau_2025}, which focus on quickly finding the active constraints in a problem and solving the projection problem based on this information. Recently, \citet{Barbeau_2025} proposed using the active-set method specifically in TO problems. However, switching among active sets during projection remains a limitation and increases complexity for multi‑constraint problems. Moreover, the PGD approach has yet to be thoroughly tested on, and made robust to, problems with multiple nonlinear constraints. 

PGD nonetheless has the potential to serve as a highly scalable, fast, \textbf{general} TO platform. We therefore develop a scalable, efficient framework for applying PGD in TO problems and show that this approach scales much better than existing general TO optimizers while exhibiting similar convergence.
In our work, we introduce a PGD optimizer that combines robustness heuristics with advanced techniques such as nonlinear conjugate gradients and spectral step-size adjustment. Most notably, we overcome the active-set search required for PGD by exploring a regularized subproblem and demonstrate that this subproblem exhibits properties that enable us to solve the projection problem without any active-set search, using a semismooth Newton method with superlinear convergence, which makes the solutions to the projection subproblem expeditious. We then evaluate this approach on a suite of benchmark problems in challenging nonlinear scenarios and demonstrate that the optimizer remains robust even in challenging problems and achieves objective values and convergence behavior comparable to the commonly used MMA solver. 

\begin{figure}[htb]
    \centering
    \includegraphics[width=\linewidth]{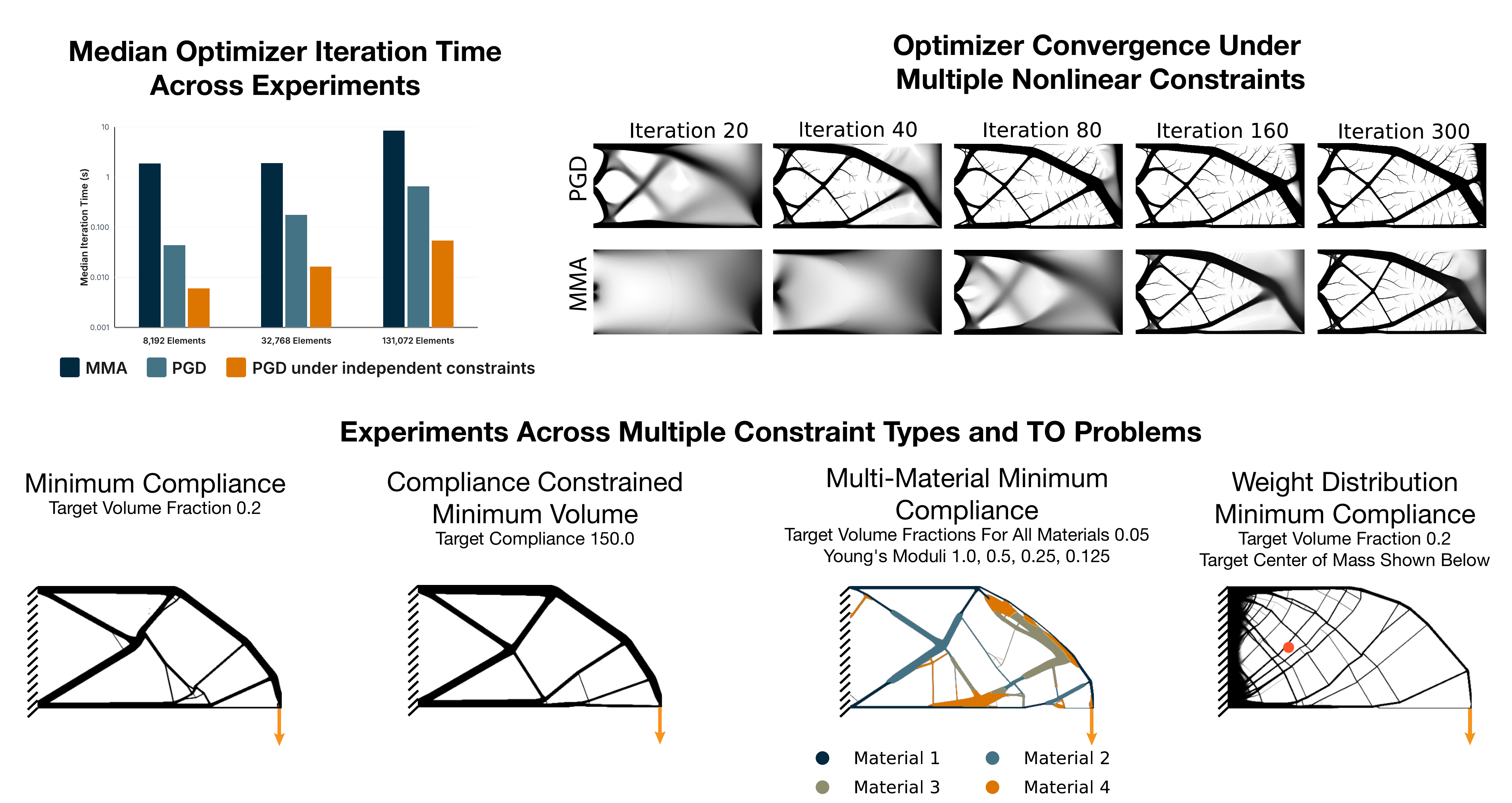}
    \caption{Top: The PGD optimizer's computational cost scales better than that of the commonly used MMA optimizer, while PGD exhibits stable convergence similar to MMA's under multiple nonlinear constraints. Bottom: Overview of the four TO benchmark problems used to evaluate convergence and robustness across solvers. Each problem is solved at three mesh resolutions, with detailed results provided in Appendix~\ref{app:results}.}
    \label{fig:problems}
\end{figure}

\paragraph{Contributions.}
This work advances the use of PGD for topology optimization through a general, efficient, and theoretically grounded framework. Its key contributions are:

\begin{itemize}
    \item \textbf{Active-set-free projection.} We solve the projection step as a strictly convex, always-feasible quadratic program with slack variables. This removes the need for active-set detection and guarantees a unique projection solution, enabling stable and scalable constrained optimization.
    
    \item \textbf{Unified projection solvers.} The framework employs a semismooth Newton method for general multi-constraint problems. It leverages the Symmetric Positive Definite (SPD) Jacobian for superlinear local convergence and a binary-search algorithm for projection in single or independent constraints, providing substantial computational savings.
    
    \item \textbf{Theoretically supported PGD algorithm.} We use insights from the $\mathcal{O}(1/\sqrt{K})$ convergence-rate analysis of PGD, where $K$ is the number of iterations, for automatic step-size relaxation with linearized constraints, and we integrate spectral step-size adaptation (without line search) and Polak--Ribi\`ere nonlinear conjugate-gradient directions to improve convergence behavior.
    
    \item \textbf{Empirical validation at scale.} Across four benchmark TO problems shown in Fig.~\ref{fig:problems} encompassing linear, nonlinear, and multi-constraint scenarios, the proposed method achieves accuracy comparable to MMA and OC while reducing per-iteration computation by one to two orders of magnitude. Code and benchmarks are publicly available.
\end{itemize}

\section{Background and Related Work}
This section reviews the scope of the TO problems we tackle, formulates the general problem concretely, and discusses some related works and nonlinear optimizers that have been developed for a similar class of problems.

\subsection{Formalizing the Topology Optimization Problems}
Structural topology optimization seeks the optimal distribution of material within a prescribed design domain to maximize or minimize a given physics-based or geometric performance metric, subject to a set of constraints.
A common formulation of TO operates over a continuous domain, which is the focus of this work. Let $\bar{\Omega} \subset \mathbb{R}^d$ represent a bounded design domain with boundary $\Gamma = \partial \bar{\Omega}$. In these approaches, a material density field, $\rho(x):\bar{\Omega}\to\{0,1\}$, becomes the design we seek to optimize. While an ideal binary solution yields a clear structural layout, obtaining such discrete solutions directly is computationally intractable. Thus, a more relaxed mapping, $\rho(x):\bar{\Omega}\to[0,1]$, is used where the density values represent the amount of material in a given part of the domain. 
This continuous representation requires evaluating physical and geometric responses over $\bar{\Omega}$, typically governed by partial differential equations describing the relevant physics. Because these PDEs rarely admit closed-form solutions for arbitrary domains, TO algorithms usually employ a discretized approximation and solve the governing equations via finite element analysis (FEA). 
In its general discrete form, the topology optimization problem can be written as:

\begin{equation}
    \begin{split}
        \min_{\boldsymbol{\rho}\in R^N}&\quad f(\boldsymbol{\rho}) \\
        \text{s.t.}& \quad l\leq \rho_i\leq u, \quad \forall i \in [N]\\ 
        &\quad g_j(\boldsymbol{\rho})\leq G_j \quad \forall j\in[m],
    \end{split}
    \label{eqn:genopt}
\end{equation}

where $\boldsymbol{\rho}\in \mathbb{R}^N$ represents the $N$ design variables (for a discretized domain mesh with $N$ elements) and $l$ and $u$ are the bounds on the design variables and $f$ is a general linear or nonlinear function of the design variables (e.g., $\mathbf{f}^T\mathbf{K}^{-1}(\boldsymbol{\rho})\mathbf{f}$ in minimum compliance, where $\mathbf{K}(\boldsymbol{\rho})$ is the FEA stiffness matrix and $\mathbf{f}$ is the load vector), and $g_j$ are linear or nonlinear functions of the design variable with upper bound $G_j$ (e.g., $\sum_{i=1}^NA_i\rho_i\leq V$ for the volume constraint used in many TO problems).

This general formulation makes no assumptions about the form of $f$ or $g_j$ beyond their dependence on the design variables and the presence of simple bounds on $\rho_i$. Equality constraints can be accommodated within the same framework and are omitted here only for brevity.

\subsection{Nonlinear Optimizers for TO}
Many prior works have focused on solving specific TO problems where the particular problem involves a specific structure or a single constraint (e.g., only volume constraint). A popular example of this is the OC approach developed for solving problems with only design variable bounds and a single volume constraint~\cite{bendsoe1995optimization}. More general approaches, such as SLP \cite{gomes2011slp}, SQP \cite{rojas2016efficient}, or IPM \cite{hoppe2002primal}, have attempted to tackle the general problem in (\ref{eqn:genopt}).

In parallel to continuous-density formulations, a broad body of work studies \emph{discrete-variable} topology optimization, where design variables are restricted to binary or multi-valued choices and the resulting problems are cast as integer programs (e.g., sequential integer programming, mixed integer linear programming (MILP), or mixed integer non-linear programming (MINLP) formulations). Representative approaches include sequential integer programming for continuum-type TO, integer-programming-based multi-material topology design, and MILP-based discrete topology optimization for truss structures, as well as global discrete truss optimization methods that rely on branch and bound and branch and cut strategies. While the algorithmic details differ from continuous projected-gradient descent, these discrete formulations likewise emphasize efficient handling of which constraints become active in large-scale problems~\cite{sipTO2019,IPmultimat2021,milpTruss2022,cutandbranch2008}.

Among these, MMA~\cite{svanberg1987method} has emerged as the most widely used optimizer in TO due to its scalability and robustness across large problem instances~\cite{ferrari2020newgeneration99line,reviewto}. Despite its success, MMA requires the solution of a subproblem that involves updating asymptotes for all design variables. As a result, its computational complexity scales primarily with the number of design variables, rather than with the typically smaller number of constraints. In contrast, the OC method is more efficient in single-constraint settings, since it reduces the optimization to a dual problem with only one dual variable. However, MMA also demands careful tuning of multiple algorithm-specific hyperparameters, which can make setup and convergence control cumbersome in practice.

More recently, \emph{Projected Gradient Descent} methods have gained attention as promising alternatives for TO due to their simplicity and favorable computational scaling, especially in large-scale or single-constraint problems~\cite{tavakoli2014multimaterial,nishioka2023inertial,Barbeau_2025}. The most recent advance by \citet{Barbeau_2025} extends PGD to multi-constraint TO problems using a modified active-set projection strategy. While effective, active-set-based projections can be difficult to solve and may lead to slow or cyclic convergence when the active set changes frequently. 

In this work, we propose a different approach: a \emph{regularized projection formulation} that eliminates the need for active-set identification entirely. Our method solves the projection subproblem directly through a convex regularization scheme, resulting in improved robustness and scalability—particularly for problems with a small number of nonlinear constraints.

\section{Projected Gradient Descent for Topology Optimization}
This section details the proposed Projected Gradient Descent framework for topology optimization and motivates our formulation's design choices.

\subsection{Projected Gradient Descent}
Given problem (\ref{eqn:genopt}), we now formulate the PGD approach. In projected gradient descent, we compute the gradients of the objective function, take a step in the negative gradient direction, and project the resulting design variables onto the feasible subset of $\mathbb{R}^N$. Let $\Pi_{\Omega}(\tilde{\boldsymbol{\rho}})$ be the projection operator that projects any given $\tilde{\boldsymbol{\rho}}\in\mathbb{R}^N$ onto the feasible set $\Omega \triangleq \left\{\boldsymbol{\rho}\mid l\leq \boldsymbol{\rho}\leq u,g_j(\boldsymbol{\rho})\leq G_j\quad \forall j\in [m]\right\}\subseteq\mathbb{R}^N$, then the update rule for projected gradient descent will be:

\begin{equation}
    \label{eqn:PGD}
    \boldsymbol{\rho}_{t+1} = \Pi_{\Omega}\left(\boldsymbol{\rho}_t-\alpha \nabla f\right).
\end{equation}

If the projection operator $\Pi_{\Omega}$ can be evaluated efficiently, this scheme provides a simple and scalable method for solving a wide range of TO problems. Moreover, the framework naturally accommodates alternative descent directions or acceleration schemes, such as momentum, conjugate gradients, or spectral updates, making PGD a versatile approach for nonlinear constrained optimization. In this work, we focus on the practically relevant regime where the number of constraints is much smaller than the number of design variables ($m \ll N$). Under this assumption, we introduce an efficient and robust formulation for solving the projection subproblem, which serves as the key computational component of our PGD-based topology optimization framework.

\subsection{Solving the Projection Problem}
To formulate the projection problem, consider the updated design-variable vector $\tilde{\boldsymbol{\rho}}=\boldsymbol{\rho}_{t}-\alpha\nabla f(\boldsymbol{\rho}_t)$, which is obtained by performing a gradient descent update on the design variables with a step size $\alpha$ (we discuss the choice of $\alpha$ in Sections \ref{section:convergence} and \ref{section:algo}). Let $\boldsymbol{\delta}$ be an adjustment vector which, when added to $\tilde{\boldsymbol{\rho}} \notin \Omega$, will bring it into the feasible set $\Omega$. We can formulate the projection as the Euclidean-projection problem:

\begin{equation}
    \label{eqn:projectio}
    \begin{split}
        \min_{\boldsymbol{\delta}\in\mathbb{R}^N}& \quad \|\tilde{\boldsymbol{\rho}} + \boldsymbol{\delta} - \tilde{\boldsymbol{\rho}}\|^2_2=\|\boldsymbol{\delta}\|_2^2\\
        \text{s.t.}& \quad l\leq \tilde{\rho}_i+ \delta_i\leq u, \quad \forall i \in [N]\\ 
        &\quad g_j(\tilde{\boldsymbol{\rho}}+\boldsymbol{\delta})\leq G_j \quad \forall j\in[m],
    \end{split}
\end{equation}

Problem (\ref{eqn:projectio}) has a strictly convex quadratic objective; however, the feasible set $\Omega$ may be nonconvex or even empty, which complicates direct computation. To address this, we adopt a simplification inspired by Sequential Linear Programming methods commonly used in topology optimization. Specifically, we linearize the nonlinear constraint functions locally, assuming that both the gradient descent update and the projection adjustment are small. 
\[
\|\alpha \tfrac{\partial f}{\partial \boldsymbol{\rho}_t}\| \ll \|\boldsymbol{\rho}_t\|, 
\qquad
\|\boldsymbol{\delta}\| \ll \|\boldsymbol{\rho}_t\|.
\]
Under these conditions, it is reasonable to approximate nonlinear constraints using their first-order Taylor expansion around $\boldsymbol{\rho}_t$. This assumption is particularly valid near local optima, where convergence behavior is stable and step sizes are small. 

Applying this linearization yields the following simplified projection subproblem:

\begin{equation}
\label{eqn:linearized}
    \begin{split}
        \min_{\boldsymbol{\delta}\in \mathbb{R}^N}& \quad \|\boldsymbol{\delta}\|_2^2\\
        \text{s.t.}& \quad l\leq \tilde{\rho}_i+ \delta_i\leq u, \quad \forall i \in [N]\\ 
        &\quad g_j(\boldsymbol{\rho}_t) + \left(\boldsymbol{\delta}-\alpha\nabla f(\boldsymbol{\rho}_t)\right)^T\nabla g_j(\boldsymbol{\rho}_t)\leq G_j \quad \forall j\in[m],
    \end{split}
\end{equation}

This formulation leads to a convex QP with a linear constraint set, which, as we will show, can be solved efficiently in most cases. We can do this by analyzing the optimality conditions of the above problem. The key insight from the optimality conditions is that the solution depends on only a small number $m$ of dual variables. Theorem~\ref{thm:solution_form} characterizes the form of this solution.

\begin{theorem}[Solution to the Linearized Projection Problem]
\label{thm:solution_form}
Let the linearized projection problem be defined as in (\ref{eqn:linearized}). A vector $\boldsymbol{\delta}^* \in \mathbb{R}^N$ is the unique optimal solution to this problem if and only if there exists a vector of dual variables $\mathbf{y}^* \in \mathbb{R}^m$ satisfying $\mathbf{y}^* \le \mathbf{0}$, such that for all $i \in \{1,\ldots,N\}$:
\begin{equation}
    \delta_i^* =
    \begin{cases}
        l-\tilde{\rho}_i & \text { if } \sum_j y_j^* (\nabla g_j(\boldsymbol{\rho}_t))_i<l-\tilde{\rho}_i, \\
        \sum_j y_j^* (\nabla g_j(\boldsymbol{\rho}_t))_i& \text { if } l-\tilde{\rho}_i \leq \sum_j y_j^* (\nabla g_j(\boldsymbol{\rho}_t))_i \leq u-\tilde{\rho}_i, \\
        u-\tilde{\rho}_i & \text { if } \sum_j y_j^* (\nabla g_j(\boldsymbol{\rho}_t))_i>u-\tilde{\rho}_i,
    \end{cases}
\end{equation}
and for all $j\in\{1,\ldots,m\}$:
\begin{align}
    &g_j(\boldsymbol{\rho}_t) + \left(\boldsymbol{\delta}^*-\alpha\nabla f(\boldsymbol{\rho}_t)\right)^T\nabla g_j(\boldsymbol{\rho}_t)\leq G_j \\
    &y_j^* \left(g_j(\boldsymbol{\rho}_t) + \left(\boldsymbol{\delta}^*-\alpha\nabla f(\boldsymbol{\rho}_t)\right)^T\nabla g_j(\boldsymbol{\rho}_t)-G_j\right)=0
\end{align}

\end{theorem}
\begin{proof}
The proof follows directly from the KKT conditions, derived in detail in Appendix~\ref{app:kkt}. The expressions for $\boldsymbol{\delta}^*$ and $\mathbf{y}^*$, along with their feasibility conditions, are exact restatements of the KKT optimality system for this convex problem.
\end{proof}

The key observation concerns how the constraint functions behave with respect to the dual variables $\mathbf{y}$. We detail this behavior in the following sections and show how the projection problem can be solved efficiently in each scenario.

\subsubsection{Solving Problems with a Single Constraint}
Many introductory TO problems involve only a single constraint, most commonly a volume constraint, a single maximum stress constraint, or a compliance constraint. Although such problems are not the most general TO setting, they are a common starting point in practice. In such cases, the optimization problem in (\ref{eqn:linearized}) reduces to a single constraint, and therefore, only one dual variable $y$ appears in the KKT system. 

It can be shown that for this case, the linearized constraint function 
\[
g(\boldsymbol{\rho}_t) + \left(\boldsymbol{\delta} - \alpha \nabla f(\boldsymbol{\rho}_t)\right)^{\!T} \nabla g(\boldsymbol{\rho}_t)
\]
is a monotone nondecreasing function of the dual variable $y$ (see Appendix~\ref{app:singlecon}). This monotonicity property enables an efficient solution of the projection subproblem through a one-dimensional binary search. 

If the constraint is inactive, then a value of $y=0$ will yield a valid solution; thus, in our algorithm, we start by checking if this is the case, and if not, we perform a binary search to find $y^*$ such that the constraint is met at equality. Algorithm \ref{algo:binary} shows how this binary search can be performed to find the optimal projection for the linearized problem.

\begin{algorithm}[H]
\caption{Single Constraint Projection Binary Search}\label{algo:binary}
\begin{algorithmic}[1]
\Require $\tilde{\boldsymbol{\rho}},\;\nabla\mathbf{g}(\boldsymbol{\rho}_t),\;\mathbf{g}(\boldsymbol{\rho}_t),\;\mathbf{G},\;\alpha\nabla f(\boldsymbol{\rho}_t),\;l,\;u,\;\text{tol}_B$
\If {$\boldsymbol{\delta}(0)$ is feasible}
    \State $y^* = 0$
    \State \textbf{Return} $y^*$
\Else
    \State $y_l =  \min_{i} \{\min\left(\frac{l-\tilde{\rho}_i}{(\nabla g(\boldsymbol{\rho}_t))_i},\frac{u-\tilde{\rho}_i}{(\nabla g(\boldsymbol{\rho}_t))_i}\right)\}$
    \State $y_u = 0$
    \State $y_m = \frac{y_u+y_l}{2}$
    \While{$|y_u-y_l|>\text{tol}_B$}
        \If {$\left(\boldsymbol{\delta}(y_m)-\alpha\nabla f(\boldsymbol{\rho}_t)\right)^T\nabla g(\boldsymbol{\rho}_t) - G + g(\boldsymbol{\rho}_t)>0$}
            \State $y_u = y_m$
        \Else
            \State $y_l = y_m$
        \EndIf
        \State $y_m = \frac{y_u+y_l}{2}$
    \EndWhile
    \State $y^* = y_m$
    \State \textbf{Return} $y^*$
\EndIf
\end{algorithmic}
\end{algorithm}

Although Algorithm~\ref{algo:binary} is as explicit as classical OC schemes, it is not an OC update: we do not use a problem-specific closed-form multiplier rule (e.g., the standard dual problem formulation of OC with heuristic move limits), and the procedure is not tailored to a volume-fraction constraint. Instead, our iterate update is a gradient step followed by projection, where feasibility is enforced through the projection subproblem derived from the KKT conditions of~(\ref{eqn:linearized}), and in scenarios where the constraint is inactive, the projection does not alter the design variables.
This approach is closely related to the work of~\citet{nishioka2023inertial}, who addressed topology optimization problems involving only a single volume constraint. In contrast, our goal is to develop a formulation that generalizes to the full class of problems described in Theorem~\ref{thm:solution_form}. To this end, we next consider two additional cases in which the solution form of the projection can be explicitly determined.

\subsubsection{Solving Problems with Independent Constraints}
In many TO problems, multiple independent constraints arise, for instance, in multi-material designs where each material is subject to its own volume constraint. In such cases, each constraint depends on a distinct subset of design variables that are independent of one another. Consequently, each constraint can be solved using the same binary search procedure introduced for the single-constraint case (Algorithm~\ref{algo:binary}). 

Because the subproblems are independent, binary searches for the corresponding dual variables can be executed independently and in parallel. This property enables significant computational acceleration, which we exploit in our implementation.

\subsubsection{Solving the General Projection Problem Efficiently}
We now turn to the most general case, where multiple \emph{non-independent} constraints are present. Prior work typically addresses this setting by identifying the active set of constraints and then solving for the corresponding dual variables based on the solution form in Theorem~\ref{thm:solution_form}. However, active-set methods can be computationally expensive and may require several iterations to determine the correct set of active constraints. Moreover, such approaches lack convergence guarantees when the problem becomes locally infeasible.

To address both active-set identification and local infeasibility, we reformulate the projection problem so that a single subproblem yields the projection even when the feasible set is empty. To do this, we propose a Newton-based root-finding method that can uniquely determine an optimal projection even when the feasible set is null. Specifically, we adjust the projection problem by introducing regularization parameters to the problem:

\begin{equation}
\label{eqn:regularized_projection}
\begin{split}
	\min_{\boldsymbol{\delta}\in \mathbb{R}^N, \mathbf{s} \in \mathbb{R}^m}& \quad \frac{1}{2}\|\boldsymbol{\delta}\|_2^2 + \frac{C}{2} \|\mathbf{s}\|_2^2 \\
	\text{s.t.}& \quad l\leq \tilde{\rho}_i+ \delta_i\leq u, \quad \forall i \in [N]\\
	&\quad g_j(\boldsymbol{\rho}_t) + \left(\boldsymbol{\delta}-\alpha\nabla f(\boldsymbol{\rho}_t)\right)^T\nabla g_j(\boldsymbol{\rho}_t) - s_j \leq G_j \quad \forall j\in[m]\\
    &\quad s_j \ge 0 \quad \forall j\in[m],
\end{split}
\end{equation}

where $C > 0$ is a large penalty parameter. This problem is a strictly convex QP and is always feasible, and the solution of this problem becomes identical to the original projection as $C\rightarrow\infty$. If we perform the same optimality criteria analysis for this projection problem, we can show that a similar form of solution based on $m$ dual variables exists with relaxed conditions on the constraints. Theorem \ref{thm:solution_form_reg} demonstrates how this regularization formulation changes the form of the solution.

\begin{theorem}[Solution to the Regularized Projection Problem]
\label{thm:solution_form_reg}
Let the regularized projection problem be defined as in (\ref{eqn:regularized_projection}). A vector $\boldsymbol{\delta}^* \in \mathbb{R}^N$ is the unique optimal solution to this problem if and only if there exists a (unique) vector of dual variables $\mathbf{y}^* \in \mathbb{R}^m$ satisfying $\mathbf{y}^* \le \mathbf{0}$, such that for all $i \in \{1,\ldots,N\}$:
\begin{equation}
    \delta_i^* =
    \begin{cases}
        l-\tilde{\rho}_i & \text { if } \sum_j y_j^* (\nabla g_j(\boldsymbol{\rho}_t))_i<l-\tilde{\rho}_i, \\
        \sum_j y_j^* (\nabla g_j(\boldsymbol{\rho}_t))_i& \text { if } l-\tilde{\rho}_i \leq \sum_j y_j^* (\nabla g_j(\boldsymbol{\rho}_t))_i \leq u-\tilde{\rho}_i, \\
        u-\tilde{\rho}_i & \text { if } \sum_j y_j^* (\nabla g_j(\boldsymbol{\rho}_t))_i>u-\tilde{\rho}_i,
    \end{cases}
\end{equation}
and for all $j\in\{1,\ldots,m\}$:
\begin{align}
    &g_j(\boldsymbol{\rho}_t) + \left(\boldsymbol{\delta}^*-\alpha\nabla f(\boldsymbol{\rho}_t)\right)^T\nabla g_j(\boldsymbol{\rho}_t)+\frac{2y_j^*}{C}\leq G_j \\
    &y_j^* \left(g_j(\boldsymbol{\rho}_t) + \left(\boldsymbol{\delta}^*-\alpha\nabla f(\boldsymbol{\rho}_t)\right)^T\nabla g_j(\boldsymbol{\rho}_t)+\frac{2y_j^*}{C}-G_j\right)=0
\end{align}


\end{theorem}
\begin{proof}
The proof follows from the detailed KKT analysis in Appendix \ref{app:kkt}. The form of $\boldsymbol{\delta}^*$ and the conditions on $\mathbf{y}^*$ and the feasibility of $\boldsymbol{\delta}^*$ are an exact restatement of the KKT conditions for this convex problem.
\end{proof}

It can be seen that when $C\rightarrow \infty$, the solution in Theorem \ref{thm:solution_form_reg} becomes identical to the form in Theorem \ref{thm:solution_form}. Thus, for a sufficiently large $C$, solving this problem will yield a solution identical to that of the true projection problem. This problem is guaranteed to be feasible, and as we will show, can be solved with just one root-finding step.

The key observation that we highlight in this problem is the fact that the Jacobian of the constraint functions $\mathbf{h}(\mathbf{y})$, where $h_j(\mathbf{y}) = g_j(\boldsymbol{\rho}_t) + \left(\boldsymbol{\delta}-\alpha\nabla f(\boldsymbol{\rho}_t)\right)^T\nabla g_j(\boldsymbol{\rho}_t)+\frac{2y_j}{C}-G_j$, will be symmetric positive definite and thus invertible, and well behaved. This allows us to solve the problem using a semismooth Newton method and obtain a projection solution directly and with the superlinear convergence associated with the Newton method. To this end, we define a system of semismooth nonlinear functions, $\boldsymbol{\Phi}(\mathbf{y})$, whose roots we obtain using a semismooth Newton iteration. These functions are defined as:

\begin{equation}
    \label{eqn:phi}
    \Phi_j(\mathbf{y}) = \max(h_j(\mathbf{y}),0) + \operatorname{sign}(\max(-h_j(\mathbf{y}),0))y_j.
\end{equation}

Solving this system of equations is equivalent to solving the projection as the root of $\Phi(\mathbf{y})$. This equivalence holds because $\boldsymbol{\Phi}(\mathbf{y})$ is constructed to strictly enforce the complementarity conditions of Theorem \ref{thm:solution_form_reg}. Specifically:
\begin{enumerate}
    \item \textbf{If the constraint is active or violated} ($h_j(\mathbf{y}) > 0$), the first term $\max(h_j(\mathbf{y}),0)$ becomes $h_j(\mathbf{y})$ while the second term vanishes (as $\max(-h_j(\mathbf{y}),0) = 0$). Consequently, the condition $\Phi_j(\mathbf{y}) = 0$ forces $h_j(\mathbf{y}) = 0$, satisfying primal feasibility.
    \item \textbf{If the constraint is inactive} ($h_j(\mathbf{y}) \le 0$), the first term vanishes and the second term becomes active (since $\max(-h_j(\mathbf{y}),0) > 0$, the sign function returns 1). This simplifies the equation to $\Phi_j(\mathbf{y}) = y_j$. Consequently, the condition $\Phi_j(\mathbf{y}) = 0$ forces $y_j = 0$, satisfying dual feasibility and complementary slackness.
\end{enumerate}
Thus, the root $\mathbf{y}^*$ corresponds uniquely to the solution satisfying the KKT conditions of the regularized problem.
We detail this choice and why the SPD nature of the Jacobian of $\mathbf{h}(\mathbf{y})$ guarantees convergence to a solution in a superlinear manner in Appendix \ref{app:newton}. Suffice it to say that for a system with a small number of constraints, $m\ll N$, obtaining the Jacobian, $\mathbf J_\Phi(\mathbf{y})$, and its inverse will be computationally inexpensive. Moreover, the full Newton solve is often unnecessary in intermediate iterations. Often only one constraint is active in intermediate iterations, as constraint violations fluctuate within the algorithm's tolerances. Given the simplicity and computational efficiency of solving a single constraint problem, in our algorithm, we propose solving the single constraint case for all constraints in parallel using binary search, and if the solution is found, avoiding the iterative semismooth Newton solver altogether. In this manner, we ensure that in some iterations where only one active constraint plays a role, we obtain the solution quickly and efficiently without the need for a more complex semi-smooth Newton method. Algorithm \ref{algo:general} shows the full projection algorithm for the general case. In practice we set $C=10^{12}$ by default; this value can be adjusted to the problem and constraint gradients. The idea of avoiding active-set detection via regularization has been explored in similar forms \cite{doi:10.1137/S1052623401383558,doi:10.1137/19M1240186}; here we provide one such framework with a line-search-based step-size determination for solving this problem.

\begin{algorithm}[H]
\caption{General Multi-Constraint Projection Solver}\label{algo:general}
\begin{algorithmic}[1]
\Require $\tilde{\boldsymbol{\rho}},\;\nabla\mathbf{g}(\boldsymbol{\rho}_t),\;\mathbf{g}(\boldsymbol{\rho}_t),\;\mathbf{G},\;\alpha\nabla f(\boldsymbol{\rho}_t),\;l,\;u,\;C,\;\text{tol}_{B},\;\text{tol}_{N},\;\text{maxiter}$
\Statex{\textit{--- Stage 1: Attempt to solve by assuming a single active constraint ---}}
\State Let $\mathbf{y}_{\text{trial}} = \mathbf{0} \in \mathbb{R}^m$.
\For{$j \in \{1, \ldots, m\}$} \Comment{This loop can be executed in parallel}
    \State Use Algorithm \ref{algo:binary} to find a candidate $y_j^*$ for constraint $j$ with tolerance $\text{tol}_{B}$.
    \State Set $y_{\text{trial},j} = y_j^*$.
    \State Compute $\boldsymbol{\delta}_{\text{trial}} = \boldsymbol{\delta}(\mathbf{y}_{\text{trial}})$ using Equation (\ref{eqn:delta_y}).
    \State \textbf{is\_feasible} = true
    \For{$k \in \{1, \ldots, m\}$}
        \If{$g_k(\boldsymbol{\rho}_t) + \boldsymbol{\delta}_{\text{trial}}^T\nabla g_k(\boldsymbol{\rho}_t) > G_k + \text{tol}_{N}$}
            \State \textbf{is\_feasible} = false; \textbf{break}
        \EndIf
    \EndFor
    \If{\textbf{is\_feasible}}
        \State \Return $\boldsymbol{\delta}_{\text{trial}}$ \Comment{Feasible projection found}
    \EndIf
\EndFor
\Statex{\textit{--- Stage 2: Solve the regularized problem via Semismooth Newton ---}}
\State Initialize $\mathbf{y}^{(0)} = \mathbf{y}_{\text{trial}}$, set $k=0$.
\While{$\|\boldsymbol{\Phi}(\mathbf{y}^{(k)})\|_{\infty} > \text{tol}_{N}$ and $k<\text{maxiter}$}
    \State Compute $\boldsymbol{\delta}(\mathbf{y}^{(k)})$ using Equation (\ref{eqn:delta_y_robust}).
    \State Compute $\mathbf{h}(\mathbf{y}^{(k)})$ using Equation (\ref{eqn:h}).
    \State Compute $\boldsymbol{\Phi}(\mathbf{y}^{(k)})$ using Equation (\ref{eqn:phi}).
    \State Compute the Jacobian $\mathbf{J}_{\Phi}(\mathbf{y}^{(k)})$ as per Appendix \ref{app:newton}.
    \State Solve the linear system $\mathbf{J}_{\Phi}(\mathbf{y}^{(k)})\boldsymbol{\Delta} = \boldsymbol{\Phi}(\mathbf{y}^{(k)})$ for $\boldsymbol{\Delta}$.
    \State Find $\alpha^* \in (0, 1]$ via line search to ensure reduction in $M = \frac{1}{2}\|\boldsymbol{\Phi}\|_2^2$ (Appendix \ref{app:newton}).
    \State $\mathbf{y}^{(k+1)} = \mathbf{y}^{(k)} - \alpha^*\boldsymbol{\Delta}$.
    \State $k \leftarrow k+1$.
\EndWhile
\State $\mathbf{y}^* = \mathbf{y}^{(k)}$.
\State Compute $\boldsymbol{\delta}^* = \boldsymbol{\delta}(\mathbf{y}^*)$ using Equation (\ref{eqn:delta_y_robust}).
\State \Return $\boldsymbol{\delta}^*$
\end{algorithmic}
\end{algorithm}

\subsubsection{A Note on Equality Constraints}
Thus far, we have considered only inequality constraints. In practice, the main challenge across all cases lies in identifying the active constraints. Equality constraints can be naturally incorporated by treating them as \emph{always active}. Consequently, they are handled identically to active inequality constraints within our framework. In the single-constraint and independent multi-constraint cases, this treatment requires no modification to the solution procedure. For the regularized projection problem, equality constraints are enforced directly by setting $\boldsymbol{\Phi}_j(\mathbf{y}) = h_j(\mathbf{y})$ and requiring this quantity to be zero, thereby ensuring that the equality constraint is always active. These adjustments integrate seamlessly with the existing formulation and introduce no additional complexity. For brevity, we omit further discussion of equality constraints and focus on inequality-constrained problems in the main body of the paper.

\subsection{Convergence Analysis and Dynamic Step Size}
\label{section:convergence}
We now analyze convergence of the proposed PGD scheme and derive a dynamic step size. Theorem \ref{thm:convergence} shows that for a small enough step size $\alpha$, the proposed projected gradient descent approach will converge at a rate $O(1/\sqrt{K)}$, if the objective function is assumed to be L-smooth. In Appendix \ref{app:converge}, we argue that in most cases for TO problems based on FEA, the assumption of L-smoothness is reasonable, so long as the resulting systems of equations are not near-singular.


\begin{theorem}[Convergence of Projected Gradient Descent with Linearized Constraints]
\label{thm:convergence}
Let the objective function $f: \mathbb{R}^N \to \mathbb{R}$ be bounded from below by $f_{\text{inf}}$ and be $L_f$-smooth with a Lipschitz constant $L_f > 0$. Consider the iterative algorithm:
$$
\boldsymbol{\rho}_{t+1} = \Pi_{\hat{\Omega}_{t+1}}\left(\boldsymbol{\rho}_t-\alpha \nabla f(\boldsymbol{\rho}_t)\right)
$$
where $\Pi_{\hat{\Omega}_{t+1}}$ is the Euclidean projection onto the linearized feasible set $\hat{\Omega}_{t+1}$ defined in Equation~(\ref{eqn:linearized}), and the step size $\alpha$ is constant.

If the step size satisfies $0 < \alpha < \frac{2}{L_f}$ and each iterate satisfies the condition $\boldsymbol{\rho}_t \in \hat{\Omega}_{t+1}$ for all $t \ge 0$, then the algorithm guarantees that the norm of the updates converges to zero, with a sublinear rate of:
$$
\min_{t=0,\dots,K-1}\|\boldsymbol{\rho}_{t+1}-\boldsymbol{\rho}_{t}\|_{2} \le \sqrt{\frac{f(\boldsymbol{\rho}_{0}) - f_{\text{inf}}}{L \cdot K}} = O\left(\frac{1}{\sqrt{K}}\right)
$$
where $L = \frac{1}{\alpha} - \frac{L_f}{2}$ is a positive constant. 

Furthermore, assuming the Linear Independence Constraint Qualification (LICQ) holds at any accumulation point $\boldsymbol{\rho}^*$ of the sequence, then $\boldsymbol{\rho}^*$ is a KKT stationary point of the original optimization problem.
\end{theorem}
\begin{proof}
The full derivation, including the required assumptions on the iterates and objective function, is provided in Appendix~\ref{app:converge}. This proof largely follows commonly established convergence analyses for PGD approaches \cite{boyd2004convex}.
\end{proof}

Based on the convergence analysis, we propose a dynamic step size assignment for the algorithm based on a local estimate of the Lipschitz constant for the objective function, obtained by tracking the gradients of the objective function \cite{boyd2004convex}:

\begin{equation}
    \alpha_{t} = \frac{\|\boldsymbol{\rho}_t - \boldsymbol{\rho}_{t-1}\|_2}{\|\nabla f(\boldsymbol{\rho}_t) - \nabla f(\boldsymbol{\rho}_{t-1})\|_2}
\end{equation}

For the first step, we choose $\alpha_0=\frac{\alpha_{\text{fallback}}}{\|\nabla f(\boldsymbol{\rho}_0)\|_{\infty}}$ so as to cap the maximum change in design variables in the first step to $\alpha_{\text{fallback}}$. This can be adjusted based on the requirements of the problem, the problem's behavior, and the design variable bounds. For our results and experiments, we use a default value of $\alpha_{\text{fallback}}=0.2$.
Over the entire process, we set an overall upper bound on the step size, denoted as $\alpha_{\text{max}}$, which brings us to the base step size:

\begin{equation}
    \label{eqn:alpha}
    \alpha_{t} = \min\left(\frac{\|\boldsymbol{\rho}_t - \boldsymbol{\rho}_{t-1}\|_2}{\|\nabla f(\boldsymbol{\rho}_t) - \nabla f(\boldsymbol{\rho}_{t-1})\|_2},\alpha_{\text{max}}\right).
\end{equation}

\subsection{Algorithm Components}
\label{section:algo}
In this section, we detail the components of the approach: the full dynamic step size, the search direction, and the safeguards used to keep the iterations stable.

\subsubsection{Spectral Dynamic Step Size}
Although a local Lipschitz estimation can be effective in most stable problems, it is well known that in PGD approaches in nonlinear optimization, spectral methods are effective at improving convergence and overall stability of the algorithm~\cite{JSSv060i03}. Given this, we further adapt the dynamic step size in our algorithm by incorporating a relatively simple spectral step size adjustment as proposed by \citet{JSSv060i03}. Let $\mathbf{s}_{t+1}=\boldsymbol{\rho}_t - \boldsymbol{\rho}_{t-1}$ and $\mathbf{y}_{t+1} = \nabla f(\boldsymbol{\rho}_{t}) - \nabla f(\boldsymbol{\rho}_{t-1})$. When $\mathbf{s}_{t+1}^T\mathbf{y}_{t+1}\leq 0$, we set $\alpha_{t+1}$ based on (\ref{eqn:alpha}), and when this does not hold we adjust the step size by setting $\alpha_{t+1}=\frac{\mathbf{s}_{t+1}^T\mathbf{s}_{t+1}}{\mathbf{s}_{t+1}^T\mathbf{y}_{t+1}}$. We also make sure that this value does not exceed either $\alpha_{\text{max}}$ or $\frac{2}{L_f}$, since step sizes above $\frac{2}{L_f}$ lack convergence guarantees (Theorem \ref{thm:convergence}). This brings us to the final spectral step size expression:

\begin{equation}
    \label{eqn:final_alpha}
    \hat\alpha_{t} = 
        \begin{cases}
            \min\left(\frac{\|\boldsymbol{\rho}_t - \boldsymbol{\rho}_{t-1}\|_2}{\|\nabla f(\boldsymbol{\rho}_t) - \nabla f(\boldsymbol{\rho}_{t-1})\|_2},\alpha_{\text{max}}\right)&\text{if} \;\;\mathbf{s}_{t+1}^T\mathbf{y}_{t+1}\leq 0\\
            \min\left(\frac{\mathbf{s}_{t+1}^T\mathbf{s}_{t+1}}{\mathbf{s}_{t+1}^T\mathbf{y}_{t+1}},\frac{2\times\|\boldsymbol{\rho}_t - \boldsymbol{\rho}_{t-1}\|_2}{\|\nabla f(\boldsymbol{\rho}_t) - \nabla f(\boldsymbol{\rho}_{t-1})\|_2},\alpha_{\text{max}}\right)& \text{if}\;\;\mathbf{s}_{t+1}^T\mathbf{y}_{t+1}> 0
        \end{cases}
\end{equation}

This is the final expression for the dynamic step size we use in our approach. Note that in practice, rather than checking $\mathbf{s}_{t+1}^T\mathbf{y}_{t+1}> 0$, we check for $\mathbf{s}_{t+1}^T\mathbf{y}_{t+1}> \epsilon_{\alpha}$ and use $\epsilon_{\alpha}=10^{-6}$.

\subsubsection{Search Directions and Projection Independence from Gradients}
The proposed PGD framework separates the projection operation from the gradient computation. This independence allows the search direction to be replaced with higher-order or quasi-Newton updates, such as BFGS, when appropriate. However, since topology optimization problems typically involve very large and sparse Hessians, we employ a nonlinear conjugate gradient approach, which avoids explicit Hessian computation. 
Specifically, we adopt the popular conjugate-gradient scheme proposed by \citet{M2AN_1969__3_1_35_0}, with a restart rule, updating the search direction as follows:

\begin{equation}
\begin{split}
    \mathbf{d}_0 =& -\nabla f(\boldsymbol{\rho}_0)\\
    \beta_t =&\max\left( \frac{\nabla f(\boldsymbol{\rho}_t)^T(\nabla f(\boldsymbol{\rho}_t)-\nabla f(\boldsymbol{\rho}_{t-1}))}{\|\nabla f(\boldsymbol{\rho}_{t-1})\|_2^2},0\right)\\
    \mathbf{d}_t =& -\nabla f(\boldsymbol{\rho}_t) + \beta_t\,\mathbf{d}_{t-1}
\end{split}
\label{eqn:CG}
\end{equation}

These search directions provide faster convergence than raw gradient descent while maintaining low computational cost.

\subsubsection{Ensuring Convergence}
Theorem \ref{thm:convergence} guarantees convergence at the stated rate only if the iterates remain feasible across iterations. Thus, to ensure stable convergence, we revert to $\alpha_{t+1}=\frac{\alpha_{\text{fallback}}}{\|\nabla f\|_{\infty}}$ when an iterate's constraint violation exceeds the Newton projection tolerance. Because early iterations naturally violate feasibility due to large steps, we apply this heuristic only after the first $t_{\text{warmup}}=50$ steps. Moreover, to simplify stability tuning, we include a global relaxation factor $\omega$ (default $1.0$, i.e., no relaxation) that scales all step sizes for ill-conditioned problems requiring smaller steps. This appears in Algorithm \ref{algo:full_pgd} below.

\subsection{The Full PGD Algorithm for TO}
In this section, we present the complete algorithm we use in our experiments, with all of the different aspects of the approach put together.
\begin{algorithm}[H]
\caption{Full PGD Algorithm for Topology Optimization}\label{algo:full_pgd}
\begin{algorithmic}[1]
\Require  $\boldsymbol{\rho}_0 \in [l, u]^N$, $\text{tol}_B$, $\text{tol}_N$, $\text{tol}$, $C$, $l$, $u$, $K_{\max}$, $\alpha_{\text{max}}$, $\alpha_{\text{fallback}}$, $t_{\text{warmup}}$, $\omega$
\Ensure Optimized design variable $\boldsymbol{\rho}^*$
\State Set step size $\alpha_0 = \min\left(\alpha_{\max}, \frac{\alpha_{\text{fallback}}}{\|\nabla f(\boldsymbol{\rho}_0)\|_\infty} \right)$
\For{$t = 0$ to $K_{\max}-1$}
    \State Compute $f(\boldsymbol{\rho}_t)$, $\mathbf{g}(\boldsymbol{\rho}_t)$, $\nabla \mathbf{g}(\boldsymbol{\rho}_t)$, $\nabla f(\boldsymbol{\rho}_{t})$
    \If{($t\geq t_{\text{warmup}}$ \textbf{and} $\|\max(\mathbf{g}(\boldsymbol{\rho}_t)-\mathbf{G},\mathbf{0})\|_\infty >\text{tol}_N$) \textbf{or} $t=0$}
        \State $\alpha_{t}=\min\left(\alpha_{\max}, \frac{\alpha_{\text{fallback}}}{\|\nabla f(\boldsymbol{\rho}_t)\|_\infty} \right)$
        
    \Else
        \State Compute $\alpha_{t}$ based on Equation \ref{eqn:final_alpha}
    \EndIf
    \State Compute $\mathbf{d}_t$ from Equation \ref{eqn:CG} \Comment{Alternates like other CG, BFGS, etc. possible}
    \State $\tilde{\boldsymbol{\rho}} = \boldsymbol{\rho}_t + \omega\,\alpha_{t} \ \mathbf{d}_t$
    \If{$m=1$ \textbf{or} constraints are independent}
        \State $\boldsymbol{\delta}_t\leftarrow$ Algorithm \ref{algo:binary}($\tilde{\boldsymbol{\rho}}, \nabla \mathbf{g}, \mathbf{g}, \mathbf{G}, \alpha_t\nabla f, l, u, \text{tol}_B$)
    \Else
        \State $\boldsymbol{\delta}_t \gets$ Algorithm~\ref{algo:general}($\tilde{\boldsymbol{\rho}}, \nabla \mathbf{g}, \mathbf{g}, \mathbf{G}, \alpha_t\nabla f, l, u, C, \text{tol}_B, \text{tol}_N, \text{maxiter}$)
    \EndIf
    \State $\boldsymbol{\rho}_{t+1} = \tilde{\boldsymbol{\rho}} + \boldsymbol{\delta}_t$
    
    \If{$\frac{\|\boldsymbol{\rho}_{t+1}-\boldsymbol{\rho}_{t}\|_2}{\|\boldsymbol{\rho}_{t+1}\|_2}\leq \text{tol}$}
        \State \Return $\boldsymbol{\rho}_{t+1}$
    \EndIf
\EndFor
\State \Return $\boldsymbol{\rho}_{K_{\max}}$
\end{algorithmic}
\end{algorithm}
Algorithm~\ref{algo:full_pgd} summarizes the complete PGD framework for topology optimization. The algorithm integrates the proposed projection solvers, spectral step-size adaptation, and nonlinear conjugate-gradient directions into a single optimization loop. While the current implementation employs a basic descent update with adaptive step sizing, more advanced update rules incorporating momentum or higher-order information can be seamlessly integrated. 
 We will briefly discuss this in experiments, but keep the simpler scheme above as our main algorithm and leave further development of more advanced update schemes to future work.
In our experiments, we use default parameters $C = 10^{12}$, $\alpha_{\text{max}} = 10^2$, $\text{tol}_B = 10^{-8}$, and $\text{tol}_N = 10^{-6}$.

In the section that follows, we will perform numerical experiments on standard TO benchmark problems, and compare our method with the algorithms most widely used in the literature in terms of both computational efficiency and convergence behavior.

\section{Numerical Verification and Experiments}
In this section, we evaluate the proposed PGD optimizer against the widely used MMA, which serves as the standard solver in many TO applications. We compare convergence behavior and computational efficiency, showing that the proposed PGD algorithm achieves convergence rates comparable to MMA while offering substantially improved computational performance.

To measure efficiency, we report wall-clock time per iteration as our primary metric, as the advantages of parallelization in the proposed PGD projection solver and implementation-specific details are not easily captured by FLOP counts or memory operations. For completeness, we also benchmark the OC method in single-constraint problems to demonstrate that PGD matches OC in computational cost while achieving convergence trends similar to MMA—highlighting its balance between simplicity and scalability.

We conduct four classes of benchmark experiments:

\begin{itemize}
    \item \textbf{Single Linear Constraint: } We run the typical minimum compliance problem with a volume fraction constraint. This problem tests how each algorithm performs on problems with a \textbf{single} constraint, which is \textbf{linear}. In this scenario, the projection is exact, and the projection reduces to the binary-search special case.
    \item \textbf{Single Nonlinear Constraint: } In this scenario, we run the inverted problem; that is, we minimize volume under a compliance constraint. This is a special case where the projection will not be exact with the linear assumption, while the single constraint benefits remain.
    \item \textbf{Multiple Independent Constraints: } In this case, we run the minimum compliance problem, but with four different materials. Each material has a different Young's modulus, and the volume fraction for each material is separately constrained. This scenario highlights the scalability of PGD when constraints are independent and the projection subproblem can be solved by binary search.
    \item \textbf{General Multiple Constraints: } Finally, we run an example of a fully general TO problem with minimum compliance as the objective, but with multiple dependent constraints. In this scenario, we impose the volume-fraction requirement and require the center of mass of the design to lie within a desired region. The latter constraint is nonlinear and not independent of the volume constraint, which represents a case where multiple constraints are present. This is a scenario where the regularized projection and semi-smooth Newton method are tested in PGD.
\end{itemize}

In each of these scenarios, we employ the popular Solid Isotropic Material with Penalization (SIMP) approach for dynamically adjusting the Young's modulus of each element in the mesh. We do not detail this in the paper for brevity and refer interested readers to the work by \citet{Bendsoe2011-tk} for more details. Suffice it to say that the FEA for the underlying problem yields a system of equations $\mathbf{K}(\boldsymbol{\rho})\mathbf{u}=\mathbf{f}$. Moreover, we use the same FEA solver, which we include with our publicly available code, for all optimizers, and use a density filter with a filter radius of $1.5$ elements (see \cite{Bendsoe2011-tk}) for the SIMP penalization scheme. For all problems, we run each optimizer for 300 steps with a SIMP penalty of $3$ and report our results for each experiment. Finally, we run all problems on regular grid meshes with resolutions $128\times64$, $256\times128$, and $512\times256$, to measure how well each approach scales with the number of design variables.

In all cases, we adopt the standard cantilever beam configuration for boundary conditions and loading, as illustrated in Fig.~\ref{fig:problems}.

\subsection{The Benchmark Problems}
Before presenting the results, we first formalize the specific optimization problems used in our benchmarks. Each of the four test cases corresponds to one of the problem classes described in the previous section, designed to assess solver performance across varying levels of nonlinearity, coupling, and constraint complexity. The formulations of these benchmark problems are summarized below, and the corresponding results are discussed in the following section.

\subsubsection{Volume-Constrained Minimum Compliance}
This problem is an example of a single linear constraint TO problem. The compliance objective is defined as $\mathbf{f}^T\mathbf{K}^{-1}(\boldsymbol{\rho})\mathbf{f}$. For a mesh with $N$ elements, the minimum compliance problem with volume constraint can be formulated as:

\begin{equation}
    \label{eqn:mincomp}
    \begin{split}
        \min_{\boldsymbol{\rho}\in R^N}&\quad \mathbf{f}^T\mathbf{K}^{-1}(\boldsymbol{\rho})\mathbf{f} \\
        \text{s.t.}& \quad 0\leq \rho_i\leq 1, \quad \forall i \in [N]\\ 
        &\quad \frac{1}{N}\sum_{i=1}^N\rho_i\leq V_f,
    \end{split}
\end{equation}

where $V_f$ is the target volume fraction. Besides the bounds, a single linear constraint remains. In these experiments, we initialize the optimizers with $\boldsymbol{\rho}=\mathbf{1}$, and set the target volume fraction to $0.2$.

\subsubsection{Compliance-Constrained Minimum Volume}
In this case, we simply invert the constraint and objective of the problem above:

\begin{equation}
    \label{eqn:mincomp}
    \begin{split}
        \min_{\boldsymbol{\rho}\in R^N}&\quad  \frac{1}{N}\sum_{i=1}^N\rho_i\\
        \text{s.t.}& \quad 0\leq \rho_i\leq 1, \quad \forall i \in [N]\\ 
        &\quad \mathbf{f}^T\mathbf{K}^{-1}(\boldsymbol{\rho})\mathbf{f}\leq C_{\text{max}},
    \end{split}
\end{equation}

where $C_{\text{max}}$ is the target maximum allowable compliance. Here we have a single nonlinear constraint. In these problems, we set a compliance limit of $C_{\text{max}} = 150$, assuming a Young's modulus of $1.0$ and a force magnitude of $1.0$.

\subsubsection{Multi-Material Volume-Constrained Minimum Compliance}
In these problems, we use multiple materials with distinct Young's moduli, and with separate volume constraints independently applied to each material. To model multiple materials, we use the penalization scheme proposed by \citet{app14020657} with the same filtering and penalty schedules as we use for SIMP. The formulation is similar to that of the previous problem, except that each material now has its own volume constraint. For a mesh with $N$ elements, and a problem with $P$ materials, the problem can be  formulated as:

\begin{equation}
    \label{eqn:mincomp}
    \begin{split}
        \min_{\boldsymbol{\rho}\in R^N}&\quad \mathbf{f}^T\mathbf{K}^{-1}(\boldsymbol{\rho})\mathbf{f} \\
        \text{s.t.}& \quad 0\leq \rho_i\leq 1, \quad \forall i \in [N]\\ 
        &\quad \frac{1}{N}\sum_{i\in M_j}\rho_i\leq V_{j},\quad \forall j\in [P]
    \end{split}
\end{equation}

where $V_j$ is the target volume fraction of material $j$ and $M_j\subset \{1,\ldots, N\}$ is the subset of design variable indices associated with material $j$. This problem has multiple independent constraints ($M_i \cap M_j = \emptyset \quad\text{if}\; i\neq j$) in this problem. In these problems, we solve for four materials with Young's moduli of $1.0,0.5,0.25,0.125$, with each material having a target volume fraction of $0.05$.

\subsubsection{Volume and Weight Distribution Constrained Minimum Compliance}
Finally, to test the general case, we solve a minimum-compliance problem with multiple interdependent constraints. For this example, we solve the minimum compliance problem with a volume constraint, and additionally require the center of mass of the resulting topology to lie within a certain region of space. This problem can be formulated as:

\begin{equation}
    \label{eqn:mincomp}
    \begin{split}
        \min_{\boldsymbol{\rho}\in R^N}&\quad \mathbf{f}^T\mathbf{K}^{-1}(\boldsymbol{\rho})\mathbf{f} \\
        \text{s.t.}& \quad 0\leq \rho_i\leq 1, \quad \forall i \in [N]\\ 
        &\quad \frac{1}{N}\sum_{i=1}^N\rho_i\leq V_f,\\
        &\quad \|\mathbf{R}(\boldsymbol{\rho}) - \mathbf{R}_t\|_2^2 \leq r_b
    \end{split}
\end{equation}

where $V_f$ is the target volume fraction, $\mathbf{R}(\boldsymbol{\rho})$ is the center of mass based on the material distribution, $\mathbf{R}_t$ is the center of the acceptable region, and $r_b$ is the maximum acceptable distance from $\mathbf{R}_t$. Here we have multiple interdependent and nonlinear constraints. This problem thus serves as a representative general TO problem. In these problems, we set the target volume fraction to $0.2$ and set the desired center of mass to $\begin{pmatrix}0.25\\0.25\end{pmatrix}$ on a domain of length $1.0$ along the x-axis and $0.5$ along the y-axis. Moreover, we set the tolerance radius $r_b=10^{-2}$ in these problems (see Fig.~\ref{fig:problems}).

\subsection{Results}
With the test problems established, we now present the results of the experiments in each scenario and compare the three different optimizers we test. For each problem, we briefly describe the exact settings and leave full details to our publicly available code for brevity. Below, we present the results of running PGD (with and without the independent constraint assumption for the multi-material case), MMA, and OC (OC is run only on the single-material and multi-material minimum-compliance problems).

\subsubsection{Solution Time and Scalability}
\begin{figure}[h]
    \centering
    \includegraphics[width=0.85\linewidth]{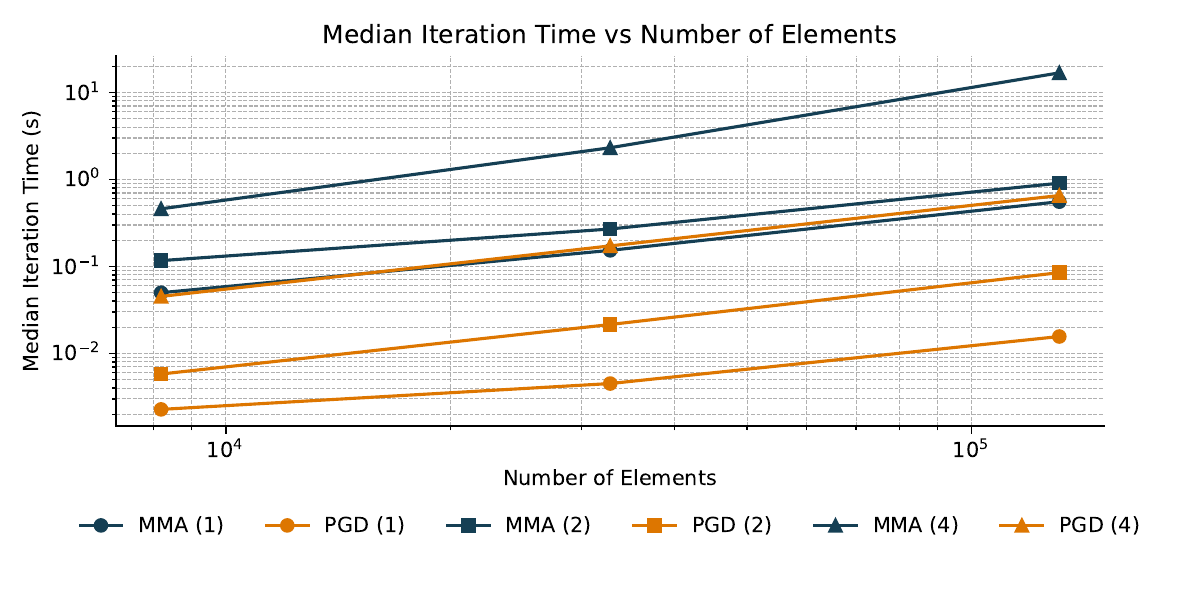}
    \caption{Scaling of iteration time with problem size and number of constraints.
    PGD achieves roughly an order-of-magnitude improvement in iteration time across all constraint counts and scales almost log-linearly with the number of elements. Notably, PGD with four constraints matches the iteration time of MMA with a single constraint, while MMA’s cost rises sharply as the number of constraints increases from two to four, indicating better scalability of the proposed approach.}
    \label{fig:iteration}
\end{figure}

Before discussing convergence and constraint handling in each case, we first examine the primary motivation for the proposed PGD algorithm: its scalability and computational efficiency relative to methods like MMA. Note that in all experiments, we find that both PGD and MMA handle linear and nonlinear constraints and converge in all cases (we discuss the nuances of this in later sections). With that in mind, we investigate the computational cost and scalability of both PGD and MMA across both the number of design variables and the number of constraints.

\begin{table*}[h]
    \centering
    \resizebox{0.8\textwidth}{!}{
    \begin{tabular}{ccccc}
        \toprule
         &  &  \multicolumn{3}{c}{\textbf{Number of Constraints}}\\
        \cmidrule(lr){3-5} 
        \textbf{Optimizer}&\textbf{  Variable Count}&  1&  2&  4\\
        \midrule
        \multirow{3}{*}{\textbf{MMA}} &  \multicolumn{1}{c|}{8192}&  0.0461&  0.1176&  1.883\\
         &  \multicolumn{1}{c|}{32768}&  0.1444&  0.2623&  1.8915\\
         &  \multicolumn{1}{c|}{131072}&  0.6188&  0.9067&  8.4764\\
        \midrule
        \multirow{3}{*}{\textbf{PGD$^{*}$}} &  \multicolumn{1}{c|}{8192}&  0.0022 (\textbf{22.35$\times$})&  0.0053 (\textbf{22.2$\times$})&  0.0437 (\textbf{43.06$\times$})\\
         &  \multicolumn{1}{c|}{32768}&  0.0041 (\textbf{35.08$\times$})&  0.0205 (\textbf{12.77$\times$})&  0.1777 (\textbf{10.64$\times$})\\
         & \multicolumn{1}{c|}{131072}& 0.0146 (\textbf{42.44$\times$})& 0.0799 (\textbf{11.35$\times$})& 0.6534 (\textbf{12.97$\times$})\\
         \midrule
         \multirow{3}{*}{\textbf{PGD}} &  \multicolumn{1}{c|}{8192}&  0.0022 (\textbf{22.35$\times$})&  - &  0.0060 (\textbf{312.11$\times$})\\
         &  \multicolumn{1}{c|}{32768}&  0.0041 (\textbf{35.08$\times$})&  - &  0.0164 (\textbf{115.23$\times$})\\
         & \multicolumn{1}{c|}{131072}& 0.0146 (\textbf{42.44$\times$})& - & 0.0537 (\textbf{157.82$\times$})\\
        \bottomrule
    \end{tabular}
    }
    \caption{Median iteration time (s) across problem sizes and constraint counts. Comparison of MMA, PGD$^{*}$ with Newton-based projection, and PGD with independent-constraint (binary-search) projection. Values in parentheses indicate the speedup of PGD relative to MMA. PGD achieves approximately ten- to forty-fold faster iterations across general problems and over one hundred-fold acceleration when constraints are independent. }
    \label{tab:iterationtime}
\end{table*}

Table \ref{tab:iterationtime} reports the median iteration times of MMA and PGD across problem sizes and numbers of constraints. The problems reported in Table \ref{tab:iterationtime} are minimum compliance (single constraint), volume- and weight-distribution-constrained minimum compliance (two constraints), and multi-material minimum compliance (four constraints). We see that in the single-constraint case, PGD speeds up optimization more than in multi-constraint scenarios. This is expected, since single-constraint problems are solved using only binary search, whereas multi-constraint problems require Newton iterations in some scenarios, as seen in the initial jump in iteration time in Fig.~\ref{fig:iteration}. More notably, in multi-constraint scenarios with independent constraints, the binary search projection solution of PGD is \emph{two orders of magnitude faster}, making it an even more efficient choice for these scenarios. Overall, we see that the PGD approach we propose provides an order-of-magnitude improvement in iteration time over MMA, especially in large-scale scenarios.

\paragraph{The Reason Behind PGD Scalability.}
While both MMA and the proposed PGD are first-order gradient-based optimization methods, PGD exhibits a lower per-iteration computational complexity. This advantage stems fundamentally from the difference between solving a sequential convex approximation subproblem and performing a projected gradient update. Thus, the computational superiority of PGD over MMA in terms of CPU time can be attributed to the arithmetic intensity and convergence properties of their respective dual subproblems in the general case (no assumption of independence of constraints):
\begin{itemize}

    \item \textbf{Operator Complexity and Arithmetic Intensity:}
    The arithmetic operations required to evaluate the primal-dual mapping within the inner loops of MMA and PGD subproblems differ significantly in instruction latency. The primal-dual mapping in MMA is derived from strictly convex rational approximations. Recovering the primal variables $\rho_i$ requires evaluating expressions of the form:
    \begin{equation}
        \rho_i(\lambda) = \frac{\sqrt{P_i(\lambda)} U_i + \sqrt{Q_i(\lambda)} L_i}{\sqrt{P_i(\lambda)} + \sqrt{Q_i(\lambda)}}
    \end{equation}
    This necessitates computationally expensive square-root and floating-point division operations for every design variable $i \in [1, N]$ at every inner iteration. Moreover, the asymptotes themselves have to be stored for all design variables and updated at each iteration, which adds computational cost.
    
    By contrast, the mapping in PGD is a metric projection onto the box constraints:
    \begin{equation}
        \rho_i(\lambda) = \text{clip}_{[l, u]} \left( \tilde{\rho}_i - \sum_{j=1}^m \lambda_j (\nabla g_j)_i \right)
    \end{equation}
    This relies almost exclusively on fused multiply-add (FMA) operations and conditional clipping. On modern hardware, the instruction latency for FMA and vector clipping is significantly lower than that of the division and square root operations required by MMA. Moreover, all of this can be implemented with two BLAS-level vectorized operations.

    \item \textbf{Spectral Properties and Jacobian Assembly:}
    The efficiency of the dual update is further governed by the spectral properties of the dual problem's Jacobian. The rational approximation in MMA introduces global curvature; every design variable $\rho_i$, regardless of its proximity to the bounds, contributes a nonzero \textit{nonlinear} term to the dual Jacobian. Consequently, the assembly cost involves a summation over all $N$ variables at every step. The projection operator, on the other hand, induces a semismooth structure where the Jacobian contribution of any variable $\rho_i$ in the saturated set (active bounds) is exactly zero. The dual system is effectively constructed solely from the free variables (where $l < \rho_i < u$). As topology optimization progresses and the design becomes binary, the size of this free set diminishes, naturally accelerating the subproblem assembly. Although our implementation does not exploit this, doing so could further reduce cost through fewer operations and less memory traffic. Nevertheless, the PGD Jacobian update relies almost exclusively on element-wise vector operations (clipping, addition, and dot products). Unlike MMA's updates and dual approximations, which involve complex (though vectorized) data dependencies, the PGD projection is trivially parallelizable.

    \item \textbf{Interior Point vs. Active Set:}
    The primary computational distinction lies in the solver hierarchy required for the subproblem. This accounts for most of PGD's efficiency gain.

    The MMA subproblem is typically solved using a Primal-Dual Interior Point Method (PDIPM). To handle inequality constraints smoothly, PDIPM relaxes the strict KKT complementarity conditions to perturbed conditions, similar to our projection regularization approach. However, the solver must iterate over a sequence of decreasing allowable perturbation parameters $\epsilon \to 0$ (continuation) until a desired tolerance is reached or the iteration count is limited. For each barrier parameter, a Newton solver is required to resolve the nonlinear system. This results in a multiplicative complexity proportional to $N_{\text{barrier}} \times N_{\text{Newton}}$. Although both rely on $O(m\times m)$ systems, this presents a major computational hurdle, which the local linearization in PGD avoids (at the cost of a lower-precision constraint approximation). PGD employs a projection operator on the tangent space of the constraints. This is mathematically equivalent to an active-set strategy without identifying the active set or using a continuation scheme on the constraint tolerances. The projection strategy identifies the active constraints implicitly through regularization, effectively collapsing the two loops of the interior-point method into a single semismooth Newton iteration. PGD typically converges in a number of steps comparable to a single inner loop of MMA, avoiding the overhead of the continuation scheme entirely via the regularized formulation.

    \item \textbf{Efficiency of Iterations with a Single Active Constraint:} Finally, as we saw in Algorithm \ref{algo:general}, the projection step checks whether any iteration can be solved with one active constraint, which binary search solves in $O(\log N)$ time. This occurs in many intermediate iterations and leads to a much lower per-iteration cost ($O(\log N)$) than Newton iterations. Beyond this, the single-constraint solution may not hold when several constraints are simultaneously active. Thus, Algorithm \ref{algo:general} uses the combined binary search solution as the initialization for projection, effectively ensuring that if constraints are independent, the Newton iterations will quickly converge, since only the slack values $\mathbf{s}$ need to be solved for. By contrast, MMA does not explicitly check for any such scenarios.
\end{itemize}

\subsubsection{Convergence and Robustness}
\begin{figure}[H]
    \centering
    \textbf{Minimum Compliance Problem}
    \parbox[t]{\linewidth}{\quad}
    \parbox[t]{0.49\linewidth}{
    \includegraphics[width=\linewidth]{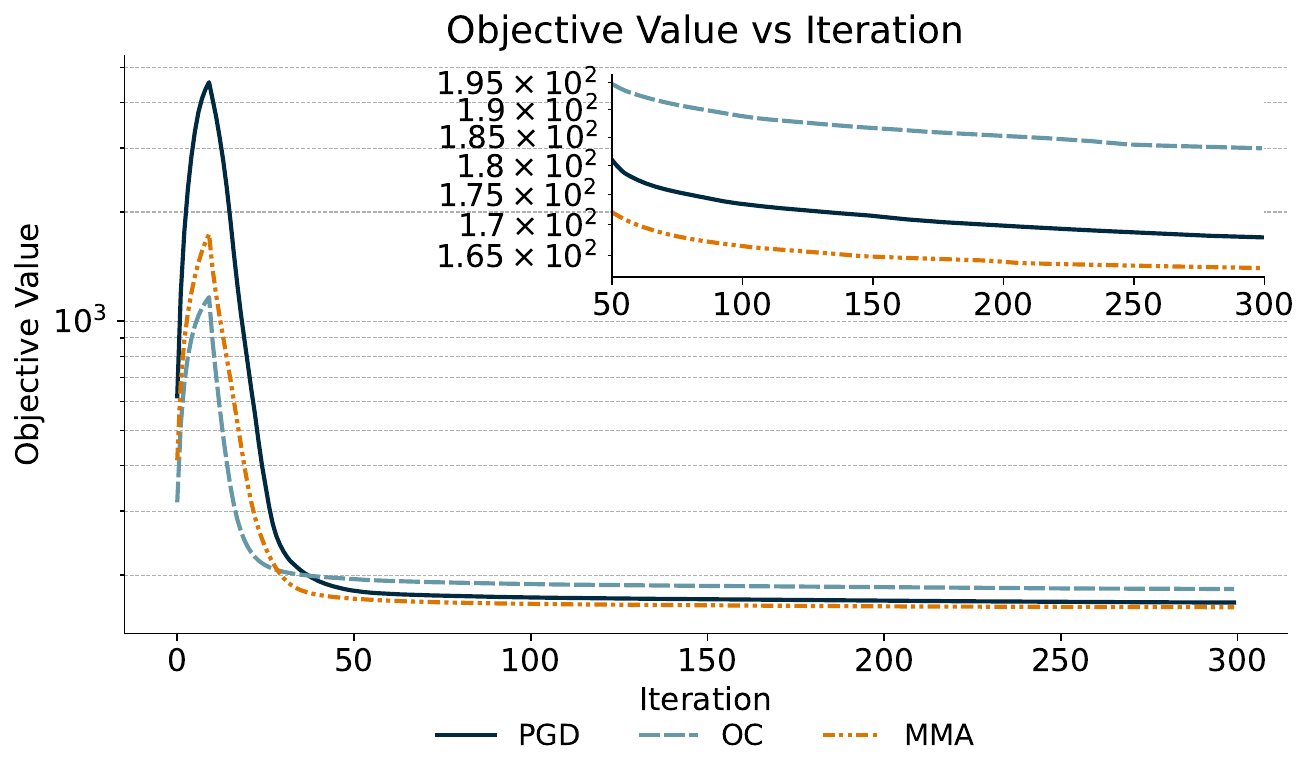}
    }
    \parbox[t]{0.49\linewidth}{
    \includegraphics[width=\linewidth]{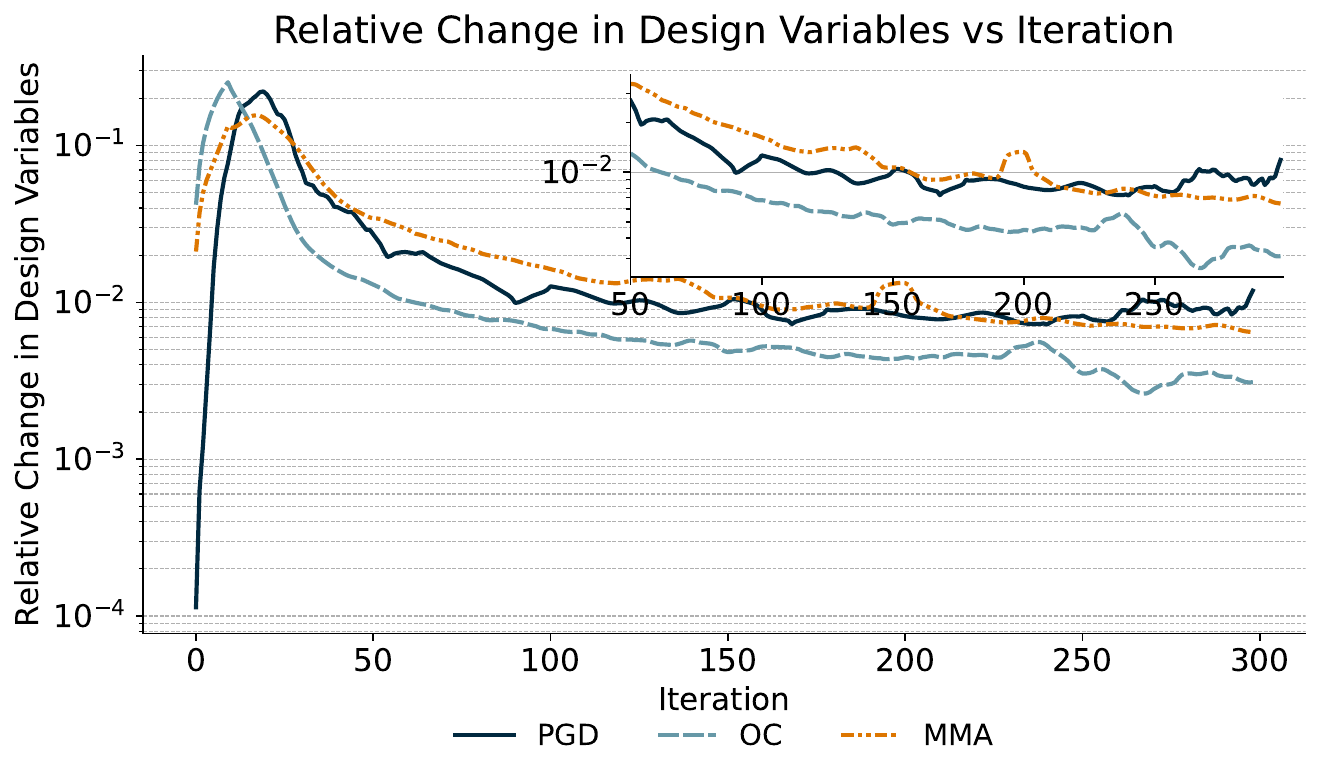}
    }
    \textbf{Compliance Constrained Minimum Volume}
    \parbox[t]{\linewidth}{\quad}
    \parbox[t]{0.49\linewidth}{
    \includegraphics[width=\linewidth]{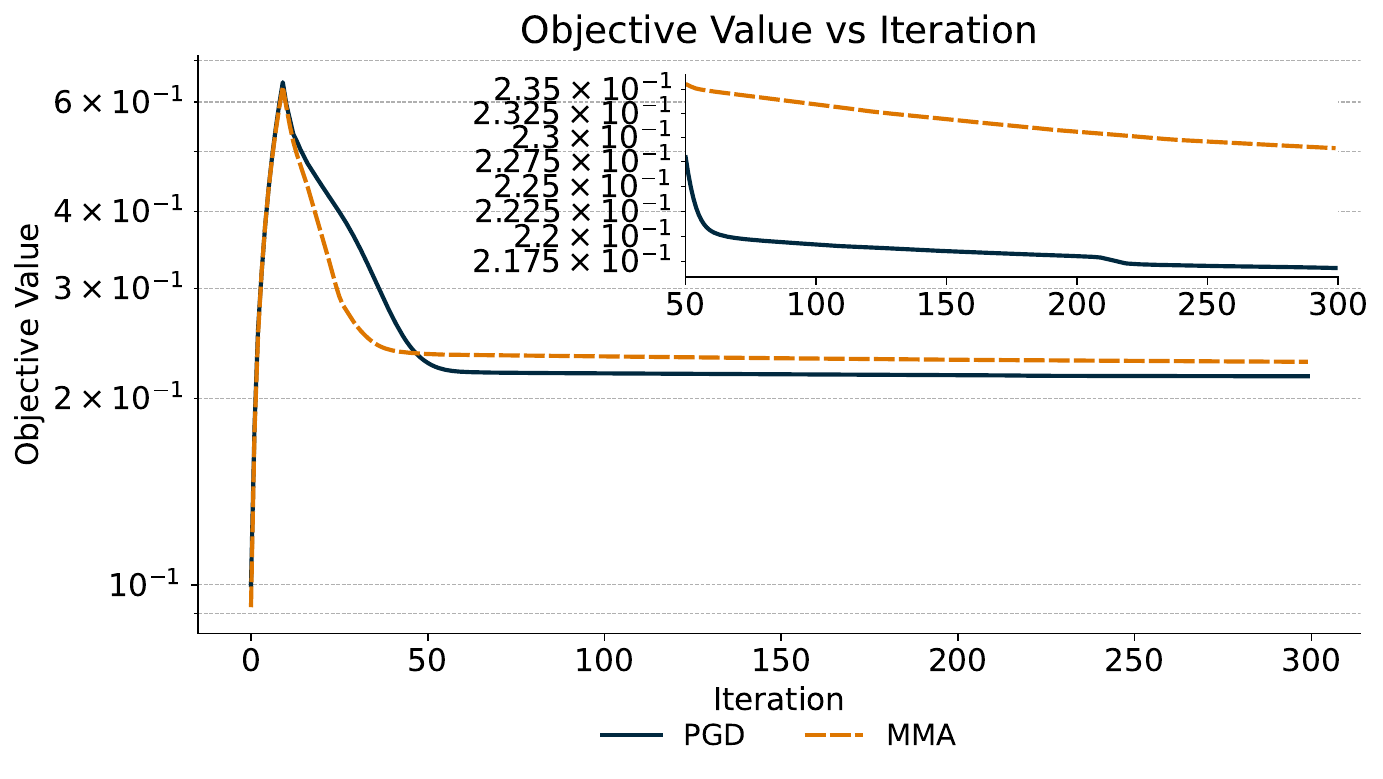}
    }
    \parbox[t]{0.49\linewidth}{
    \includegraphics[width=\linewidth]{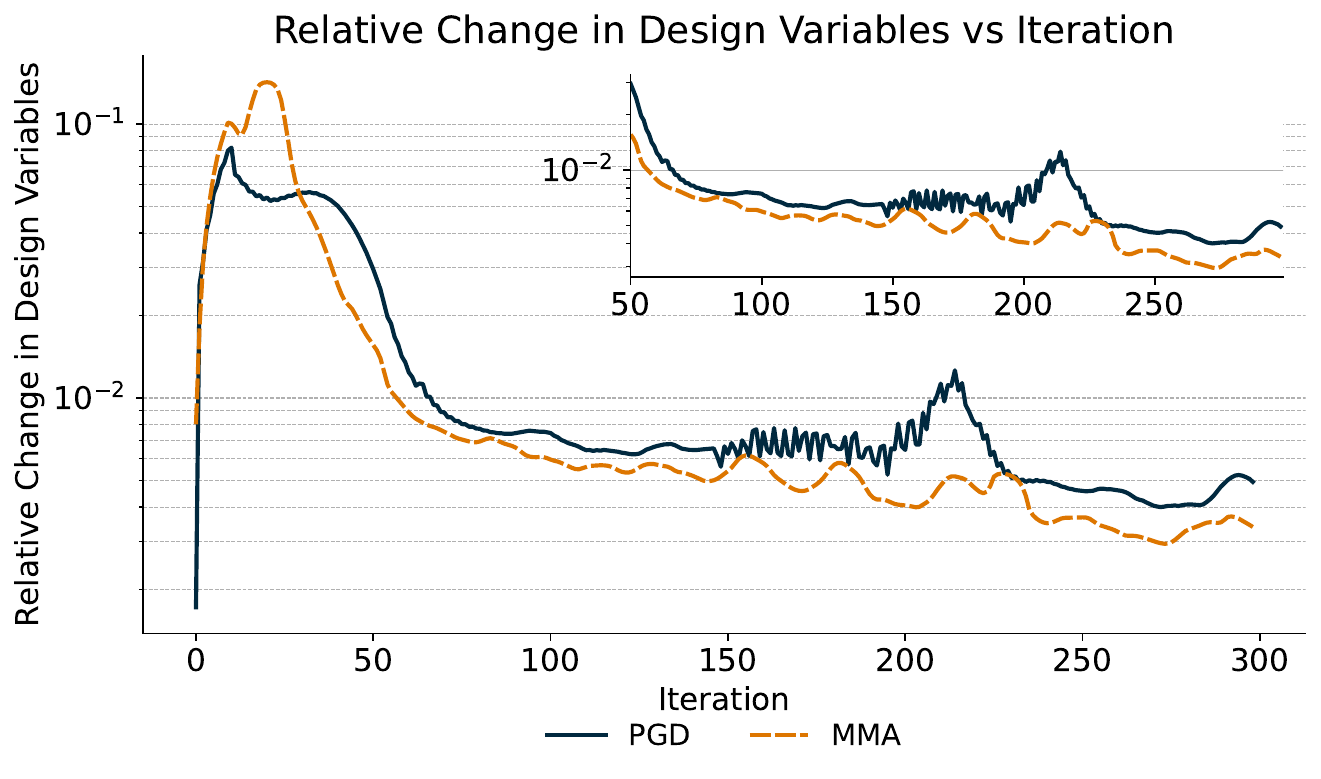}
    }
    \textbf{Multi-Material Minimum Compliance}
    \parbox[t]{\linewidth}{\quad}
    \parbox[t]{0.49\linewidth}{
    \includegraphics[width=\linewidth]{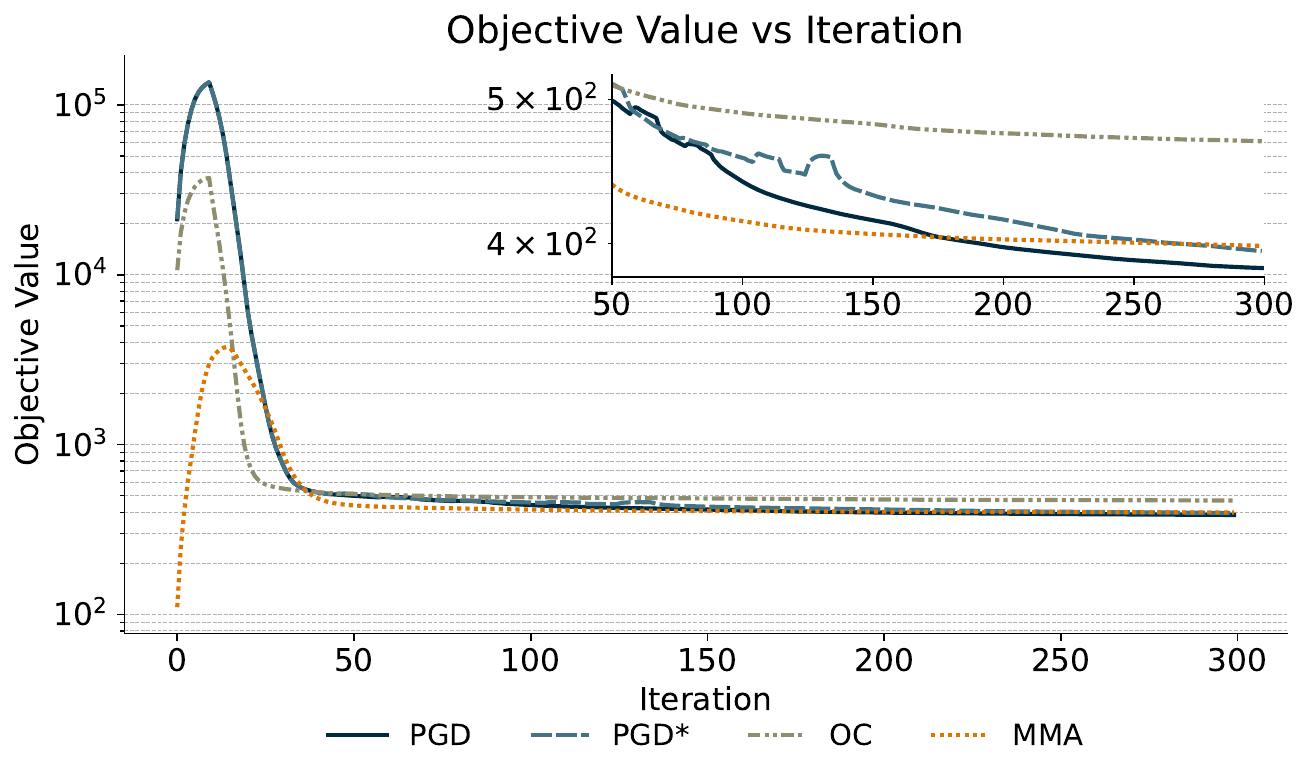}
    }
    \parbox[t]{0.49\linewidth}{
    \includegraphics[width=\linewidth]{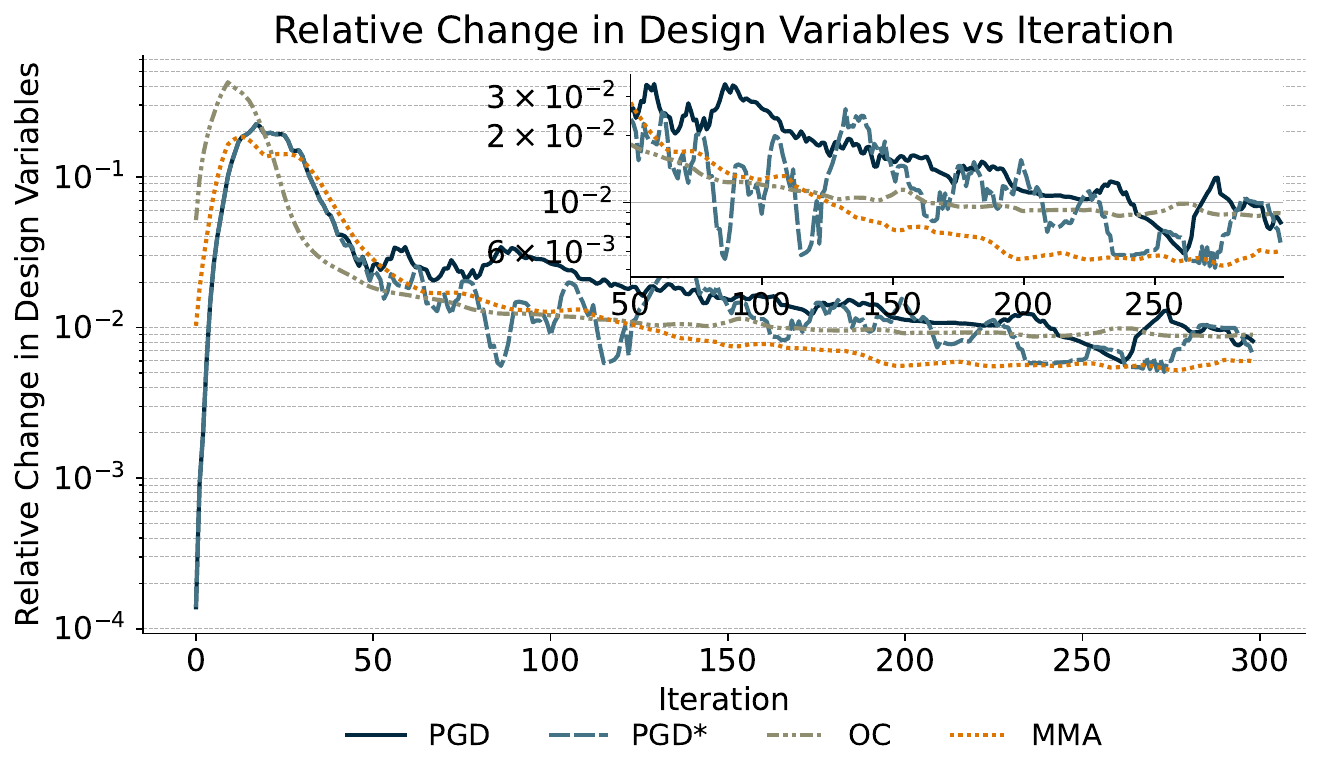}
    }
    \textbf{Volume and Weight Distribution Constrained Minimum Compliance}
    \parbox[t]{\linewidth}{\quad}
    \parbox[t]{0.49\linewidth}{
    \includegraphics[width=\linewidth]{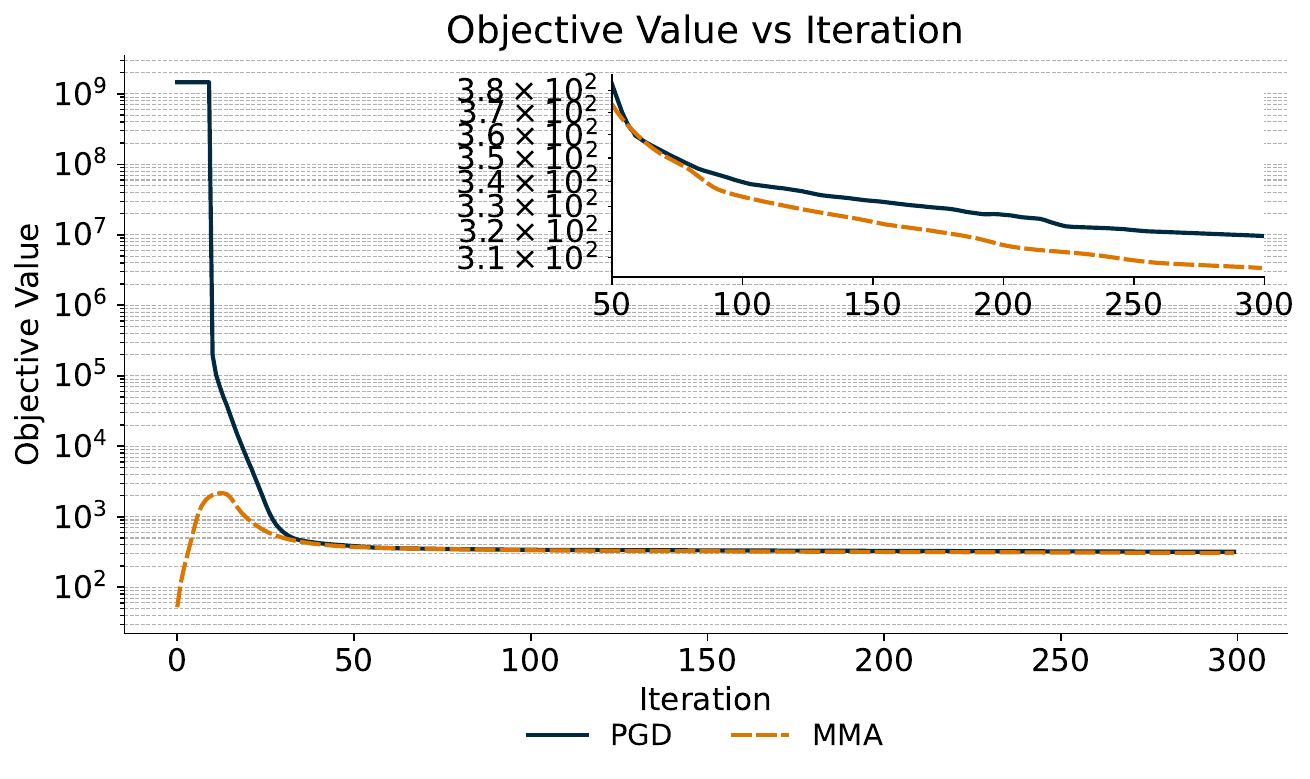}
    }
    \parbox[t]{0.49\linewidth}{
    \includegraphics[width=\linewidth]{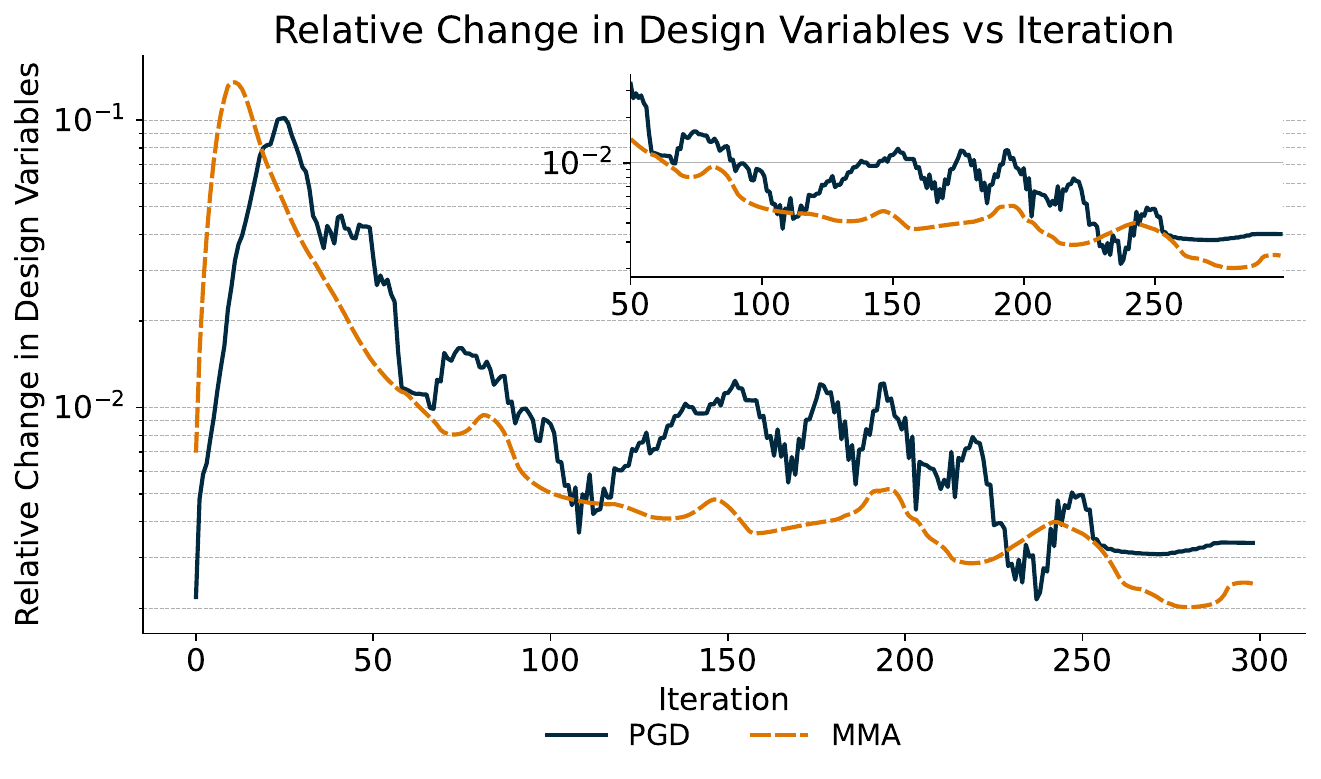}
    }
    \caption{Convergence behavior at fine mesh resolution for each benchmark problem. Each row shows objective evolution (left) and relative design-variable change (right) for single- and multi-constraint problems with both linear and nonlinear constraints. Across all four scenarios, PGD, MMA, and OC exhibit nearly identical convergence trajectories, confirming that the proposed PGD framework is broadly applicable and numerically stable across diverse constraint types. PGD$^*$ denotes PGD using the Newton-based projection instead of binary search in cases with independent constraints; the default PGD employs the faster binary-search projection, and PGD$^*$ is shown for completeness.}
    \label{fig:convergence}
\end{figure}
We provide the full set of results for all experiments in Appendix \ref{app:results}, and only discuss the main findings in this section. Fig.~\ref{fig:convergence} shows that PGD converges stably on problems with both linear and nonlinear constraints, with convergence rates similar to those of MMA and OC. This confirms that the sequential linear approximation of constraints does not destabilize or slow convergence in PGD.

Besides the general stability of solvers, it is important to note the overall convergence behavior of each approach. Qualitatively, when looking at the convergence plots in Fig.~\ref{fig:convergence}, we observe that MMA and PGD in most cases exhibit similar rates of convergence and converge to designs with similar objective values. This suggests that both approaches provide stable convergence, while PGD offers an order-of-magnitude faster iteration time.

\subsubsection{Constraint Handling}
Another essential part of optimizing TO problems is adhering to constraints. In our experiments, we track deviations from the constraint boundary across iterations to assess how accurately each optimizer adheres to the problem constraints. For brevity, we do not include all these figures across problems, and refer interested readers to Appendix \ref{app:results}. In all problems, the optimizers show stable and consistently low constraint violations, with PGD exhibiting better adherence than MMA in most experiments.

\begin{figure}[H]
    \centering
    \textbf{Optimized Topologies}
    \includegraphics[width=\linewidth]{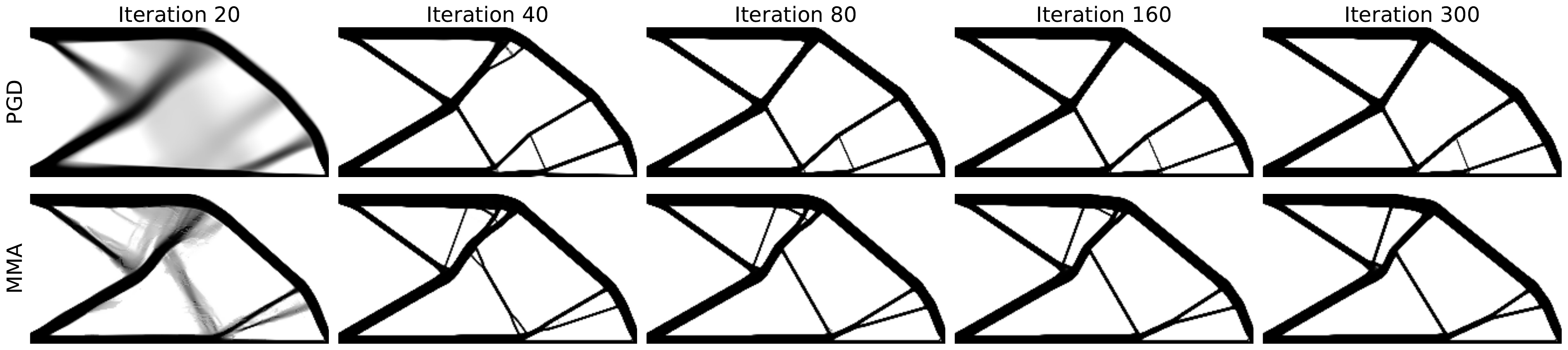}
    \textbf{Constraint Violation Plot}
    \includegraphics[trim={0 0 0 1.8em},clip,width=\linewidth]{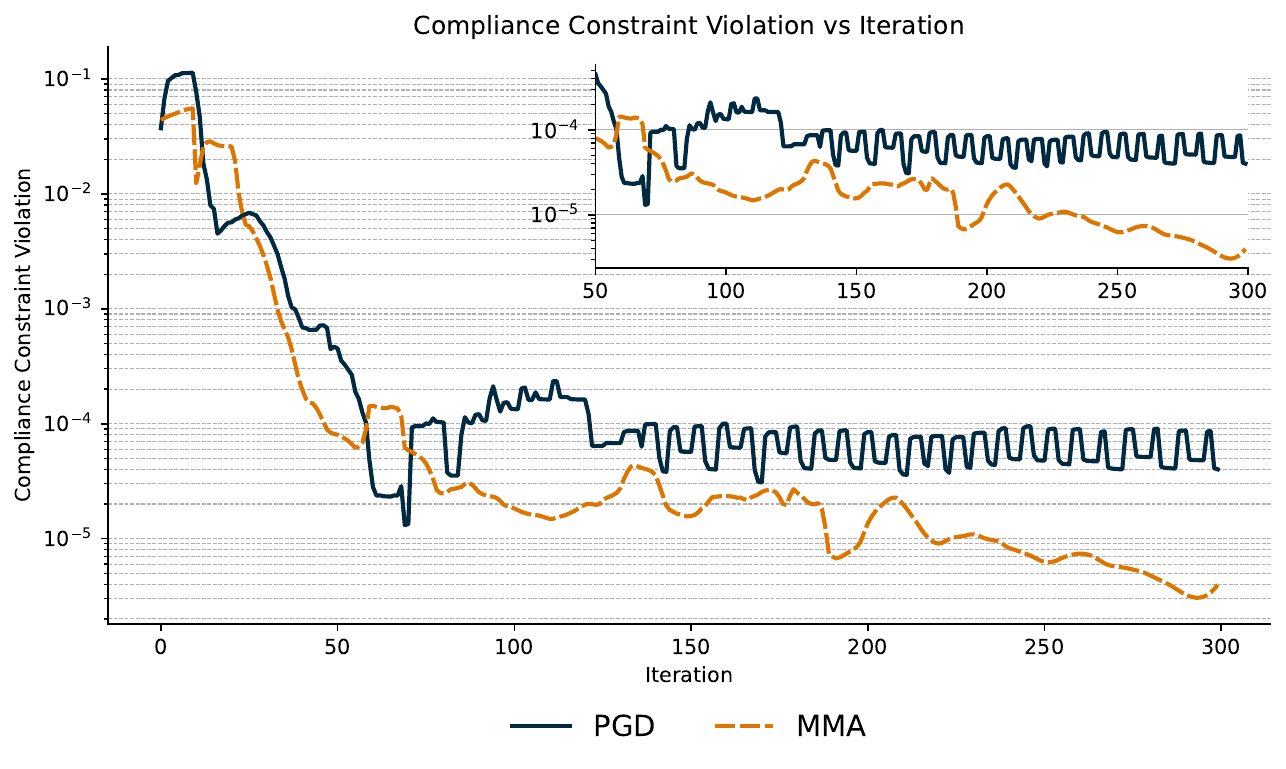}
    \textbf{Convergence Plots}
    \parbox[t]{\linewidth}{\quad}
    \parbox[t]{0.495\linewidth}{
    \includegraphics[width=\linewidth]{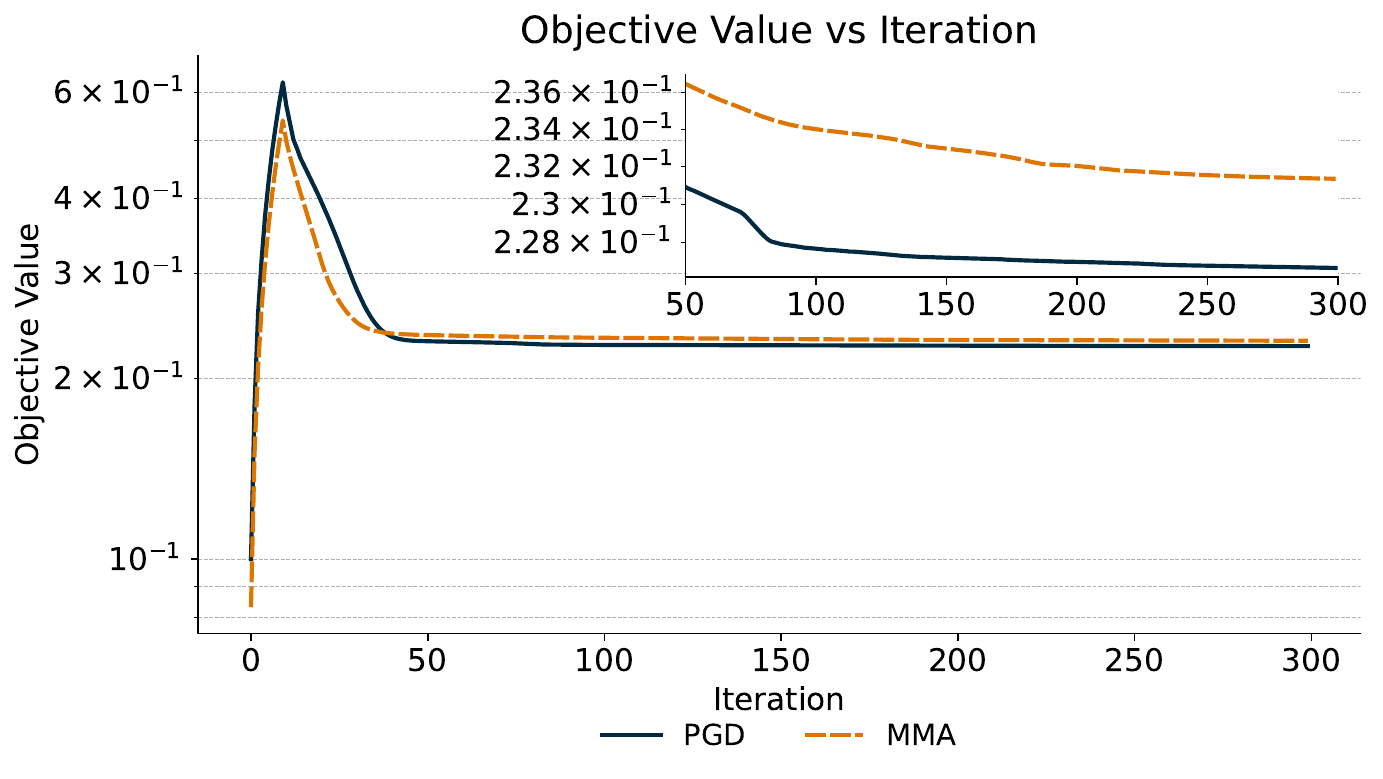}
    }
    \parbox[t]{0.495\linewidth}{
    \includegraphics[width=\linewidth]{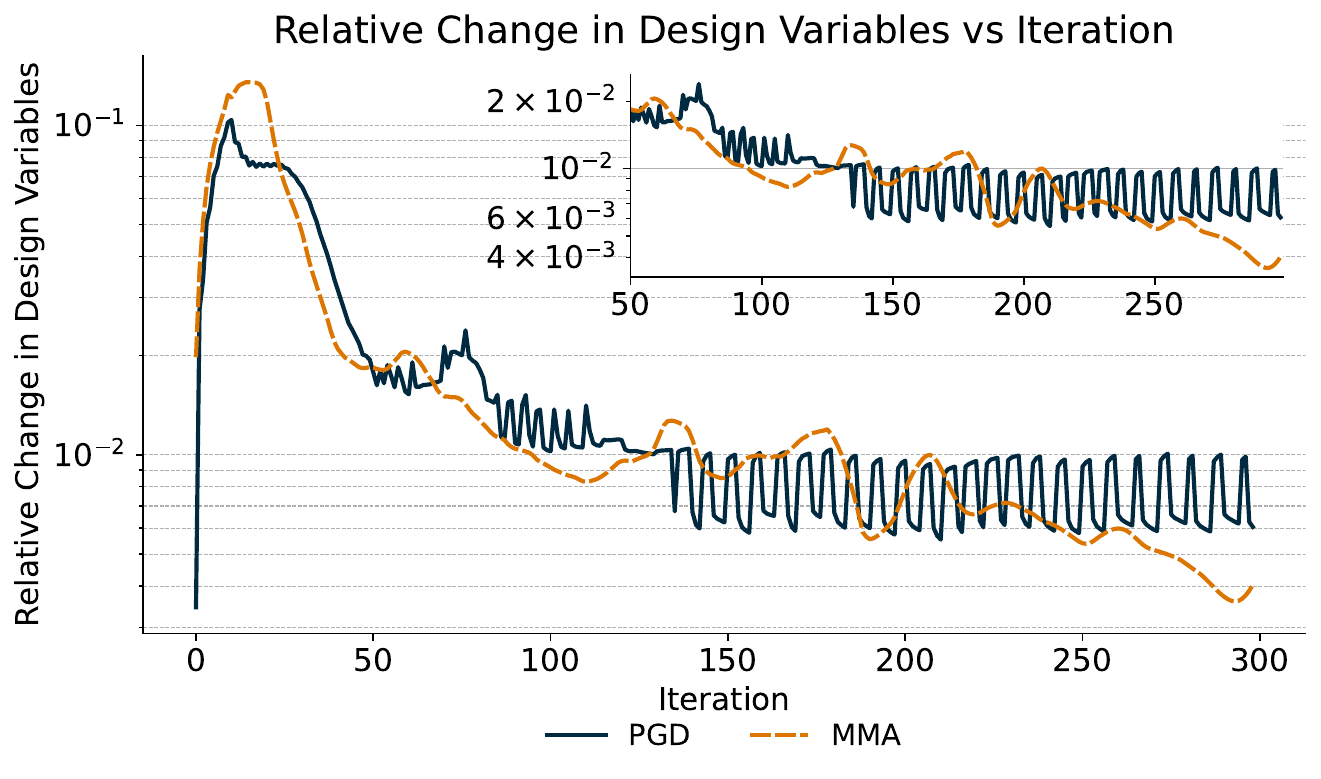}
    }
    \caption{An example of linear approximation of the constraint leading to step size relaxation in our algorithm, keeping the problem stable and convergence behavior robust---as seen visually and in convergence plots---without the need to adjust the global step size relaxation from the start.}
    \label{fig:constraint_oscillation}
\end{figure}

However, in the compliance-constrained minimum volume problem at coarse and medium grid resolutions, the simulation, and hence the compliance, is more sensitive to individual design variables. In this regime the post-warm-up step-size heuristic is triggered for PGD, and the constraint violations and steps become oscillatory. We show an example of this in Fig.~\ref{fig:constraint_oscillation} for the medium resolution run of this problem. Despite the oscillations, we observe that our heuristic-based approach, which relaxes the step sizes when violations are too high, maintains the constraint violation at a low level and does not hinder stable convergence in the objective or design variables. This is evident in the convergence plots shown in Fig.~\ref{fig:constraint_oscillation} as well as the visualization of the designs across iterations. This example illustrates a scenario where sequential linear approximations may become problematic and highlights the importance of incorporating the heuristic step relaxation into the algorithm. Moreover, one could relax all step sizes in the optimizer to stabilize this behavior; however, the inclusion of automatic step relaxation makes the overall algorithm more robust in cases where such oscillations are not known a priori.

\subsubsection{Stress Testing under Multiple Nonlinear Constraints}
\begin{figure}[H]
    \centering
    \includegraphics[width=\linewidth]{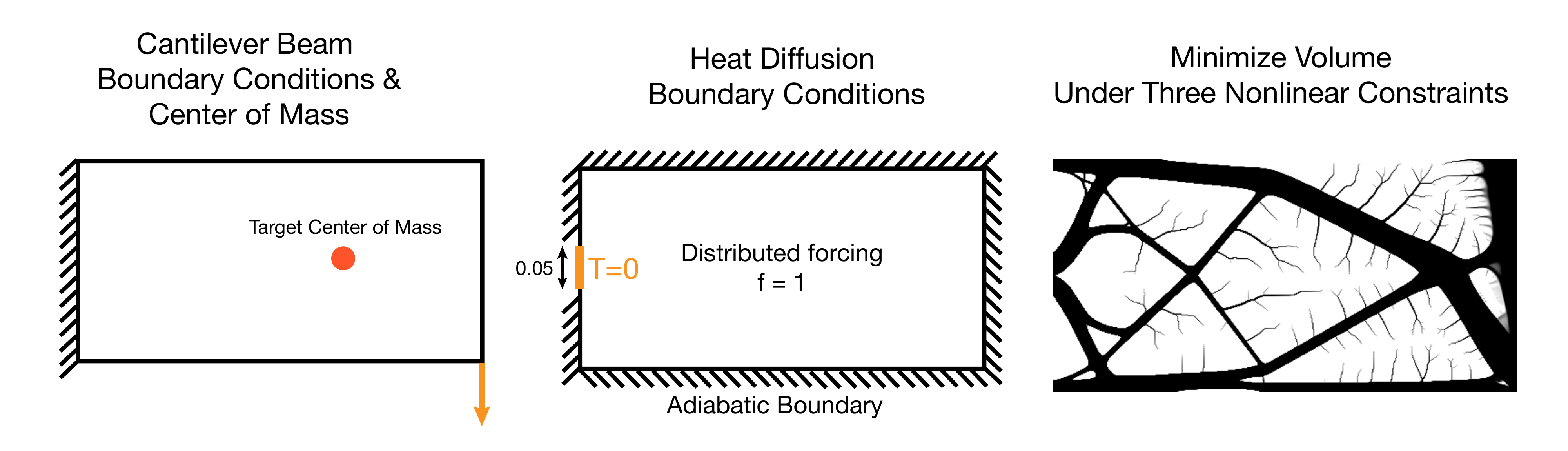}
    \caption{Overview of the stress-test problem with multiple nonlinear constraints.}
    \label{fig:nonlinearexample}
\end{figure}
Finally, beyond the preceding experiments, we evaluate PGD on problems with multiple nonlinear constraints. To do this, we combine three decoupled nonlinearities and solve a substantially more difficult TO problem. This example is designed to stress-test nonlinear optimization, not to represent a practical TO problem.

Specifically, we formulate a minimum volume problem under three different nonlinearities. In this problem, the objective is to minimize the total volume of material, while keeping the center of mass within a desired region and limiting both elastic and thermal compliance. In this way, we combine a geometric nonlinearity with two decoupled physics-driven nonlinearities, to demonstrate how PGD remains stable even in highly nonlinear regimes. The boundary conditions for linear elasticity are identical to the cantilever beam problem we have been focusing on so far, and the thermal compliance boundary conditions are set as shown in Fig.~\ref{fig:nonlinearexample} with a thermal conductivity of $k=1.0$ for the material regions of the topology and a distributed source totaling $1.0$ applied over the entire domain. Overall, this toy problem can be formulated as:

\begin{equation}
    \label{eqn:toynonlinear}
    \begin{split}
        \min_{\boldsymbol{\rho}\in R^N}&\quad  \frac{1}{N}\sum_{i=1}^N\rho_i\\
        \text{s.t.}& \quad 0\leq \rho_i\leq 1, \quad \forall i \in [N]\\ 
        &\quad \mathbf{f}_{\text{LE}}^T\mathbf{K}_{\text{LE}}^{-1}(\boldsymbol{\rho})\mathbf{f}_{\text{LE}}\leq C_{\text{LE,max}}\\ 
        &\quad \mathbf{f}_{\text{TD}}^T\mathbf{K}_{\text{TD}}^{-1}(\boldsymbol{\rho})\mathbf{f}_{\text{TD}}\leq C_{\text{TD,max}}\\
        &\quad \|\mathbf{R}(\boldsymbol{\rho}) - \mathbf{R}_t\|_2^2 \leq r_b,
    \end{split}
\end{equation}

where the subscripts ${\text{LE}}$ and ${\text{TD}}$ denote the linear-elasticity and thermal-diffusion FEA systems. In this problem, we set compliance limits of $C_{\text{LE,max}} = 150.0$ for linear elasticity and $C_{\text{TD,max}}=5.0$ for thermal diffusion, and set the target center of mass to be  $\begin{pmatrix}0.55\\0.25\end{pmatrix}$ with a tolerance radius $r_b=0.01$. We then run both MMA and PGD on this problem for 300 steps. We run this experiment only on the fine mesh setting of $512 \times 256$.

\begin{figure}[h]
    \centering
    \textbf{Optimized Topologies}
    \includegraphics[width=\linewidth]{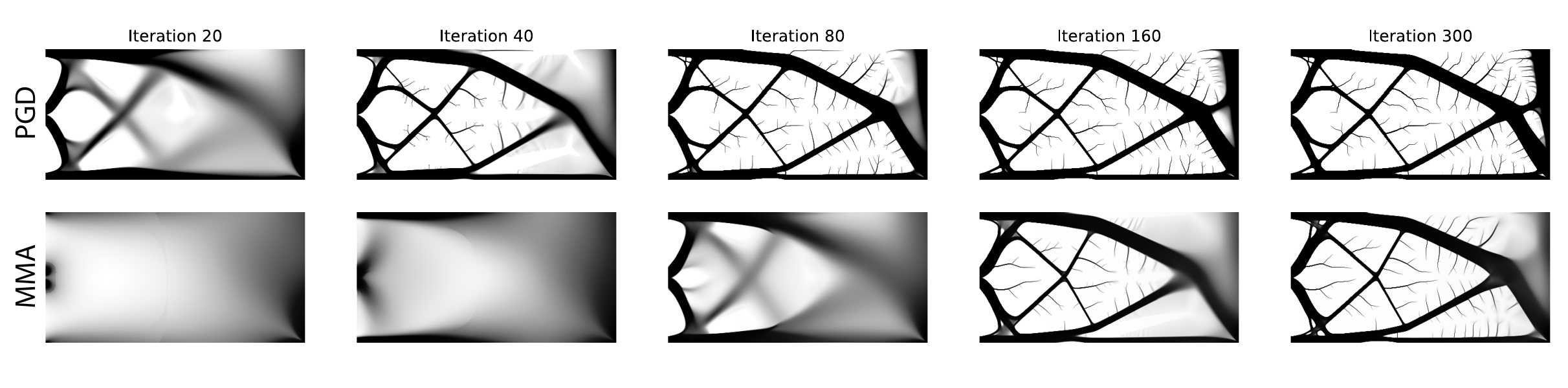}
    \textbf{Constraint Violation Plots}
    \parbox[t]{\linewidth}{\quad}
    \parbox[t]{0.495\linewidth}{\includegraphics[trim={0 0 0 2.8em},clip,width=\linewidth]{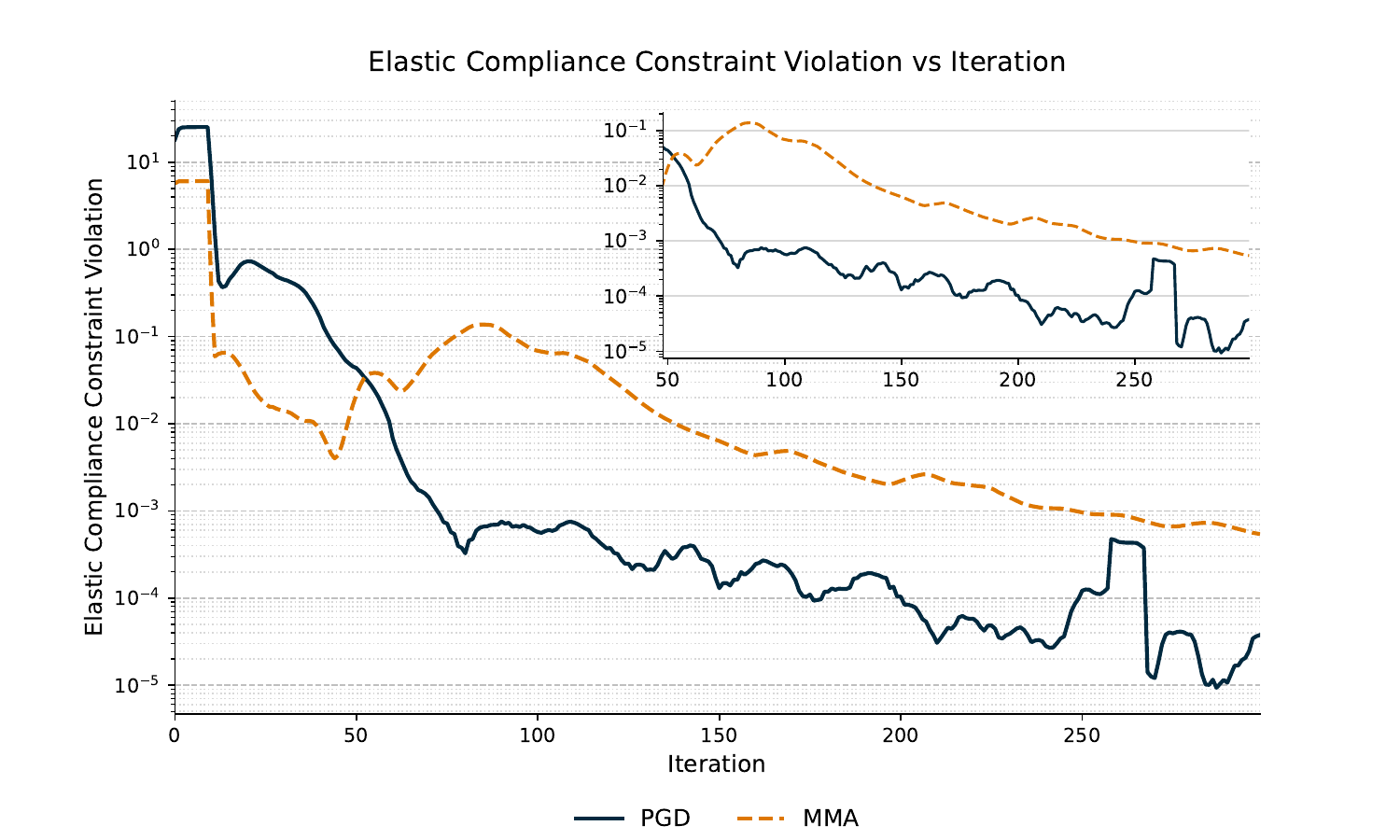}}
    \parbox[t]{0.495\linewidth}{\includegraphics[trim={0 0 0 1.8em},clip,width=\linewidth]{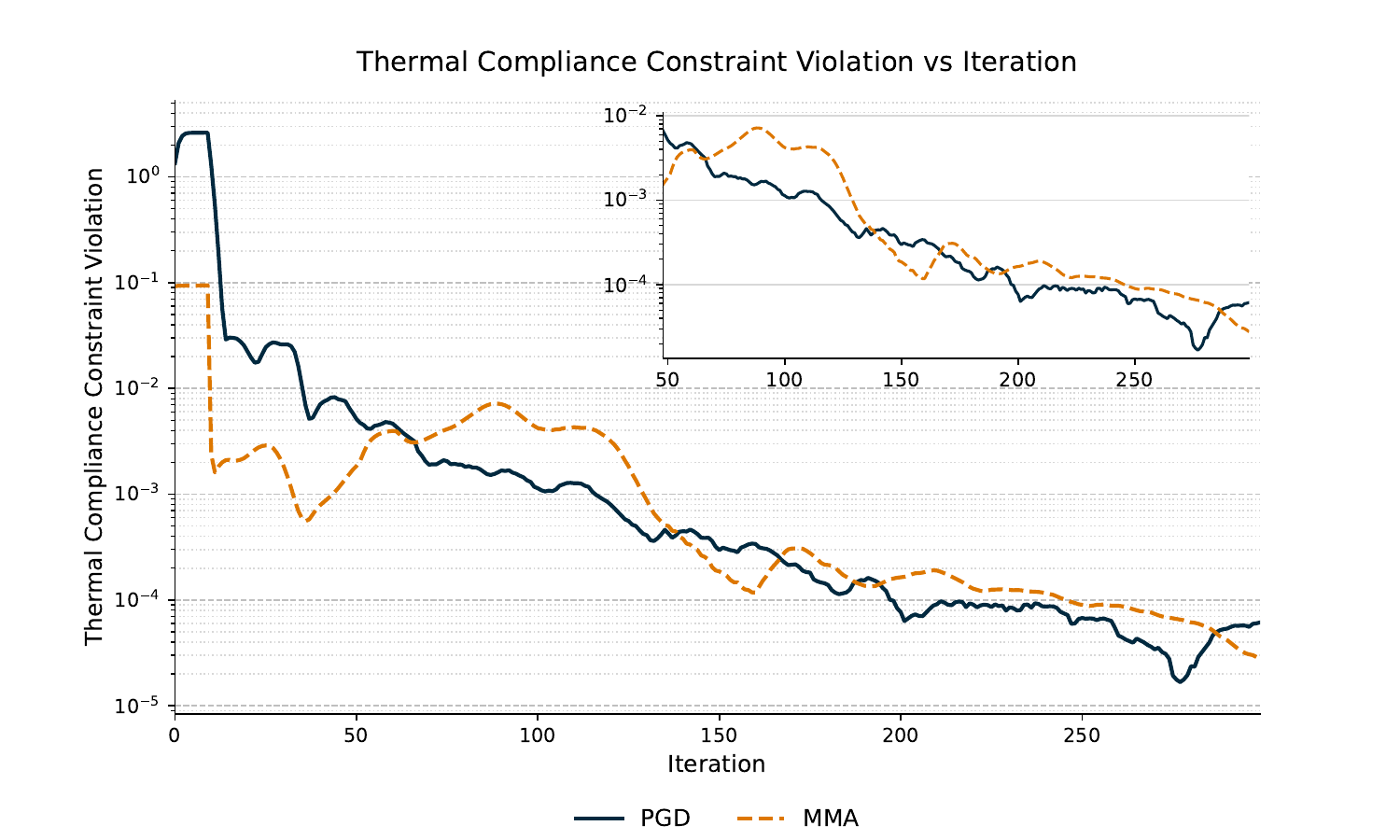}}
    \parbox[t]{0.495\linewidth}{\includegraphics[trim={0 0 0 1.8em},clip,width=\linewidth]{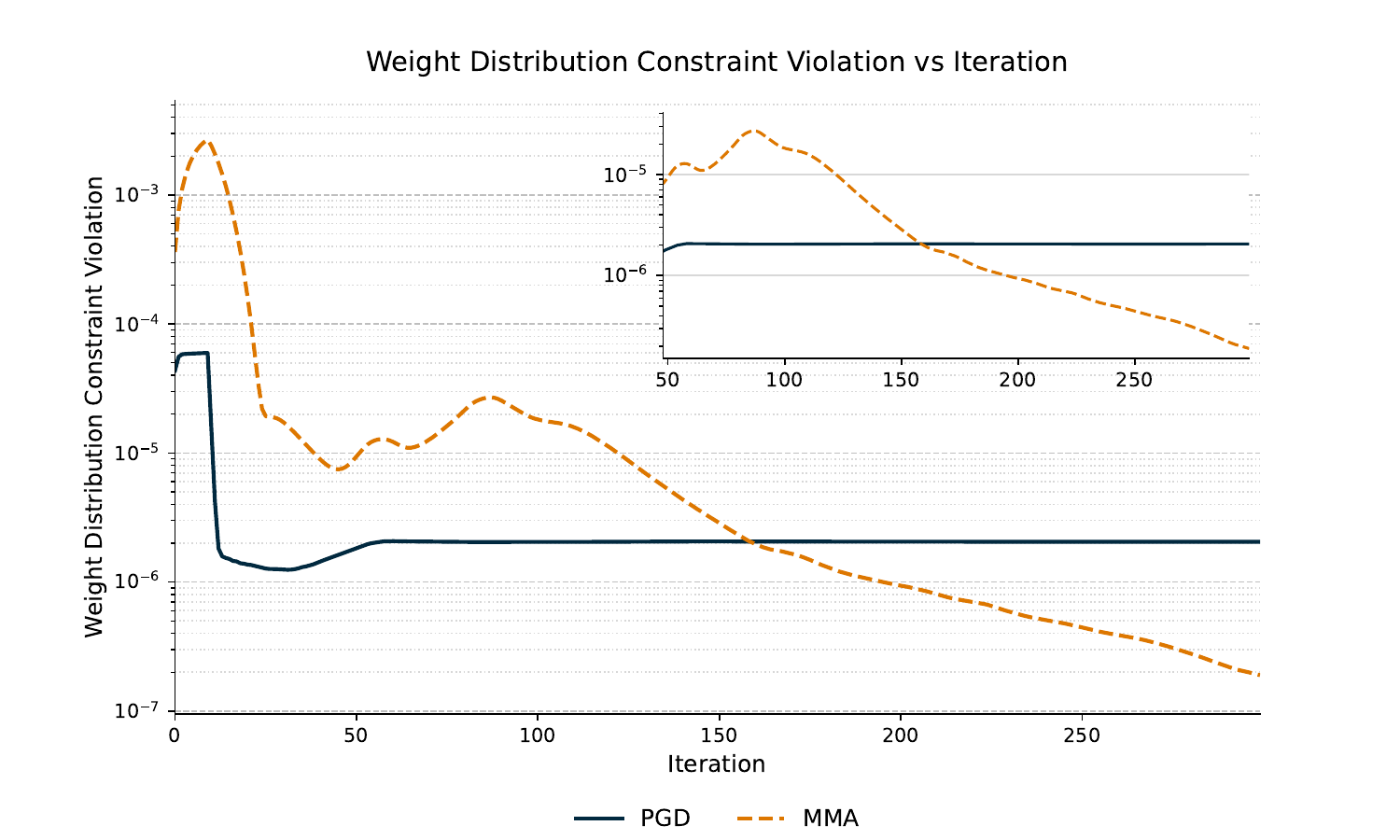}}
    \parbox[t]{\linewidth}{\quad}
    \textbf{Convergence Plots}
    \parbox[t]{\linewidth}{\quad}
    \parbox[t]{0.495\linewidth}{
    \includegraphics[width=\linewidth]{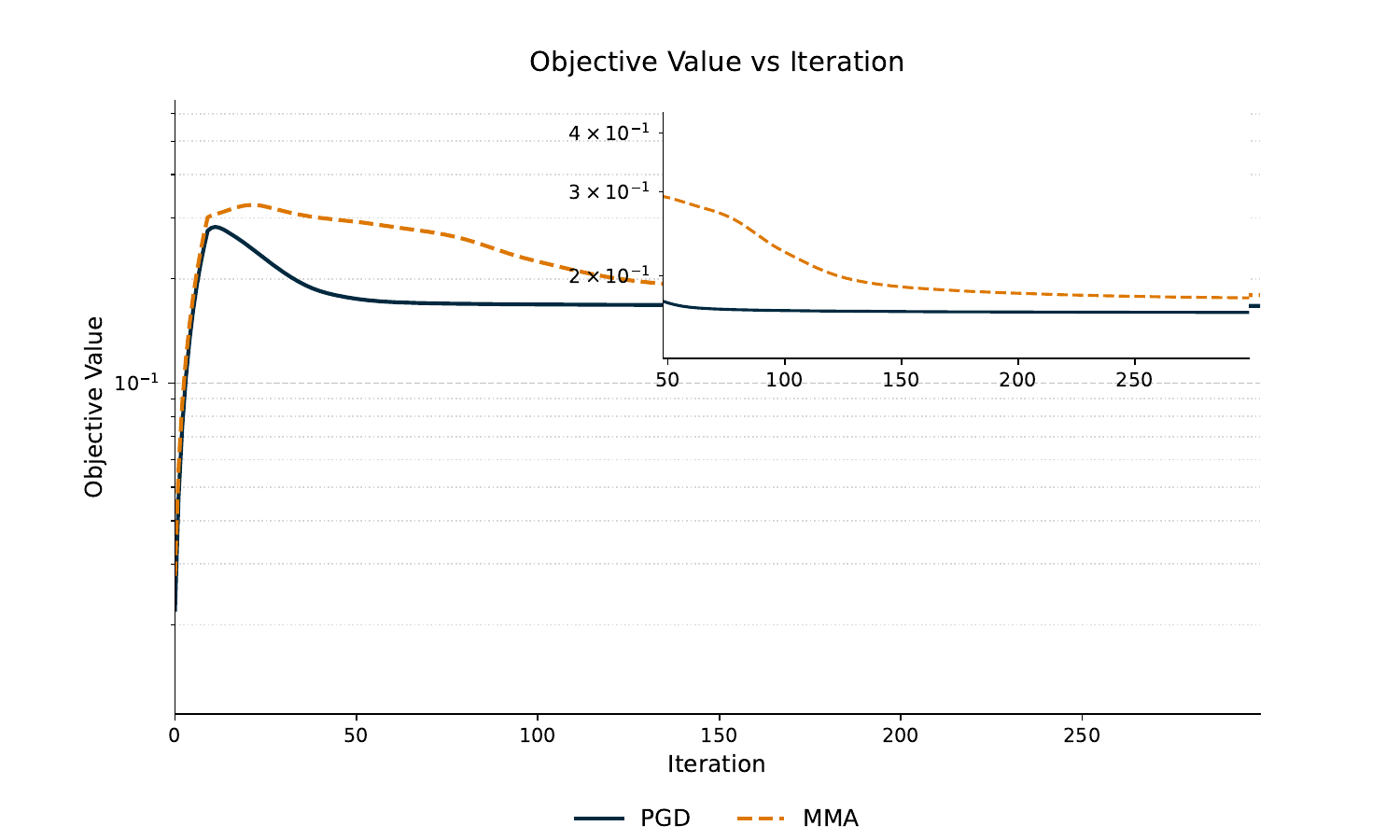}
    }
    \parbox[t]{0.495\linewidth}{
    \includegraphics[width=\linewidth]{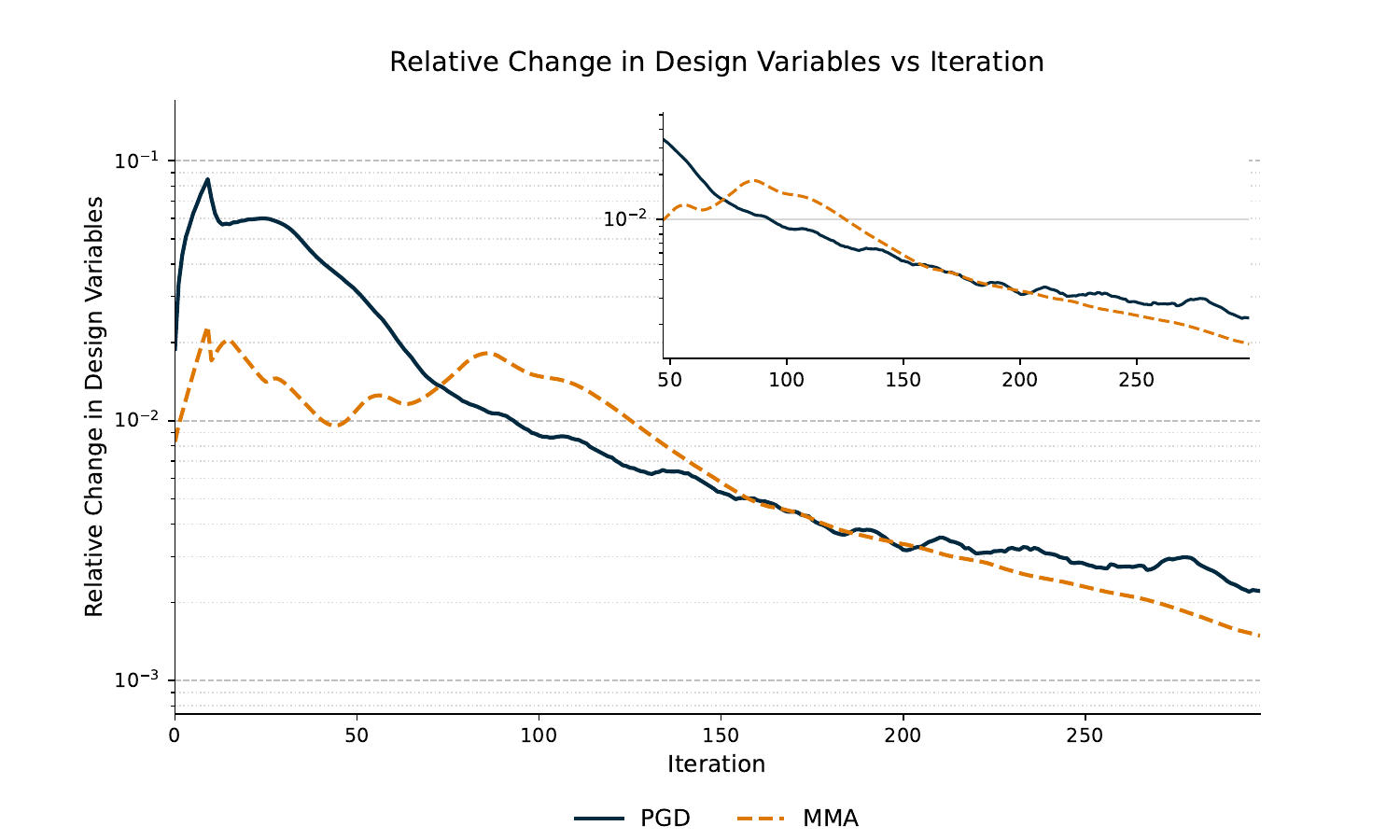}
    }
    \caption{Results of the multi-physics, multi-constraint stress test. MMA and PGD behave similarly quantitatively, with MMA converging slightly more slowly. Even in this highly nonlinear problem, PGD handles multiple nonlinear constraints effectively.}
    \label{fig:nonlinresults}
\end{figure}

Fig.~\ref{fig:nonlinresults} shows that both MMA and PGD converge more slowly to binary-density topologies than in the previous problems; both exhibit similar quantitative convergence and constraint handling, and PGD converges somewhat faster than MMA in this example. This indicates that even in highly nonlinear regimes, PGD remains a viable approach for handling multiple nonlinear constraints. Here, similar to prior experiments, we see PGD performing significantly better in terms of compute time with a median iteration time of $0.0795$ s compared to $0.9479$ s for MMA.

\section{Conclusion}
In this work, we propose a comprehensive projected gradient descent framework for constrained optimization, with a particular focus on its application to topology optimization. The central contribution lies in reformulating the projection step, which is traditionally difficult for multiple nonlinear constraints and requires identifying the active set at each iteration. This reformulation into a regularized projection problem replaces the explicit active-set determination with a strictly convex quadratic program augmented by penalty terms on constraint violations. This regularization ensures well-posedness and guarantees the existence and uniqueness of the solution, even in the presence of infeasible or near-singular constraint sets. We derive the associated Karush–Kuhn–Tucker conditions and establish the strict monotonicity of the resulting operator, enabling the use of a semismooth Newton method for efficient and globally convergent solution of the projection subproblem. Moreover, we show that under certain conditions---a single active constraint, or multiple independent constraints---the projection can be solved far faster with a binary search, which our method incorporates. 

Beyond the algorithmic reformulation, we provide a theoretical analysis of convergence under standard smoothness assumptions. Specifically, we prove that the proposed PGD scheme converges sublinearly for sufficiently small step sizes when the objective function is L-smooth. To enhance convergence in practice, we propose a dynamic step-size rule based on local Lipschitz-constant estimates, combined with a spectral step-size scheme, to ensure robustness across problem scales and conditioning. Moreover, we see that convergence requires iterates to be feasible across iterations, which for nonlinear constraints could become problematic; thus, we introduce a heuristic relaxation to automatically handle cases where such patterns emerge. The empirical results demonstrate that, when applied across topology optimization problems, the PGD method achieves convergence behavior and final compliance comparable to---and in some cases better than---those of established approaches such as MMA and OC, while offering an order-of-magnitude improvement in scalability and speed.

Our findings suggest several broader implications. The proposed regularized PGD algorithm provides a general-purpose tool for solving high-dimensional, constrained optimization problems that exhibit similar structural properties to TO, namely, those involving nonlinear objectives and constraints with constant bounds on design variables. Its capacity to handle infeasible iterates makes it suitable for a range of domains beyond topology optimization. Furthermore, the modular nature of the approach allows integration with higher-order or even stochastic methods, potentially enabling more advanced optimization approaches. Moreover, the reformulated regularized projection may be solved even more efficiently in the future by taking advantage of highly optimized convex programming packages and algorithms, which we do not directly investigate in this paper.

In summary, this paper establishes a theoretically grounded and computationally efficient framework for projected gradient descent with regularized projections, extending both the analytical understanding and practical usability of PGD-type methods in nonlinear constrained TO problems. Future research can focus on better search directions and step-size adjustment schemes, e.g., incorporating momentum-based approaches. Notably, the increased efficiency and scalability of such an optimizer could enable the solution of large-scale nonlinear problems, allowing dataset generation for deep-learning approaches, which require large amounts of data and have shown promising results in recent works~\cite{regenwetter2022deep, nobari2025optimize, nie2021topologygan, nito, maze2022topodiff, giannone2023aligning,HU2024103639}. Finally, this framework, although evaluated here on various nonlinear constraint scenarios, has not yet been applied to a wider range of TO problems with advanced structural and manufacturing constraints; it therefore requires more detailed investigation and possibly refinement to be adopted widely by TO and structural optimization researchers. We believe the results here demonstrate the value of such an adoption, especially given that, based on the nature of the proposed algorithms, one would expect an even greater improvement in speed if highly parallelized hardware platforms such as GPUs and TPUs were to be utilized for optimization. To encourage further investigation of the optimizer, we release our GPU- and CPU-accelerated FEA and topology-optimization platform at \href{https://github.com/ahnobari/pyFANTOM}{https://github.com/ahnobari/pyFANTOM}.

\section{Statements and Declarations}
Below are the relevant statements and declarations related to this work:

\begin{itemize}
    \item \textbf{Funding: }The authors did not receive support from any organization for the submitted work.
    \item \textbf{Conflict of Interest: }The authors have no relevant financial or non-financial interests to disclose.
    \item\textbf{Author contributions: } All authors contributed to the study conception and design. Material preparation, data collection, and analysis were performed by Amin Heyrani Nobari. The first draft of the manuscript was written by Amin Heyrani Nobari, and all authors commented on previous versions of the manuscript. All authors read and approved the final manuscript.
    \item \textbf{Data Availability and Replication of Results: }The code and platform used for experiments in this paper are available at \href{https://github.com/ahnobari/pyFANTOM}{https://github.com/ahnobari/pyFANTOM}.
    \item \textbf{Compliance with Ethical Standards: } Not applicable.
\end{itemize}

\bibliography{sn-bibliography-v2}
\clearpage
\begin{appendices}
\section{KKT Analysis of the Projection Problem}
\label{app:kkt}
\noindent
In this section, we will discuss the optimality conditions for the projection problem and provide a detailed derivation of the proposed approach.
Before we continue solving the problem, we will formulate the Karush–Kuhn–Tucker~(KKT) conditions for the projection problem (\ref{eqn:linearized}) and refer to these KKT conditions in our solution:

\begin{align}
    & g_j(\boldsymbol{\rho}_t) + \left(\boldsymbol{\delta}-\alpha\nabla f(\boldsymbol{\rho}_t)\right)^T\nabla g_j(\boldsymbol{\rho}_t)-G_j\leq 0\quad \forall j \in \{1,\ldots,m\},\label{kkt1}\\[1mm]
    & l-\tilde{\rho}_i-\delta_i \leq 0, \quad \forall i\in\{1, \ldots, N\},\label{kkt2}\\[1mm]
    & \delta_i + \tilde{\rho}_i-u \leq 0, \quad \forall i\in\{1, \ldots, N\},\label{kkt3}\\[1mm]
    & 2 \delta_i+\sum_j\lambda_j (\nabla g_j(\boldsymbol{\rho}_t))_i +\nu_i-\mu_i=0 \quad \forall i\in\{1, \ldots, N\},\label{kkt4}\\[1mm]
    & \mu_i\left(l-\tilde{\rho}_i-\delta_i\right)=0 \quad \forall i\in\{1, \ldots, N\},\label{kkt5}\\[1mm]
    & \nu_i\left(\delta_i-u+ \tilde{\rho}_i\right)=0 \quad \forall i\in\{1, \ldots, N\},\label{kkt6}\\[1mm]
    & \lambda_j \left((\boldsymbol{\delta}-\alpha\nabla f(\boldsymbol{\rho}_t))^T\nabla g_j(\boldsymbol{\rho}_t) - G_j + g_j(\boldsymbol{\rho}_t)\right) = 0 \quad \forall j \in \{1,\ldots,m\},\label{kkt7}\\[1mm]
    & \lambda_j \geq 0, \quad \mu_i \geq 0, \quad \nu_i \geq 0 \quad \forall i\in\{1, \ldots, N\},\forall j \in \{1,\ldots,m\}.\label{kkt8}
\end{align}

We can now perform a simple analysis of this problem to limit the solution space to some extent. From (\ref{kkt4}) we have the stationary condition:

\begin{equation}
    \delta_i = -\frac{\sum_j\lambda_j (\nabla g_j(\boldsymbol{\rho}_t))_i +\nu_i-\mu_i}{2}.
\end{equation}

Given (\ref{kkt5}), (\ref{kkt6}), (\ref{kkt8}), and the fact that $l<u$, when $\delta_i< -\frac{1}{2}\sum_j\lambda_j (\nabla g_j(\boldsymbol{\rho}_t))_i$ we must have $\nu_i>0$ and $\mu_i=0$. Similarly, when $\delta_i> -\frac{1}{2}\sum_j\lambda_j (\nabla g_j(\boldsymbol{\rho}_t))_i$, we must have $\mu_i>0$ and $\nu_i=0$. Therefore we know:

\begin{align}
    \delta_i< -\frac{1}{2}\sum_j\lambda_j (\nabla g_j(\boldsymbol{\rho}_t))_i \rightarrow \delta_i = u - \tilde{\rho}_i \label{box1},\\
    \delta_i> -\frac{1}{2}\sum_j\lambda_j (\nabla g_j(\boldsymbol{\rho}_t))_i \rightarrow \delta_i = l - \tilde{\rho}_i \label{box2}.
\end{align}

We can see that for the terms in the stationary condition pertaining to the inequality constraints, we have three possibilities:

\begin{itemize}
    \item \textbf{Case 1:} $-\frac{1}{2}\sum_j\lambda_j (\nabla g_j(\boldsymbol{\rho}_t))_i> u-\tilde{\rho}_i$, which from (\ref{kkt3}) yields $\delta_i<-\frac{1}{2}\sum_j\lambda_j (\nabla g_j(\boldsymbol{\rho}_t))_i$ and thereby, from (\ref{box1}), $\delta_i = u-\tilde{\rho}_i$.
    \item \textbf{Case 2:} $-\frac{1}{2}\sum_j\lambda_j (\nabla g_j(\boldsymbol{\rho}_t))_i< l-\tilde{\rho}_i$, which from (\ref{kkt2}) yields $\delta_i>-\frac{1}{2}\sum_j\lambda_j (\nabla g_j(\boldsymbol{\rho}_t))_i$ and thereby, from (\ref{box2}), $\delta_i = l-\tilde{\rho}_i$.
    \item \textbf{Case 3:} $l-\tilde{\rho}_i \leq -\frac{1}{2}\sum_j\lambda_j (\nabla g_j(\boldsymbol{\rho}_t))_i \leq u-\tilde{\rho}_i$, which then means that if $\delta_i<-\frac{1}{2}\sum_j\lambda_j (\nabla g_j(\boldsymbol{\rho}_t))_i$, then $\delta_i = u-\tilde{\rho}_i$, which means $\delta_i\geq-\frac{1}{2}\sum_j\lambda_j (\nabla g_j(\boldsymbol{\rho}_t))_i$, which contradicts the assumption. Similarly if $\delta_i>-\frac{1}{2}\sum_j\lambda_j (\nabla g_j(\boldsymbol{\rho}_t))_i$ we will have a contradiction, which leaves only $\delta_i=-\frac{1}{2}\sum_j\lambda_j (\nabla g_j(\boldsymbol{\rho}_t))_i$.
\end{itemize}

It is now evident that a three-case breakdown can give us a candidate solution for the projection. Letting $y_j = -\frac{\lambda_j}{2}$ for all $j\in \{1,\ldots,m\}$, a candidate solution to the problem can take the form (given $\mathbf y \in \mathbb{R}^m$):

\begin{equation}
\label{eqn:delta_y}
\delta_i(\mathbf{y})= 
    \begin{cases}
        l-\tilde{\rho}_i & \text { if } \sum_j y_j (\nabla g_j(\boldsymbol{\rho}_t))_i<l-\tilde{\rho}_i, \\[1mm] 
         \sum_j y_j (\nabla g_j(\boldsymbol{\rho}_t))_i& \text { if } l-\tilde{\rho}_i \leq \sum_j y_j (\nabla g_j(\boldsymbol{\rho}_t))_i \leq u-\tilde{\rho}_i, \\[1mm] 
        u-\tilde{\rho}_i & \text { if } \sum_j y_j (\nabla g_j(\boldsymbol{\rho}_t))_i>u-\tilde{\rho}_i .
    \end{cases}
\end{equation}

Now let:

\begin{align}
    \lambda_j &= -2y_j,\\[1mm]
    \mu_i&=2\max(0,l-\tilde{\rho}_i-\sum_j y_j (\nabla g_j(\boldsymbol{\rho}_t))_i),\\[1mm]
    \nu_i&=2\max(0,\sum_j y_j (\nabla g_j(\boldsymbol{\rho}_t))_i-u+\tilde{\rho}_i).
\end{align}

If there exists a $\mathbf{y}^*\in \mathbb{R}^m$ with $\mathbf{y}^*\leq \mathbf{0}$ such that the choice of $\lambda_j = -2y_j$ above satisfies (\ref{kkt1}) and (\ref{kkt7}), we can show that the above choices for $\boldsymbol{\delta}$, $\mu_i$, and $\nu_i$ satisfy all other KKT conditions. We can look at the inequality breakdowns of the proposed solution to prove this:

\begin{itemize}
    \item \textbf{Case 1:} $\sum_j y_j (\nabla g_j(\boldsymbol{\rho}_t))_i< l-\tilde{\rho}_i$, then $\delta_i = l-\tilde{\rho}_i$, $\mu_i= 2 \left(l-\tilde{\rho}_i-\sum_j y_j (\nabla g_j(\boldsymbol{\rho}_t))_i\right)$, $\nu_i = 0$, which satisfy (\ref{kkt2})-(\ref{kkt6}) as well as (\ref{kkt8}).
    \item \textbf{Case 2:} $\sum_j y_j (\nabla g_j(\boldsymbol{\rho}_t))_i> u-\tilde{\rho}_i$, then $\delta_i = u-\tilde{\rho}_i$, $\mu_i=0$, $\nu_i = 2\left(\sum_j y_j (\nabla g_j(\boldsymbol{\rho}_t))_i-u+\tilde{\rho}_i\right)$, which satisfy (\ref{kkt2})-(\ref{kkt6}) as well as (\ref{kkt8}).
    \item \textbf{Case 3:} $l-\tilde{\rho}_i \leq \sum_j y_j (\nabla g_j(\boldsymbol{\rho}_t))_i \leq u-\tilde{\rho}_i$, then $\delta_i=\sum_j y_j (\nabla g_j(\boldsymbol{\rho}_t))_i$, $\mu_i=\nu_i=0$, which also satisfy (\ref{kkt2})-(\ref{kkt6}) as well as (\ref{kkt8}).
\end{itemize}

Thus, by convexity, $\boldsymbol{\delta}^*$ defined by (\ref{eqn:delta_y}) computed at $\mathbf{y}^*$ is the solution to the problem. Hence the problem reduces to finding $\mathbf{y}^*\in \mathbb{R}^m$ with $\mathbf{y}^*\leq \mathbf{0}$ such that the choice of $\lambda_j = -2y_j$ satisfies (\ref{kkt1}) and (\ref{kkt7}). We will now propose an approach for solving this problem in the special case of one active constraint, as well as a regularized approach for the general problem.

\subsection{Single Constraint Solution}
\label{app:singlecon}
Let $m=1$ in the projection subproblem defined in (\ref{eqn:linearized}). Then the solution form introduced in (\ref{eqn:delta_y}) will take the simple form:

\begin{equation}
\label{eqn:delta_y_1}
\delta_i(y)= 
    \begin{cases}
        l-\tilde{\rho}_i & \text { if }  y (\nabla g(\boldsymbol{\rho}_t))_i<l-\tilde{\rho}_i, \\[1mm] 
         y(\nabla g(\boldsymbol{\rho}_t))_i& \text { if } l-\tilde{\rho}_i \leq y(\nabla g(\boldsymbol{\rho}_t))_i \leq u-\tilde{\rho}_i, \\[1mm] 
        u-\tilde{\rho}_i & \text { if } y (\nabla g(\boldsymbol{\rho}_t))_i>u-\tilde{\rho}_i .
    \end{cases}
\end{equation}

Given the fact that we chose $\lambda = -2y$, from (\ref{kkt1}) and (\ref{kkt7}) we have:

\begin{align}
    y =  0 \Rightarrow \lambda=0 \Rightarrow  (\boldsymbol{\delta}-\alpha\nabla f(\boldsymbol{\rho}_t))^T\nabla g(\boldsymbol{\rho}_t) - G + g(\boldsymbol{\rho}_t)\leq 0 \label{inactive},\\
    y<0 \Rightarrow \lambda>0 \Rightarrow  (\boldsymbol{\delta}-\alpha\nabla f(\boldsymbol{\rho}_t))^T\nabla g(\boldsymbol{\rho}_t) - G + g(\boldsymbol{\rho}_t) = 0\label{active}.
\end{align}

Hence, if $y=0$ yields a feasible solution, it is optimal by convexity, with $\boldsymbol{\delta}$ given by (\ref{eqn:delta_y_1}). Otherwise, the inequality constraint is active. Letting $\Gamma_i = (\nabla g(\boldsymbol{\rho}_t))_i\,\delta_i(y)$, substituting (\ref{eqn:delta_y_1}) into (\ref{active}) for the active-constraint case gives:

\begin{equation}
     \sum_{i}\Gamma_i - \hat{G} = 0,
     \label{eqn:activeequality}
\end{equation}

where $\hat{G} = G - g(\boldsymbol{\rho}_t)+\alpha \nabla f(\boldsymbol{\rho}_t)^T\nabla g(\boldsymbol{\rho}_t)$. Note that for $\boldsymbol{\Gamma}$ we have:

\begin{equation}
\label{eqn:Delta}
\Gamma_i(y)= 
    \begin{cases}
        (\nabla g(\boldsymbol{\rho}_t))_i(l-\tilde{\rho}_i) & \text { if }  y (\nabla g(\boldsymbol{\rho}_t))_i<l-\tilde{\rho}_i, \\[1mm] 
         y\left((\nabla g(\boldsymbol{\rho}_t))_i\right)^2& \text { if } l-\tilde{\rho}_i \leq y(\nabla g(\boldsymbol{\rho}_t))_i \leq u-\tilde{\rho}_i, \\[1mm] 
        (\nabla g(\boldsymbol{\rho}_t))_i(u-\tilde{\rho}_i) & \text { if } y (\nabla g(\boldsymbol{\rho}_t))_i>u-\tilde{\rho}_i .
    \end{cases}
\end{equation}

It is clear that $\Gamma_i(y)$ is piecewise linear and nondecreasing in $y$; hence the left-hand side of (\ref{eqn:activeequality}) is as well. This means that if there exists a $y^*\in\mathbb{R}$ with $y^*\leq0$, it can be found through an efficient binary search in the range \(\left[\min_{i} \{\min\left(\frac{l-\tilde{\rho}_i}{(\nabla g(\boldsymbol{\rho}_t))_i},\frac{u-\tilde{\rho}_i}{(\nabla g(\boldsymbol{\rho}_t))_i}\right)\}, 0 \right]\). Thus, for the case of a single constraint (e.g., any nonlinear topology optimization under only a volume constraint), the projection can be solved by first testing if $y=0$ yields a feasible solution, and if not, performing a binary search in the aforementioned range. This approach is inspired by the algorithm of \citet{Pardalos1990AnAF}.

\subsection{Efficiently Solving the General Problem}
In the general case, we must find all $m$ dual values $y_j$. There are three scenarios that we consider in our approach, each of which can be handled efficiently:

\begin{itemize}
    \item \textbf{Single Constraint}: If a single constraint is present, we can simply obtain $y^*$ through a binary search as described in the prior section.
    \item \textbf{Multiple Independent Constraints}: If a given problem has $m$ constraints that each involve a subset of the design variables $\varphi_j\subset\boldsymbol{\rho}$, which are non-overlapping, i.e., $\varphi_k\cap\varphi_j = \varnothing \: \text{for all}\: j\neq k$, then each constraint will only involve the specific subset of variables and thus the same analysis as for a single constraint holds, and all $y^*_j$ values can be obtained by $m$ binary searches, which can be done in parallel.
    \item \textbf{Single Active Constraint}: In many real-world problems, typically only one constraint is active. This occurs when one constraint holds with equality while the others are strictly satisfied (a common occurrence in high-dimensional numerical problems such as TO). Given this, one can start with the assumption that the active set only contains one constraint and find each constraint's $y^*_j$ value independently by binary search (efficiently parallelizable), and if one such solution is feasible, the projection problem is solved by convexity.
\end{itemize}

The final case is when the problem involves multiple active constraints. In this scenario, we must find which constraints are active and solve the system of nonlinear equations to find the vector $\mathbf{y}^*$, whose entries are zero for inactive constraints and which yields a feasible $\boldsymbol{\delta}(\mathbf{y^*})$. Noting that in practice, we do not know if the problem is even feasible, we must construct an approach that is robust to uncertainties and makes finding the active set and dual variable values efficient. Thus, we take a regularization approach and rewrite the projection problem in an equivalent form regularized by constraint violations.

\subsubsection{The Regularized Projection Problem}
We introduce a non-negative slack variable, $s_j \ge 0$, for each of the $m$ general constraints. This variable measures the amount by which a constraint is violated. The new, regularized optimization problem is then formulated as:
\begin{equation}
\label{eqn:regularized}
\begin{split}
	\min_{\boldsymbol{\delta}\in \mathbb{R}^N, \mathbf{s} \in \mathbb{R}^m}& \quad \frac{1}{2}\|\boldsymbol{\delta}\|_2^2 + \frac{C}{2} \|\mathbf{s}\|_2^2 \\
	\text{s.t.}& \quad l\leq \tilde{\rho}_i+ \delta_i\leq u, \quad \forall i \in [N]\\
	&\quad g_j(\boldsymbol{\rho}_t) + \left(\boldsymbol{\delta}-\alpha\nabla f(\boldsymbol{\rho}_t)\right)^T\nabla g_j(\boldsymbol{\rho}_t) - s_j \leq G_j \quad \forall j\in[m]\\
    &\quad s_j \ge 0 \quad \forall j\in[m],
\end{split}
\end{equation}
where $C > 0$ is a large penalty parameter. This problem is a strictly convex QP and is always feasible. As $C\rightarrow\infty$ and if the original problem (\ref{eqn:linearized}) is feasible, the solution to this problem will have $\mathbf{s}=\mathbf{0}$, recovering the original projection. Otherwise, a solution that optimally balances minimizing the projection distance and the constraint violations will emerge. This enables solving the projection directly, avoiding the potentially expensive active-set identification. The proper value for $C$ can be determined based on the constraint tolerances and the constraint gradients.

\subsubsection{Optimality Conditions and Solution Form}
In this section, we derive the optimality conditions for the regularized projection problem defined in (\ref{eqn:regularized}). The KKT conditions for the problem are:
\begin{align}
    & g_j(\boldsymbol{\rho}_t) + \left(\boldsymbol{\delta}-\alpha\nabla f(\boldsymbol{\rho}_t)\right)^T\nabla g_j(\boldsymbol{\rho}_t) - s_j - G_j \le 0, \quad \forall j \in [m] \label{rkkt1} \\
    & l - \tilde{\rho}_i - \delta_i \le 0, \quad \forall i\in[N] \label{rkkt2} \\
    & \delta_i + \tilde{\rho}_i - u \le 0, \quad \forall i\in[N] \label{rkkt3} \\
    & -s_j \le 0, \quad \forall j \in [m] \label{rkkt4} \\
    & \delta_i+\sum_j\lambda_j (\nabla g_j(\boldsymbol{\rho}_t))_i +\nu_i-\mu_i=0, \quad \forall i\in[N] \label{rkkt5} \\
    & Cs_j - \lambda_j - \eta_j = 0, \quad \forall j\in[m] \label{rkkt6} \\
    & \mu_i\left(l-\tilde{\rho}_i-\delta_i\right)=0, \quad \forall i\in[N] \label{rkkt7} \\
    & \nu_i\left(\delta_i-u+ \tilde{\rho}_i\right)=0, \quad \forall i\in[N] \label{rkkt8} \\
    & \lambda_j \left(g_j(\boldsymbol{\rho}_t) + \left(\boldsymbol{\delta}-\alpha\nabla f(\boldsymbol{\rho}_t)\right)^T\nabla g_j(\boldsymbol{\rho}_t) - s_j - G_j\right) = 0, \quad \forall j \in [m] \label{rkkt9} \\
    & \eta_j s_j = 0, \quad \forall j \in [m] \label{rkkt10} \\
    & \lambda_j \ge 0, \quad \mu_i \ge 0, \quad \nu_i \ge 0, \quad \eta_j \ge 0 \quad \forall i,j \label{rkkt11}
\end{align}
As before, let $y_j = -\lambda_j/2$, then the analysis of the box constraints on $\boldsymbol{\delta}$ allows us to express $\boldsymbol{\delta}$ as the same piecewise function in (\ref{eqn:delta_y}).

From (\ref{rkkt6}), (\ref{rkkt10}), and (\ref{rkkt11}), we can deduce a direct relationship between $s_j$ and $\lambda_j$. From (\ref{rkkt6}), we have $Cs_j = \lambda_j + \eta_j$. If $\eta_j>0$, then (\ref{rkkt10}) gives $s_j=0$ and thus $\lambda_j=-\eta_j<0$, which contradicts (\ref{rkkt11}); hence $\eta_j>0$ is impossible unless $s_j=0$, in which case $\lambda_j=\eta_j=0$. If instead $s_j>0$, then (\ref{rkkt10}) implies $\eta_j=0$, so (\ref{rkkt6}) yields $s_j =\frac{\lambda_j }{C}=\frac{-2y_j}{C}$. This also holds in the case of $s=0$, as we saw before; therefore we can deduce that in general:

\begin{equation}
    s_j(\mathbf{y}) = \frac{-2y_j}{C}.
\end{equation}

The problem now reduces to finding the vector $\mathbf{y} \le \mathbf{0}$ that satisfies the complementarity condition (\ref{rkkt9}). Let us define a function $\mathbf{h}(\mathbf{y})$:
\begin{equation}
    \begin{split}
        h_j(\mathbf{y}) &= g_j(\boldsymbol{\rho}_t) + \left(\boldsymbol{\delta}-\alpha\nabla f(\boldsymbol{\rho}_t)\right)^T\nabla g_j(\boldsymbol{\rho}_t) - s_j(\mathbf{y}) - G_j \\&= g_j(\boldsymbol{\rho}_t) + \left(\boldsymbol{\delta}-\alpha\nabla f(\boldsymbol{\rho}_t)\right)^T\nabla g_j(\boldsymbol{\rho}_t) + \frac{2y_j}{C} - G_j.
    \end{split}
\end{equation}
Thus, the solution is found by finding $\mathbf{y} \le \mathbf{0}$ such that $h_j(\mathbf{y}) \le \mathbf{0}$ and $y_j h_j(\mathbf{y}) = 0$ for all $j$.

To summarize, the solution to this convex QP is fully characterized by its KKT conditions. This analysis reveals that the entire problem can be reduced to finding a unique vector of dual variables $\mathbf{y} \in \mathbb{R}^m$. The optimal adjustment vector $\boldsymbol{\delta}^*$ and slack variables $\mathbf{s}^*$ are functions of $\mathbf{y}^*$:
\begin{equation}
\label{eqn:delta_y_robust}
\delta_i^*(\mathbf{y}^*)=\begin{cases}
l-\tilde{\rho}_i & \text { if } \sum_j y_j^* (\nabla g_j(\boldsymbol{\rho}_t))_i<l-\tilde{\rho}_i, \\
\sum_j y_j^* (\nabla g_j(\boldsymbol{\rho}_t))_i& \text { if } l-\tilde{\rho}_i \leq \sum_j y_j^* (\nabla g_j(\boldsymbol{\rho}_t))_i \leq u-\tilde{\rho}_i, \\
u-\tilde{\rho}_i & \text { if } \sum_j y_j^* (\nabla g_j(\boldsymbol{\rho}_t))_i>u-\tilde{\rho}_i,
\end{cases}
\end{equation}
\begin{equation}
\label{eqn:s_y_robust}
    s_j^*(\mathbf{y}^*) = -\frac{2y_j^*}{C}.
\end{equation}
Thus,  $\mathbf{y}^* \in \mathbb{R}^m$ is a unique solution such that for all $j \in \{1, \ldots, m\}$:
\begin{enumerate}
    \item $y_j \le 0$
    \item $h_j(\mathbf{y}) \le 0$
    \item $y_j h_j(\mathbf{y}) = 0$
\end{enumerate}
where the function $\mathbf{h}(\mathbf{y})$ is defined as:
\begin{equation}
    \label{eqn:h}
    h_j(\mathbf{y}) =  \boldsymbol{\delta}(\mathbf{y})^T\nabla g_j(\boldsymbol{\rho}_t) + \frac{2y_j}{C} - \hat{G}_j,
\end{equation}
where $\hat{G}_j = G_j - g_j(\boldsymbol{\rho}_t) + \alpha \nabla f(\boldsymbol{\rho}_t)^T\nabla g_j(\boldsymbol{\rho}_t)$.
We can show that the function $\mathbf{h}(\mathbf{y})$ is strictly monotone with respect to $\mathbf{y}$, and thus guarantees the existence of a unique $\mathbf{y}^*\in\mathbb{R}^m$ which solves the regularized projection problem. We can prove this by analyzing its Jacobian matrix, $\mathbf{J}_{\mathbf{h}}$.
\begin{equation}
    (J_{\mathbf{h}})_{jk} = \frac{\partial h_j}{\partial y_k} = \sum_i (\nabla g_j(\boldsymbol{\rho}_t))_i \frac{\partial \delta_i}{\partial y_k} + \frac{2}{C}\mathbb{I}_{jk},
\end{equation}
where $\mathbb{I}_{jk}$ is the Kronecker delta. The derivative $\frac{\partial \delta_i}{\partial y_k}$ is $\chi_i(\mathbf{y}) (\nabla g_k(\boldsymbol{\rho}_t))_i$, where $\chi_i(\mathbf{y})$ is an indicator function that is 1 if $l-\tilde{\rho}_i \leq \sum_j y_j (\nabla g_j(\boldsymbol{\rho}_t))_i \leq u-\tilde{\rho}_i$ and 0 otherwise (See \ref{eqn:delta_y_robust}). The Jacobian is therefore:

\begin{equation}
    (J_{\mathbf{h}})_{jk} = \sum_i \chi_i(\mathbf{y}) (\nabla g_j(\boldsymbol{\rho}_t))_i (\nabla g_k(\boldsymbol{\rho}_t))_i + \frac{2}{C}\mathbb{I}_{jk}.
\end{equation}

In matrix form, let $\mathbf{A}$ ($m\times N$) be the matrix of constraint gradients and $\mathbf{D}(\mathbf{y})$ be the diagonal matrix of indicators $\chi_i(\mathbf{y})$ ($N\times N$). The Jacobian is:
\begin{equation}
\label{eqn:reg_jacobian}
    \mathbf{J}_{\mathbf{h}}(\mathbf{y}) = \mathbf{A} \mathbf{D}(\mathbf{y}) \mathbf{A}^T + \frac{2}{C}\mathbf{I}.
\end{equation}
The matrix $\mathbf{A}\mathbf{D}\mathbf{A}^T$ is symmetric and positive semi-definite. The matrix $\frac{2}{C}\mathbf{I}$ is symmetric and positive definite (since $C>0$), making the sum strictly positive definite and therefore showing $\mathbf{h}$ strictly monotone. This property guarantees that the problem has a unique solution $\mathbf{y}^*$, ensuring the robustness of the regularized formulation.

The original (unregularized) projection problem yields a similar Jacobian; the derivation is analogous and omitted for brevity:

\begin{equation}
\label{eqn:unreg_jacobian}
    \mathbf{J}_{\mathbf{h}}(\mathbf{y}) = \mathbf{A}_{\mathcal{J}} \mathbf{D}(\mathbf{y}) \mathbf{A}_{\mathcal{J}}^T.
\end{equation}

where ${\mathcal{J}}$ is the set of active constraints. This yields a symmetric positive semi-definite matrix, which implies that even with a feasible subproblem with all active constraints, the solution to the dual problem is not guaranteed to be unique, unless $\mathbf{J}_{\mathbf{h}}(\mathbf{y})$ is invertible. This can only be guaranteed if $\text{rank}(\mathbf{A}_{\mathcal{J}})= |{\mathcal{J}}|$, thus requiring the constraints to be linearly independent.

Finally, we must determine the appropriate bounds for $C$ given the gradients of the constraints and a constraint tolerance. Let $\tau$ be a desired tolerance on the constraints. Then $C$ can be set to a value:

\begin{equation}
    C = \frac{2\min_j\min_{i} \{\min\left(\frac{l-\tilde{\rho}_i}{(\nabla g(\boldsymbol{\rho}_t))_i},\frac{u-\tilde{\rho}_i}{(\nabla g(\boldsymbol{\rho}_t))_i}\right)\}}{\tau}
\end{equation}

This is the theoretical bound on constraint violation; however, in practice, smaller values of $C$ may result in $\|\mathbf{s}\|_2$ being zero. In our approach, we keep $C$ constant at $10^{12}$ and solve the subproblems with this penalty value. In most cases, a sufficiently large $C$ will enable solving the projection nearly perfectly. However, for infeasible problems, one can adjust $C$ to establish a balance between constraint violation and projection distance.

\subsubsection{Solving the Regularized Problem}
\label{app:newton}
To solve the regularized problem as described in the prior section, we will exploit the monotonicity established in Appendix~\ref{app:kkt} to solve the problem and find the optimal projection efficiently. First, we define a function $\boldsymbol{\Phi}(\mathbf{y})$:

\begin{equation}
    \Phi_j(\mathbf{y}) = \max(h_j(\mathbf{y}),0) + \operatorname{sign}(\max(-h_j(\mathbf{y}),0))y_j.
\end{equation}

Note that the root of this function, i.e., $\boldsymbol{\Phi}(\mathbf{y})=0$, corresponds to the solution of the regularized projection problem (see prior section). Therefore, by finding the solution to $ \Phi_j(\mathbf{y}) = \mathbf{0}$, we can effectively solve the optimization problem. 

To do this, we can use a semismooth Newton iteration. First, $\Phi_j(\mathbf{y})$ is semismooth in $\mathbf{y}$: $\mathbf{h(y)}$ is piecewise linear (see (\ref{eqn:h})) and hence semismooth, and the max and sign functions in $\boldsymbol{\Phi}$ are also piecewise linear and semismooth. Therefore $\boldsymbol{\Phi}$ is semismooth, and the properties of semismooth Newton methods, such as superlinear local convergence, apply to this problem. Furthermore, we demonstrated the monotonicity of $\mathbf{h}(\mathbf{y})$, which will be useful in guaranteeing the global convergence of the semismooth Newton method we propose here. Before we continue, let us derive the Jacobian of $\boldsymbol{\Phi}$, which we will use in our Newton method. Given the iterate $\mathbf{y}^{(k)}$ of the proposed algorithm at iteration $k$, we can compute $\boldsymbol{\delta}(\mathbf{y}^{(k)})$ using (\ref{eqn:delta_y_robust}) and then $\mathbf{h}(\mathbf{y}^{(k)})$ using (\ref{eqn:h}). Given these we can also compute $\mathbf{J}_{\mathbf{h}}(\mathbf{y}^{(k)})$ using (\ref{eqn:reg_jacobian}). Now let $\chi_{\Phi,j}(\mathbf{y}^{(k)})$ be the indicator function for the active set, meaning $\chi_{\Phi,j}(\mathbf{y})=1$ if $h_j(\mathbf{y}^{(k)})>0$ and $\chi_{\Phi,j}(\mathbf{y})=0$ otherwise. Let $\mathbf{D}_{\boldsymbol{\Phi}}$ be the diagonal matrix of $\chi_{\Phi,j}(\mathbf{y}^{(k)})$ ($m\times m$) then the Jacobian of $\boldsymbol{\Phi}(\mathbf{y}^{(k)})$ can be written as:

\begin{equation}
    \mathbf{J}_{\Phi}(\mathbf{y}^{(k)})=\mathbf{D}_{\boldsymbol{\Phi}}\mathbf{J}_{\mathbf{h}}(\mathbf{y}^{(k)}) + \mathbf{I} - \mathbf{D}_{\boldsymbol{\Phi}}
\end{equation}

Since $\mathbf{J}_{\mathbf{h}}(\mathbf{y}^{(k)})$ is invertible (given it is SPD), $ \mathbf{J}_{\Phi}(\mathbf{y}^{(k)})$ is also invertible and thus one can update the dual values with the following update rule:

\begin{equation}
    \mathbf{y}^{(k+1)} = \mathbf{y}^{(k)} - \gamma\; \mathbf{J}_{\Phi}(\mathbf{y}^{(k)})^{-1}\mathbf{\Phi} = \mathbf{y}^{(k)} - \gamma \boldsymbol{\Delta},
\end{equation}

where $0<\gamma\leq1$ is a step size for the update. Note that $\gamma=1$ is equivalent to a pure Newton update; however, a semismooth Newton scheme only guarantees local convergence; therefore, we propose finding $\gamma$ by line search to ensure global convergence. Let $M(\mathbf{y}) = \frac{1}{2} \|\mathbf{\Phi}(\mathbf{y})\|^2_2$ be a merit function measuring closeness to the solution (at which $M(\mathbf{y})=0$). We determine the step size for the semismooth Newton method by a line search on $M$, aligned with the standard Wolfe conditions. This gives us the proposed line search formulated as follows:

\begin{align}
    \gamma^*=&\max_{\gamma} \gamma \\
    \text{s.t.}\quad &M(\mathbf{y}- \gamma\boldsymbol{\Delta})\leq M(\mathbf{y}) - c_1\,\gamma\,\nabla M(\mathbf{y})^T\boldsymbol{\Delta}\\
    & \nabla M(\mathbf{y}-\gamma \boldsymbol{\Delta})^T\boldsymbol{\Delta}\leq c_2 \nabla M(\mathbf{y})^T\boldsymbol{\Delta}\\
    & 0<\gamma\leq1
\end{align}

where $\nabla M(\mathbf{y})=\mathbf{J}_{\Phi}(\mathbf{y})^T\boldsymbol{\Phi}$. Noting that the cost of determining $\boldsymbol{\Phi}$ and $\mathbf{J}_{\Phi}(\mathbf{y}^{(k)})$ is relatively low given the linearized approximation, this line search can be carried out using an efficient binary-search scheme. The existence of $\gamma^*$ is guaranteed by the fact that $\nabla M^T\mathbf{\Delta}=\|\mathbf{\Phi}\|_2^2>0$ for all $\mathbf{\Phi}\neq\mathbf{0}$, so that $-\mathbf{\Delta}$ is a descent direction for $M$ (this can be shown with simple algebra skipped here for brevity). In our approach, we set $c_1=10^{-4}$ and $c_2=0.9$.

If $m$ is large, the required Jacobian inverse makes this approach prohibitively expensive; we therefore assume a small number of constraints, as is common in TO.

\subsection{Convergence Analysis}
\label{app:converge}
Let the objective function $f$ and constraint functions $g_j ,\; \forall j\in[m]$ in (\ref{eqn:linearized}) be $L_f$-smooth, namely that $\|\nabla f(\mathbf{a})-\nabla f(\mathbf{b})\|_2\leq L_f\|\mathbf{a}-\mathbf{b}\|_2$ for all $\mathbf{a},\mathbf{b}\in\mathbb{R}^N$, then we have the following:

\begin{equation}
\label{eqn:quad_limit}
\begin{aligned}
f(\mathbf{b})-f(\mathbf{a}) & =\int_{0}^{1}\langle\nabla f(\mathbf{a}+\theta \cdot(\mathbf{b}-\mathbf{a})), \mathbf{b}-\mathbf{a}\rangle \mathrm{d} \theta \\
& =\left(\int_{0}^{1}\langle\nabla f(\mathbf{a}+\theta \cdot(\mathbf{b}-\mathbf{a}))-\nabla f(\mathbf{a}), \mathbf{b}-\mathbf{a}\rangle \mathrm{d} \theta\right)+\langle\nabla f(\mathbf{a}), \mathbf{b}-\mathbf{a}\rangle\\
& \leq\left(\int_{0}^{1}\|\nabla f(\mathbf{a}+\theta \cdot(\mathbf{b}-\mathbf{a}))-\nabla f(\mathbf{a})\|_{2} \cdot\|\mathbf{b}-\mathbf{a}\|_{2} \mathrm{~d} \theta\right)+\langle\nabla f(\mathbf{a}), \mathbf{b}-\mathbf{a}\rangle \\
& \leq\left(\int_{0}^{1} \theta L_f\|\mathbf{b}-\mathbf{a}\|_{2}^{2} \mathrm{~d} \theta\right)+\langle\nabla f(\mathbf{a}), \mathbf{b}-\mathbf{a}\rangle \\
& =\frac{L_f}{2}\|\mathbf{b}-\mathbf{a}\|_{2}^{2}+\nabla f(\mathbf{a})^T(\mathbf{b}-\mathbf{a})
\end{aligned}
\end{equation}

This is true for both $f$ and $g_j$. Now let $\boldsymbol{\rho}_{t+1} = \Pi_{\Omega}(\boldsymbol{\rho}_{t}-\alpha\nabla f)=\boldsymbol{\rho}_{t}-\alpha\nabla f + \boldsymbol{\delta}^*$ be the update at each iteration, where $\boldsymbol{\delta}^*$ is the solution of (\ref{eqn:linearized}). We can use (\ref{eqn:quad_limit}) to obtain:

\begin{equation}
\begin{aligned}
    f(\boldsymbol{\rho}_{t+1}) - f(\boldsymbol{\rho}_{t}) &\leq \frac{L_f}{2}\|\boldsymbol{\rho}_{t+1}-\boldsymbol{\rho}_{t}\|_{2}^{2}+\nabla f(\boldsymbol{\rho}_{t})^T(\boldsymbol{\rho}_{t+1}-\boldsymbol{\rho}_{t})\\
    & \leq \frac{L_f}{2}\|\boldsymbol{\rho}_{t+1}-\boldsymbol{\rho}_{t}\|_{2}^{2}-
    \frac{1}{\alpha}(\boldsymbol{\rho}_{t+1}-\boldsymbol{\rho}_{t}-\boldsymbol{\delta}^*)^T(\boldsymbol{\rho}_{t+1}-\boldsymbol{\rho}_{t})\\
    &\leq (\frac{L_f}{2}-\frac{1}{\alpha}) \|\boldsymbol{\rho}_{t+1}-\boldsymbol{\rho}_{t}\|_{2}^{2} + \frac{1}{\alpha} \boldsymbol{\delta}^{*T}(\boldsymbol{\rho}_{t+1}-\boldsymbol{\rho}_{t})
\end{aligned}
\end{equation}

Now, assuming $\boldsymbol{\rho}_t\in\hat{\Omega}_{t+1}$, meaning the iterate at iteration $t$ lies in the linearized convex feasible set $\hat{\Omega}_{t+1}$, then by convexity of $\hat{\Omega}_{t+1}$, we have $\boldsymbol{\delta}^{*T}(\boldsymbol{\rho}_{t+1}-\boldsymbol{\rho}_{t})\leq 0$, since if $\boldsymbol{\rho}_{t+1}\in\hat{\Omega}_{t+1}$, then (\ref{eqn:linearized}) will yield $\boldsymbol{\delta}^*=\mathbf 0$ and otherwise the correction to the convex set from a point outside the set will yield $\boldsymbol{\delta}^{*T}(\boldsymbol{\rho}_{t+1}-\boldsymbol{\rho}_{t})\leq 0$. Thus, we have:

\begin{equation}
    f(\boldsymbol{\rho}_{t+1}) - f(\boldsymbol{\rho}_{t}) \leq (\frac{L_f}{2}-\frac{1}{\alpha}) \|\boldsymbol{\rho}_{t+1}-\boldsymbol{\rho}_{t}\|_{2}^{2}
\end{equation}

Now let $L=\frac{1}{\alpha}-\frac{L_f}{2}$, for a small enough $\alpha <\frac{2}{L_f}$, $L>0$ and we can sum the iteration over $K$ steps, yielding the following inequality:

\begin{equation}
     L\sum_{t=0}^{K-1} \|\boldsymbol{\rho}_{t+1}-\boldsymbol{\rho}_{t}\|_{2}^{2}\leq f(\boldsymbol{\rho}_{0}) - f(\boldsymbol{\rho}_{K}).
\end{equation}

Now, assuming $f$ is bounded below on the box-constrained domain, namely $f(\boldsymbol{\rho})\geq f_{\text{inf}}$, we have:

\begin{equation}
    \sum_{t=0}^{K-1} \|\boldsymbol{\rho}_{t+1}-\boldsymbol{\rho}_{t}\|_{2}^{2}\leq \frac{f(\boldsymbol{\rho}_{0}) - f_{\text{inf}}}{L},
\end{equation}

which means that there must exist one iteration in which $\|\boldsymbol{\rho}_{t+1}-\boldsymbol{\rho}_{t}\|_{2}^{2}\leq \frac{f(\boldsymbol{\rho}_{0}) - f_{\text{inf}}}{L\,K}$. In other words:

\begin{equation}
    \min_t\|\boldsymbol{\rho}_{t+1}-\boldsymbol{\rho}_{t}\|_{2}\leq\sqrt{\frac{f(\boldsymbol{\rho}_{0}) - f_{\text{inf}}}{L\,K}},
\end{equation}

which implies that the Euclidean norm of the design variables update converges at a rate of $O(\frac{1}{\sqrt{K}})$.

Note that in our analysis, we made two crucial assumptions. First, we assumed that the design variables at each iteration are in the feasible set. This assumption will hold for linear constraints at all times, since the convex set for the projection is constant throughout the iterations. Noting that the most common constraints in TO problems, namely volume constraints, are linear, this assumption is valid in most cases. However, for the linearized approximation of nonlinear constraints, this may not always be true, but for a well-behaved problem the updates shrink over iterations, making the linear approximations progressively more accurate, so the assumption holds after the initial large updates. To prevent non-convergence in our approach, if the constraint violations are above the tolerance for the Newton iteration, we fall back to the same minimum step size as used in the starting iteration. The second assumption that was made here was the L-smooth nature of the objective function. We argue that this assumption should be true in most TO problems based on FEA solutions. In the section that follows, we formalize this argument.

\subsection{On the L-Smooth Assumption of the Objective Function}
In most TO problems, the objective function is defined based on solutions from an FEA solver. Thus, it is safe to assume that the objective function can be formulated as a polynomial expression of the solution to the FEA problem. Therefore, if we can show L-smoothness of the solution to the FEA problem, it can be shown that, in most cases, the objective function will also be L-smooth. Let $K(\boldsymbol{\rho})$ be the FEA system matrix for design variables $\boldsymbol{\rho}$. Let the solution to this system of equations be $u=K^{-1}f$, where $f$ is the forcing term, assumed to be constant here for simplicity. Then, if we show that for any arbitrary $q\in\mathbb{R}^N$, the function $\psi(\boldsymbol{\rho})=q^Tu=q^TK^{-1}f$ is $L_f$-smooth, then it can be easily inferred that any multi-linear polynomial function of $u$ will be $L_f$-smooth as well (inferred from the smoothness of the polynomial form itself). To show this, we formulate the Hessian of the proposed test function $\psi$. The gradient of the function can be written as:

\begin{equation}
    \frac{\partial \psi}{\partial \rho_i}=-q^T K^{-1} \frac{\partial K}{\partial \rho_i} K^{-1}f
\end{equation}

given the identity $\frac{\partial K^{-1}}{\partial \rho_i}=- K^{-1} \frac{\partial K}{\partial \rho_i} K^{-1}$ we have:

\[
\frac{\partial^2 \psi}{\partial \rho_i \partial \rho_j}  = q^T\left[K^{-1} \frac{\partial^2 K}{\partial \rho_j \partial \rho_i} K^{-1}-K^{-1} \frac{\partial K}{\partial \rho_j} K^{-1} \frac{\partial K}{\partial \rho_i} K^{-1}-K^{-1} \frac{\partial K}{\partial \rho_i} K^{-1} \frac{\partial K}{\partial \rho_j} K^{-1}\right] f.
\]

We can assume elements in $K$ are smooth functions of $\boldsymbol{\rho}$, which is commonly the case; for example, in the SIMP approach the elements of $K$ are polynomial (penalized) functions of the densities, which sometimes also involve a smoothed Heaviside projection. Assuming further that these functions are defined and finite over the box-constrained $\boldsymbol{\rho}$, we expect:

\begin{align}
    &\|\frac{\partial K}{\partial \rho_i}\|\leq C_1,\quad \forall i\in[N]\\
    &\|\frac{\partial^2 K}{\partial \rho_j \partial \rho_i}\|\leq C_2,\quad \forall i,j\in[N].
\end{align}

Combined with the fact that we assume $K^{-1}$ exists for all $\boldsymbol{\rho}\in\Omega$, and the feasible set $\Omega$ is non-empty and bounded, $\|K^{-1}\|$ will also be bounded within the feasible space of the problem, namely $\|K^{-1}\|\leq C_3$. Thus we can say that the Hessian matrix elements are bounded as well, $\left|\frac{\partial^2 \psi}{\partial \rho_i \partial \rho_j}\right|\leq C_4$, making the objective function $L_f$-smooth:

\begin{equation}
    \|\nabla \psi(\mathbf{a})-\nabla \psi(\mathbf{b})\|_2\leq L_\psi\|\mathbf{a}-\mathbf{b}\|_2, \quad \forall \mathbf{a},\mathbf{b}\in \Omega,
\end{equation}

for some constant $L_\psi$. In general, although the analysis above is true when $K$ is not ill-conditioned, if in any part of the feasible space the resulting system becomes ill-conditioned, the Lipschitz constant can be very large, making the step size prohibitively small for the optimization. Given this and the fact that finding the constant for any arbitrary TO problem is not trivial, we take a more pragmatic approach to setting the step size in our algorithm. We use the local approximation $\hat L_f=\frac{\|\nabla f(\boldsymbol{\rho}_t)-\nabla f(\boldsymbol{\rho}_{t-1})\|}{\|\boldsymbol{\rho}_t-\boldsymbol{\rho}_{t-1}\|}$ and set $\alpha=\frac{1}{\hat L_f}$ at each iteration $t$. Furthermore, when a problem is ill-conditioned, we fall back to a step size that limits the maximum change in the design variables.

\clearpage
\subsection{Full Experimental Results}
\label{app:results}
In this section, we provide detailed plots and visualizations of the results of each optimization problem. In the main body, we highlight only the primary findings.

The following conventions apply to every figure in this section. Design snapshots are taken at five approximately log-spaced iterations. All per-iteration plots show a 10-iteration moving average to remove noise, and the inset repeats the same axes after 50 iterations, where the solvers have stabilized and the plot is scaled more suitably.

\subsubsection{Volume-Constrained Minimum Compliance}
The following subsections give the full results for each mesh resolution.

\paragraph{Coarse Mesh Results}
Here we provide figures for the results of running each optimizer for the coarse $128 \times 64$ mesh.

\begin{figure}[H]
    \centering
    \includegraphics[width=\linewidth]{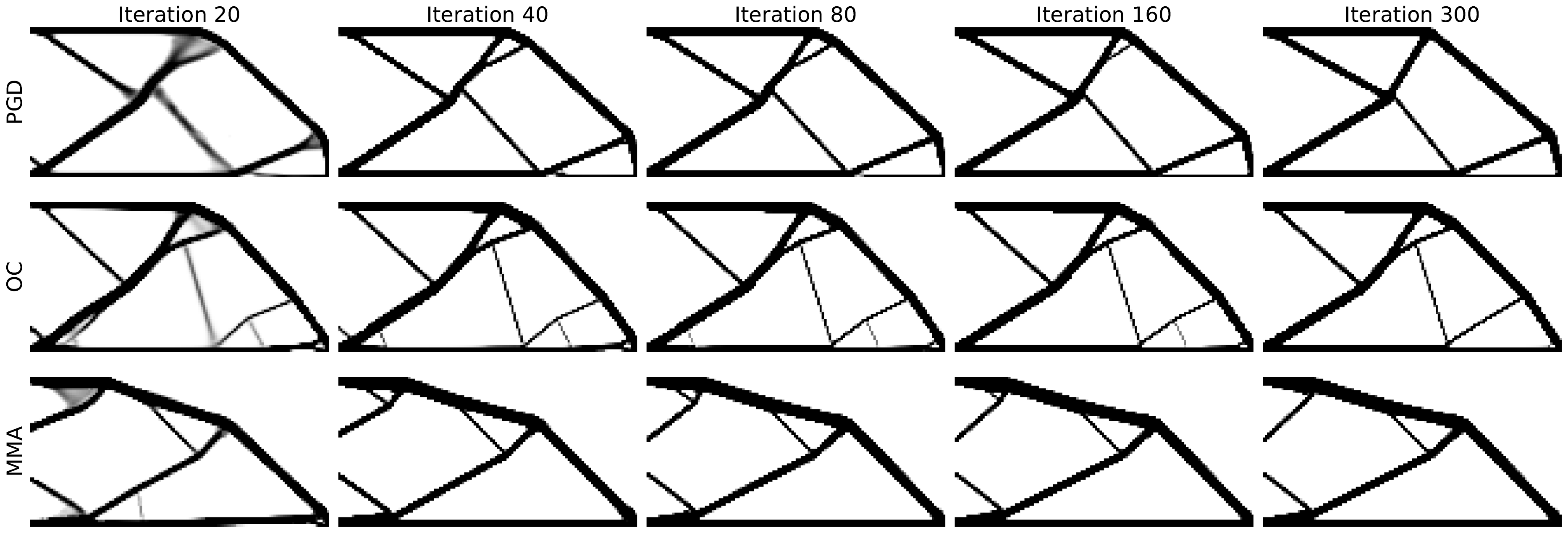}
    \caption{Designs produced by each optimizer for the volume-constrained minimum compliance problem on the cantilever beam with a volume fraction target of $0.2$, coarse $128 \times 64$ mesh.}
    \label{fig:mincomp_coarse_designs}
\end{figure}

\begin{figure}[H]
    \centering
    \includegraphics[width=\linewidth]{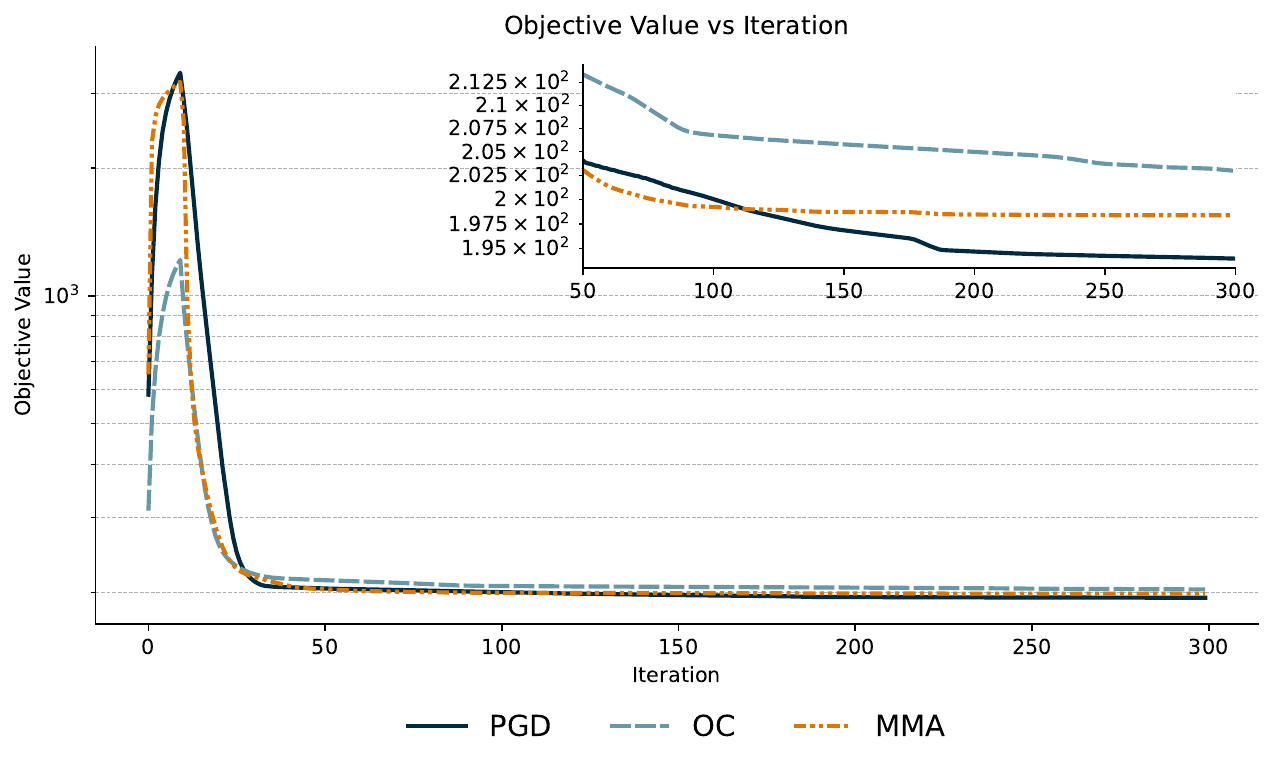}
    \caption{The value of the objective function per iteration for all solvers: volume-constrained minimum compliance problem, coarse $128 \times 64$ mesh.}
    \label{fig:mincomp_coarse_obj}
\end{figure}

\begin{figure}[H]
    \centering
    \includegraphics[width=\linewidth]{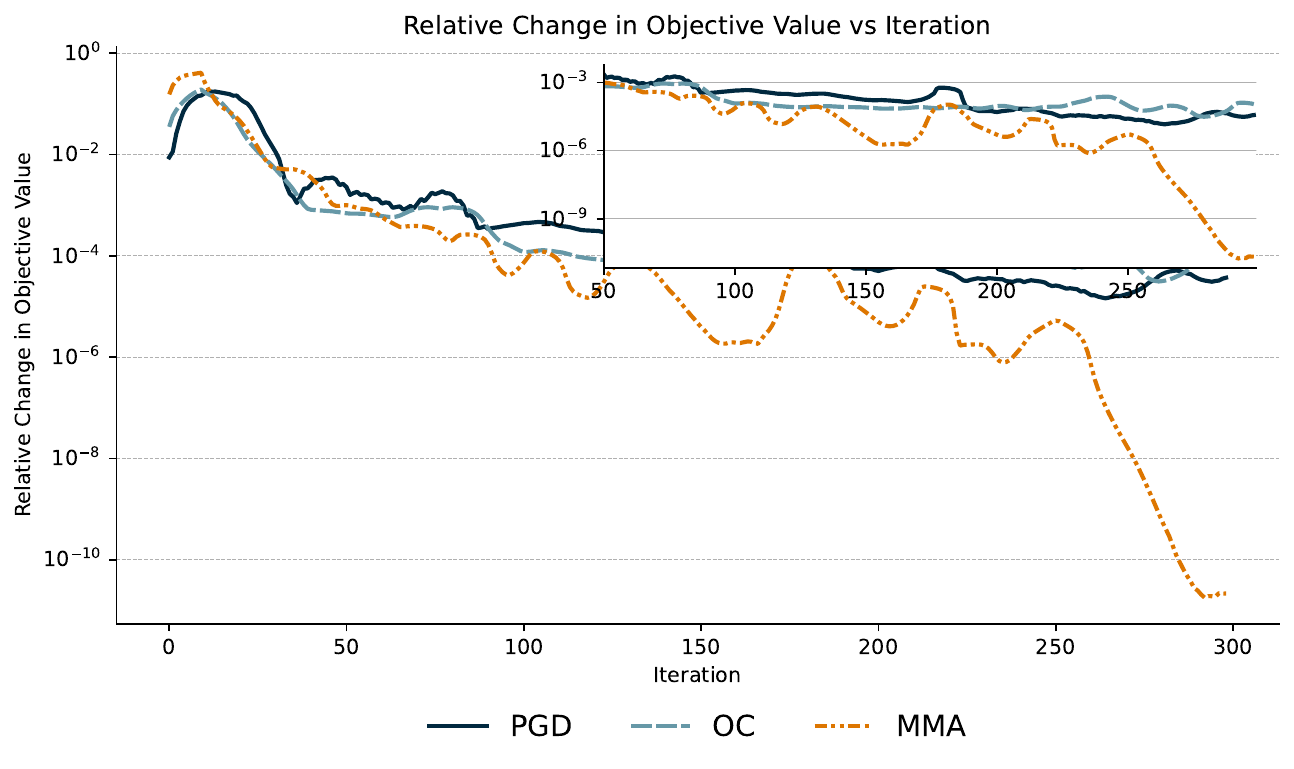}
    \caption{The relative change in the objective function per iteration for all solvers: volume-constrained minimum compliance problem, coarse $128 \times 64$ mesh.}
    \label{fig:mincomp_coarse_rel_obj}
\end{figure}

\begin{figure}[H]
    \centering
    \includegraphics[width=\linewidth]{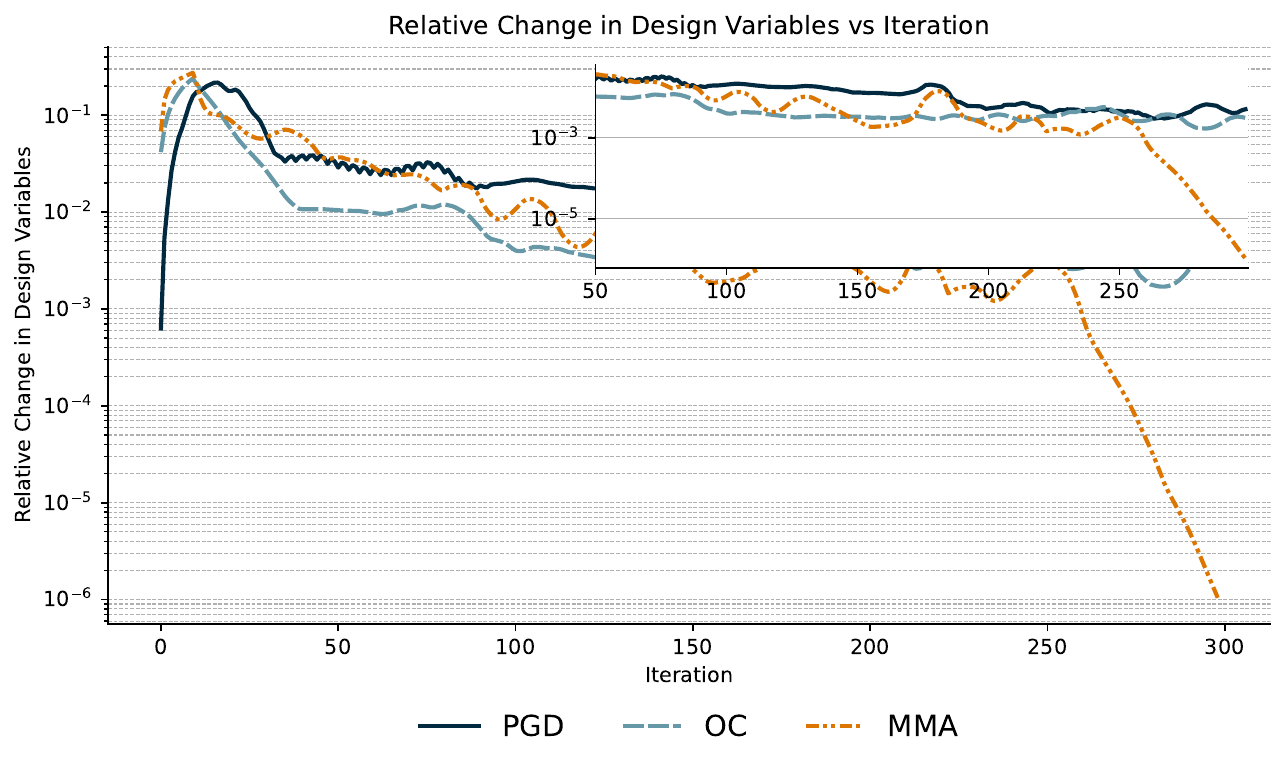}
    \caption{The relative change in the design variable norm per iteration for all solvers: volume-constrained minimum compliance problem, coarse $128 \times 64$ mesh.}
    \label{fig:mincomp_coarse_rel_change}
\end{figure}

\begin{figure}[H]
    \centering
    \includegraphics[width=\linewidth]{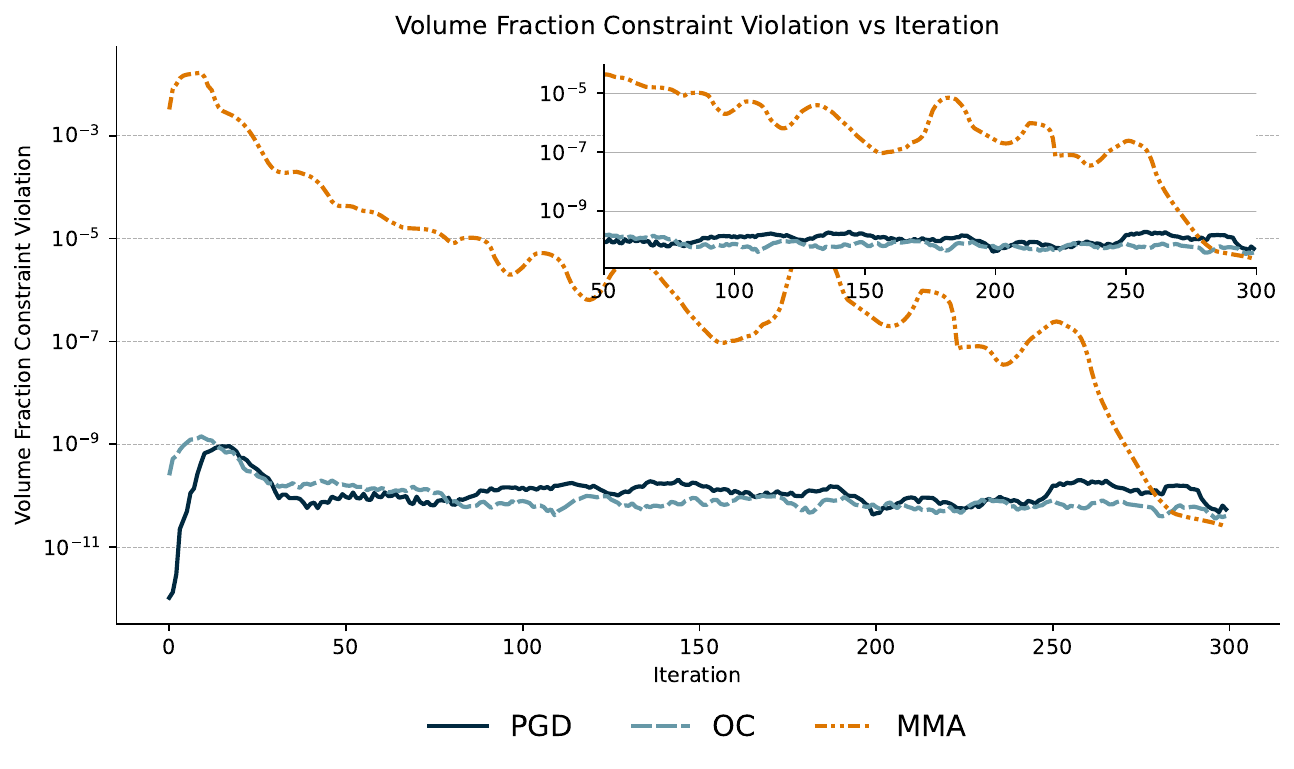}
    \caption{The constraint violation per iteration for all solvers: volume-constrained minimum compliance problem, coarse $128 \times 64$ mesh.}
    \label{fig:mincomp_coarse_violation}
\end{figure}

\paragraph{Medium Mesh Results}
Here we provide figures for the results of running each optimizer for the medium $256 \times 128$ mesh.

\begin{figure}[H]
    \centering
    \includegraphics[width=\linewidth]{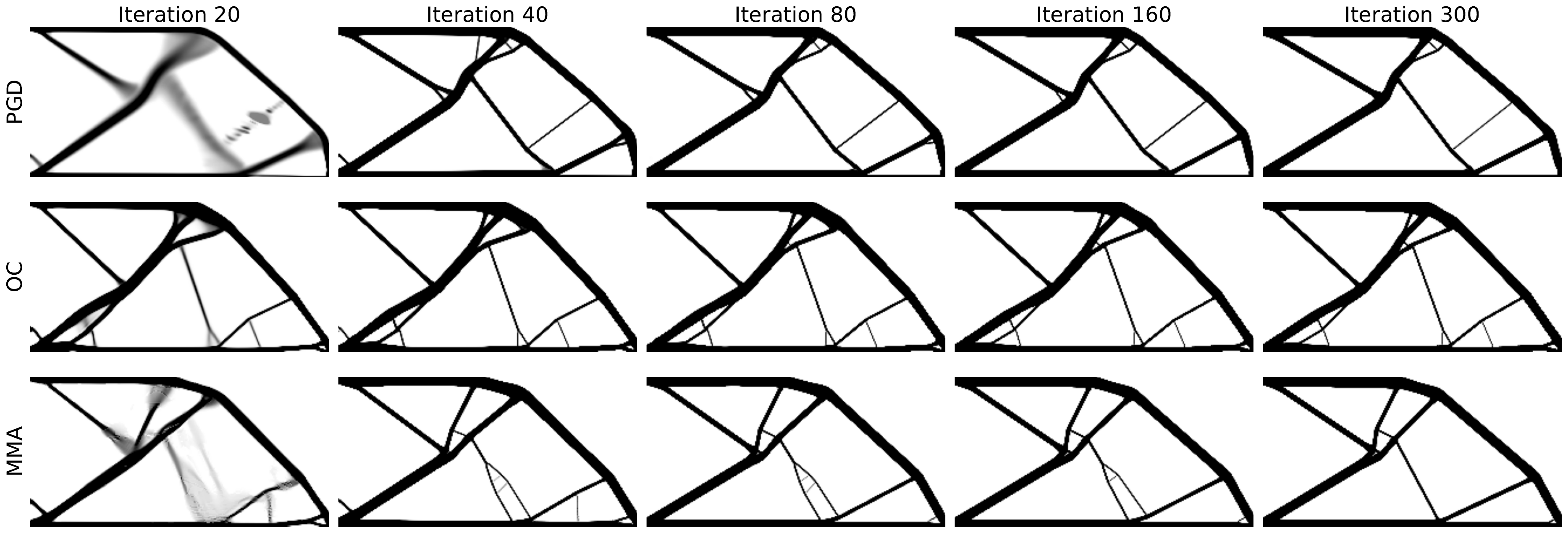}
    \caption{Designs produced by each optimizer for the volume-constrained minimum compliance problem on the cantilever beam with a volume fraction target of $0.2$, medium $256 \times 128$ mesh.}
    \label{fig:mincomp_med_designs}
\end{figure}

\begin{figure}[H]
    \centering
    \includegraphics[width=\linewidth]{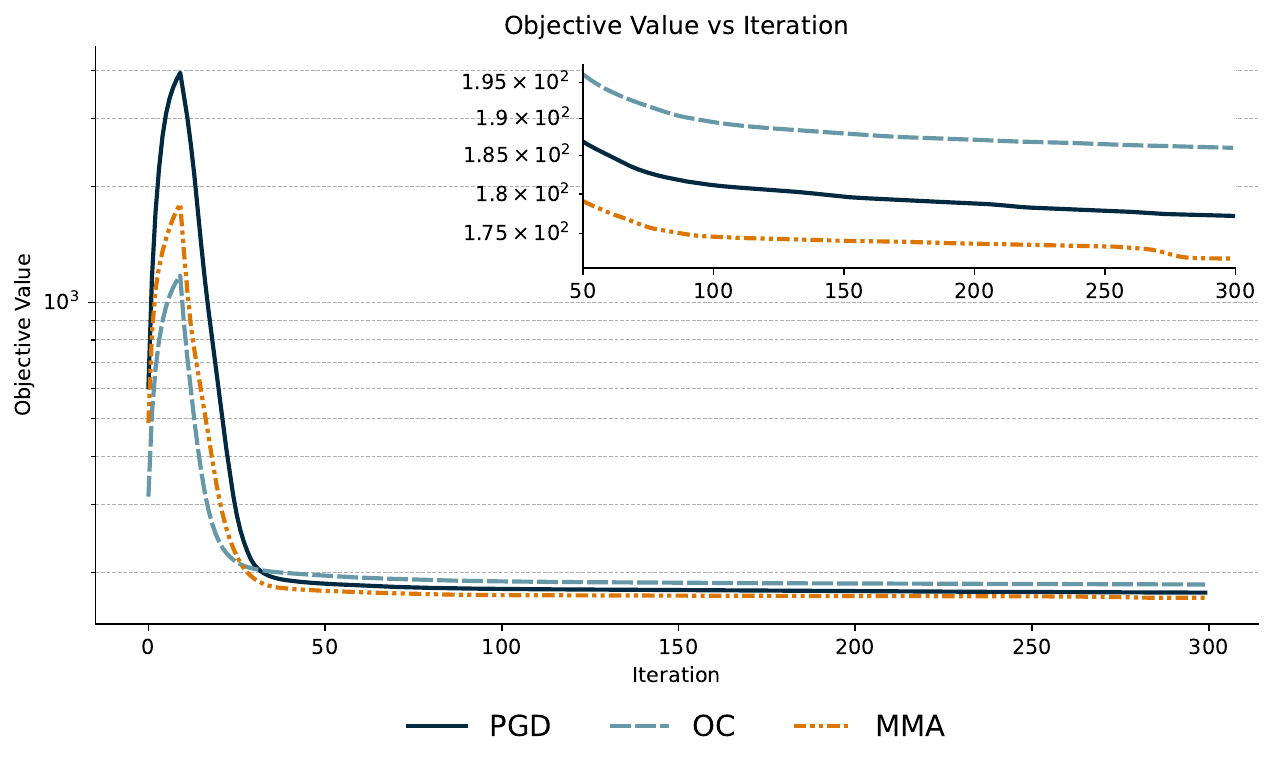}
    \caption{The value of the objective function per iteration for all solvers: volume-constrained minimum compliance problem, medium $256 \times 128$ mesh.}
    \label{fig:mincomp_med_obj}
\end{figure}

\begin{figure}[H]
    \centering
    \includegraphics[width=\linewidth]{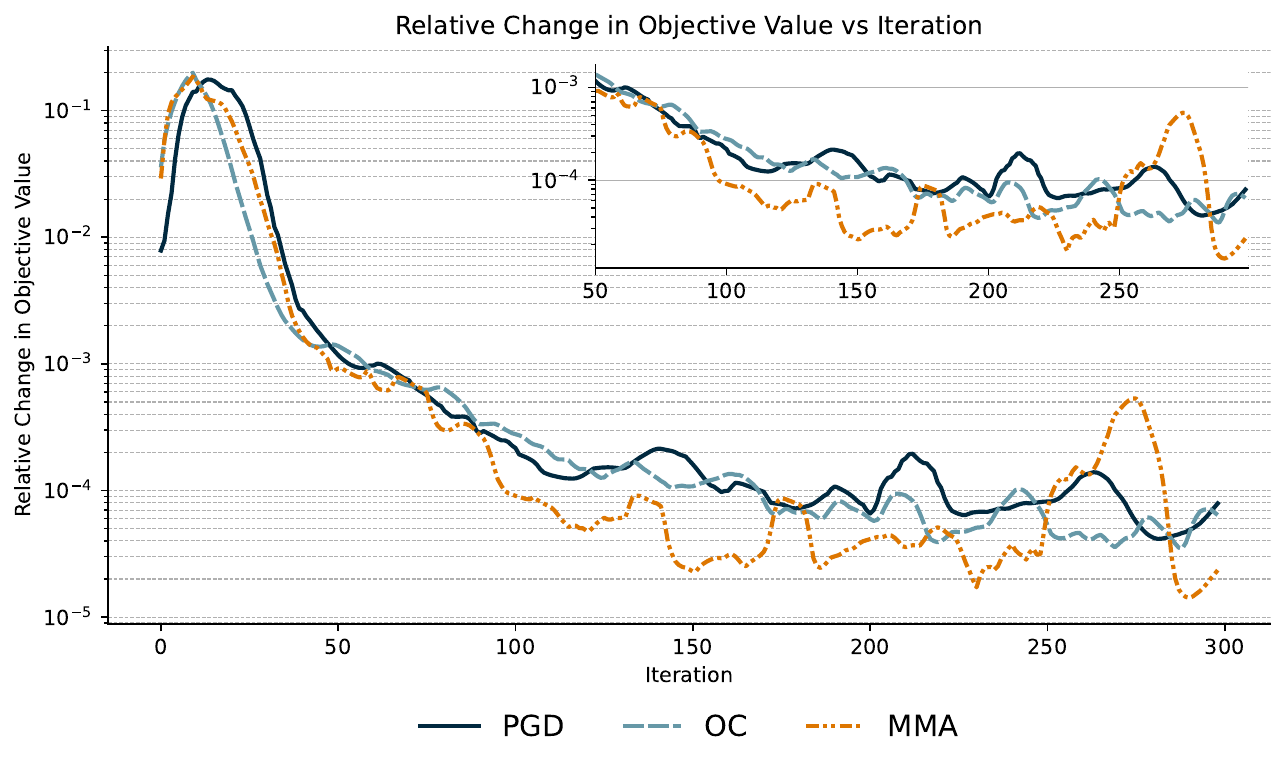}
    \caption{The relative change in the objective function per iteration for all solvers: volume-constrained minimum compliance problem, medium $256 \times 128$ mesh.}
    \label{fig:mincomp_med_rel_obj}
\end{figure}

\begin{figure}[H]
    \centering
    \includegraphics[width=\linewidth]{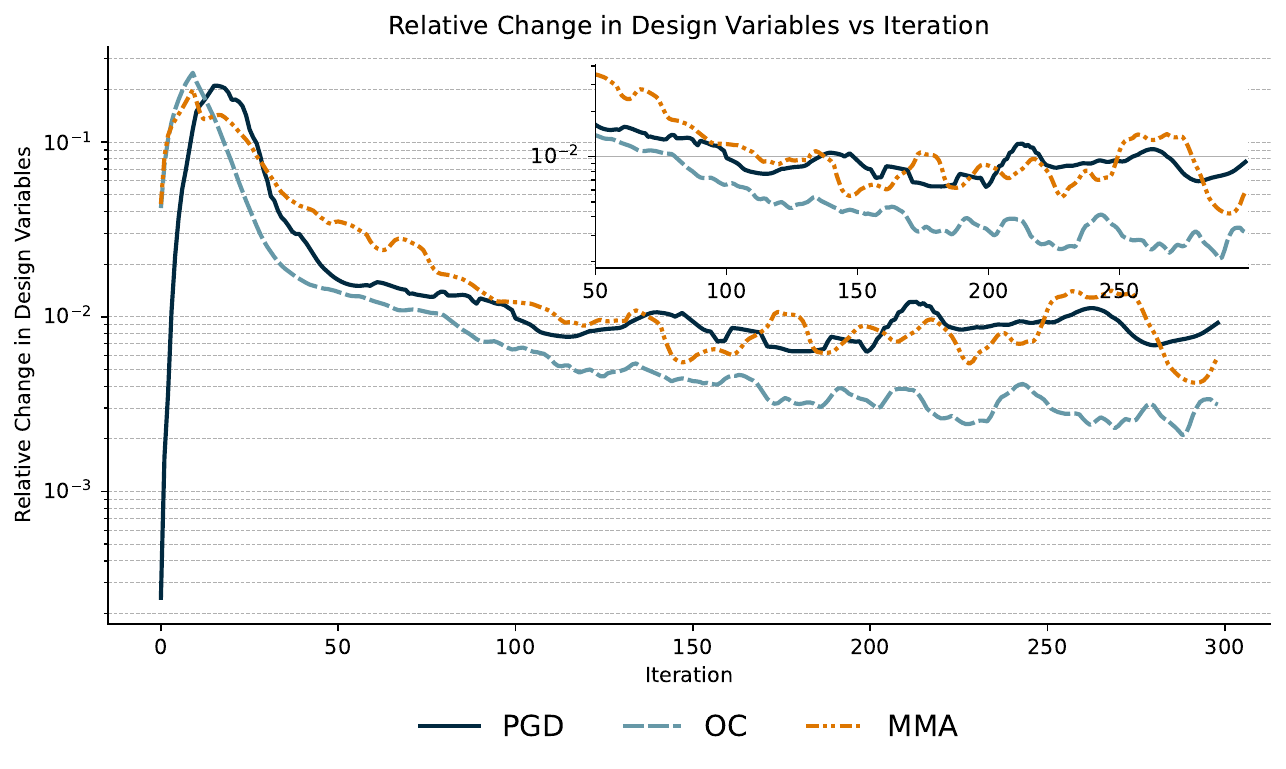}
    \caption{The relative change in the design variable norm per iteration for all solvers: volume-constrained minimum compliance problem, medium $256 \times 128$ mesh.}
    \label{fig:mincomp_med_rel_change}
\end{figure}

\begin{figure}[H]
    \centering
    \includegraphics[width=\linewidth]{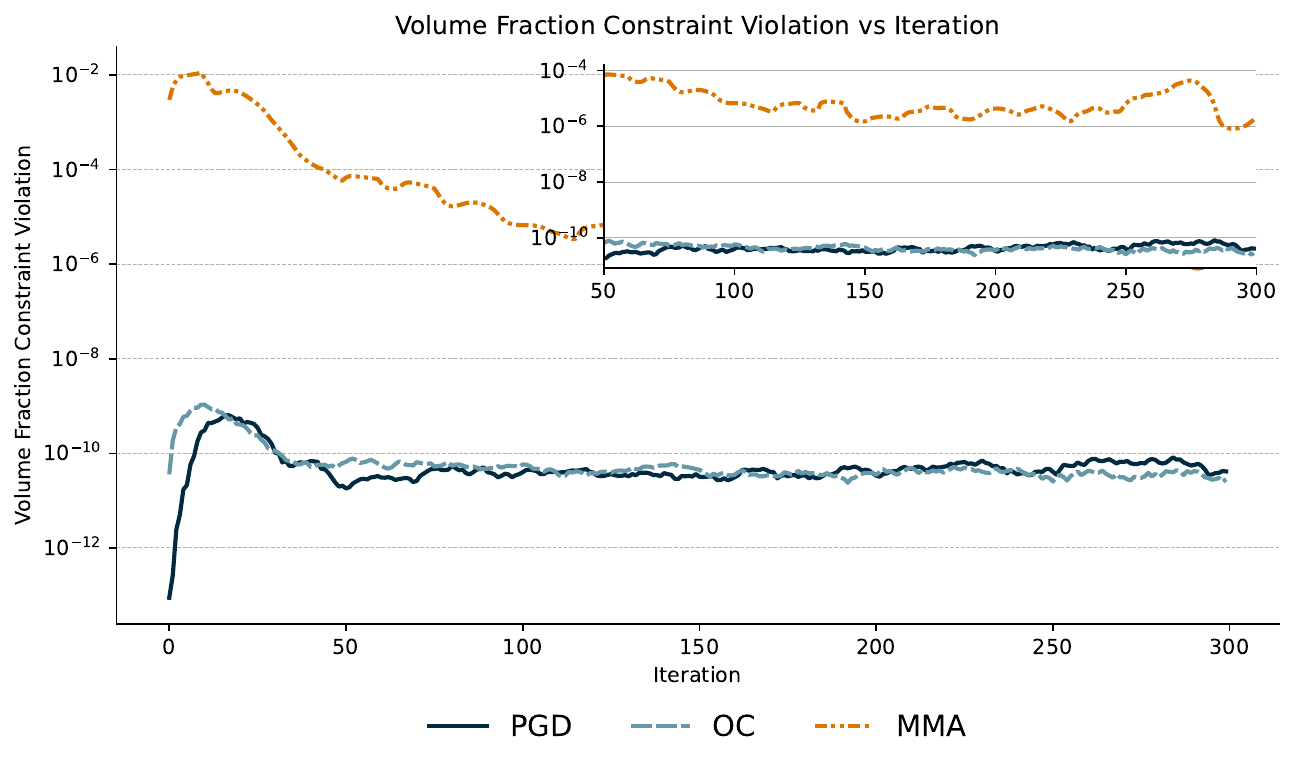}
    \caption{The constraint violation per iteration for all solvers: volume-constrained minimum compliance problem, medium $256 \times 128$ mesh.}
    \label{fig:mincomp_med_violation}
\end{figure}

\paragraph{Fine Mesh Results}
Here we provide figures for the results of running each optimizer for the fine $512 \times 256$ mesh.

\begin{figure}[H]
    \centering
    \includegraphics[width=\linewidth]{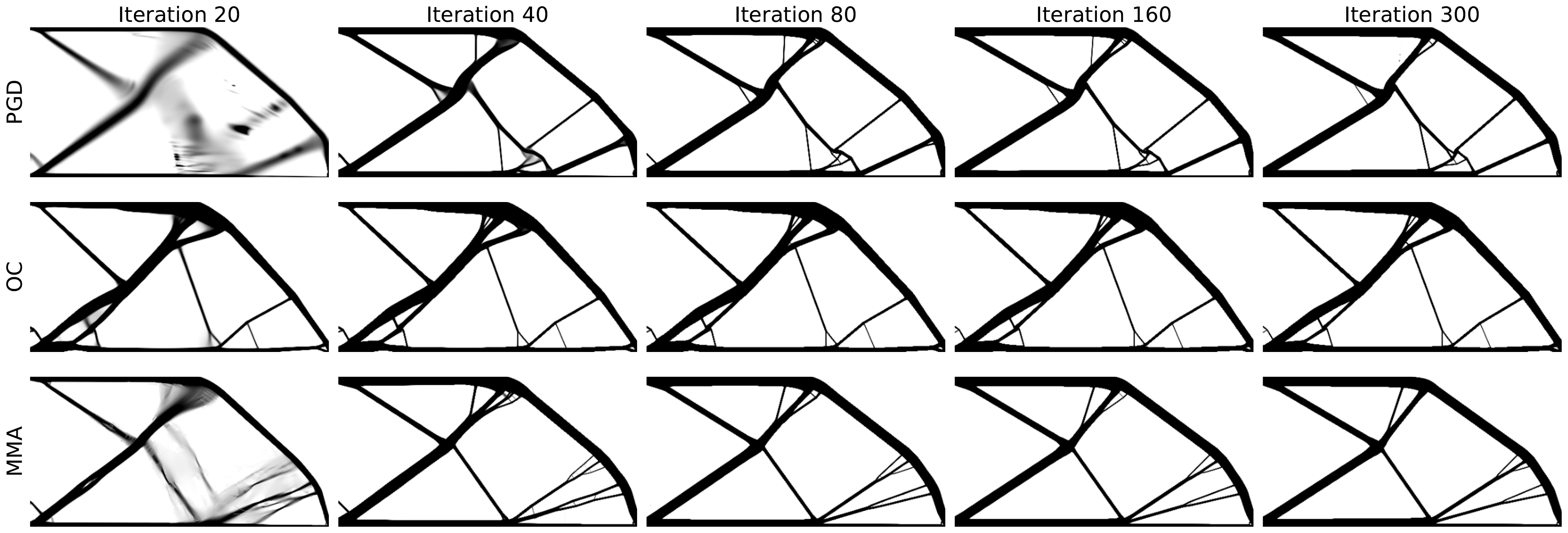}
    \caption{Designs produced by each optimizer for the volume-constrained minimum compliance problem on the cantilever beam with a volume fraction target of $0.2$, fine $512 \times 256$ mesh.}
    \label{fig:mincomp_fine_designs}
\end{figure}

\begin{figure}[H]
    \centering
    \includegraphics[width=\linewidth]{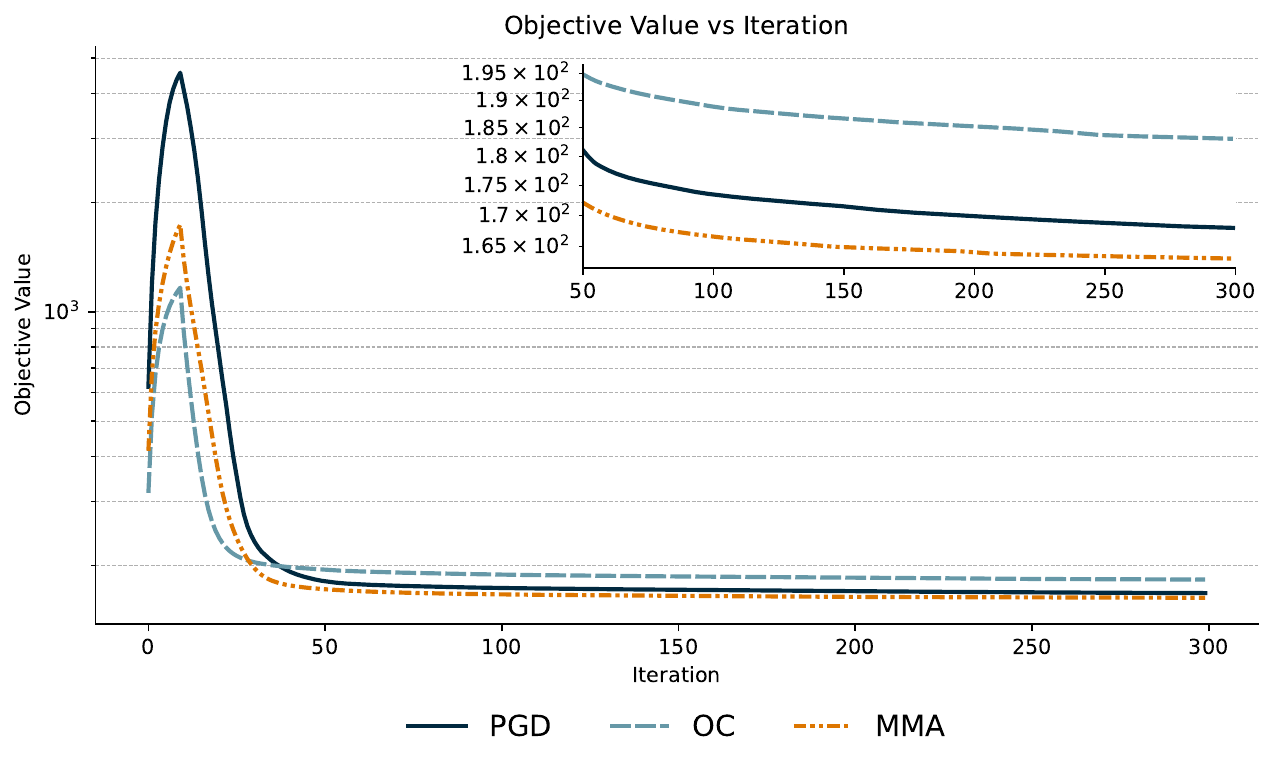}
    \caption{The value of the objective function per iteration for all solvers: volume-constrained minimum compliance problem, fine $512 \times 256$ mesh.}
    \label{fig:mincomp_fine_obj}
\end{figure}

\begin{figure}[H]
    \centering
    \includegraphics[width=\linewidth]{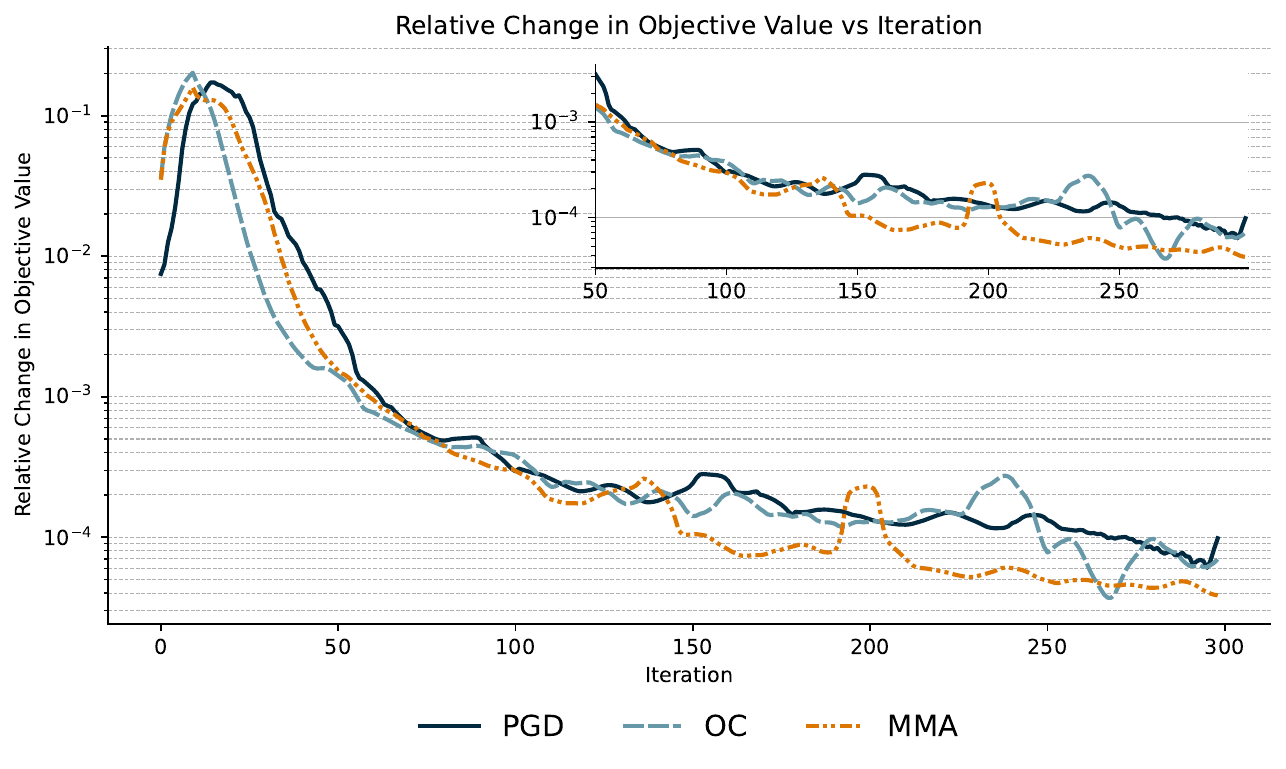}
    \caption{The relative change in the objective function per iteration for all solvers: volume-constrained minimum compliance problem, fine $512 \times 256$ mesh.}
    \label{fig:mincomp_fine_rel_obj}
\end{figure}

\begin{figure}[H]
    \centering
    \includegraphics[width=\linewidth]{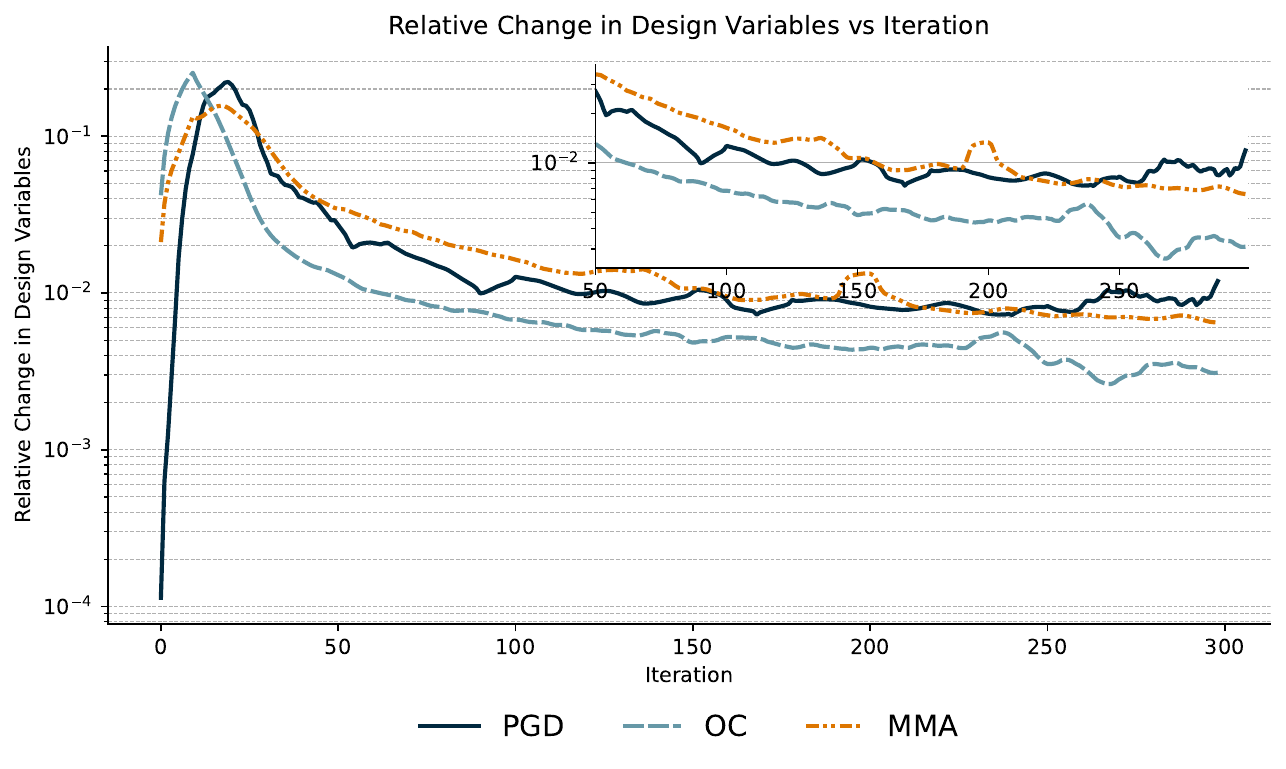}
    \caption{The relative change in the design variable norm per iteration for all solvers: volume-constrained minimum compliance problem, fine $512 \times 256$ mesh.}
    \label{fig:mincomp_fine_rel_change}
\end{figure}

\begin{figure}[H]
    \centering
    \includegraphics[width=\linewidth]{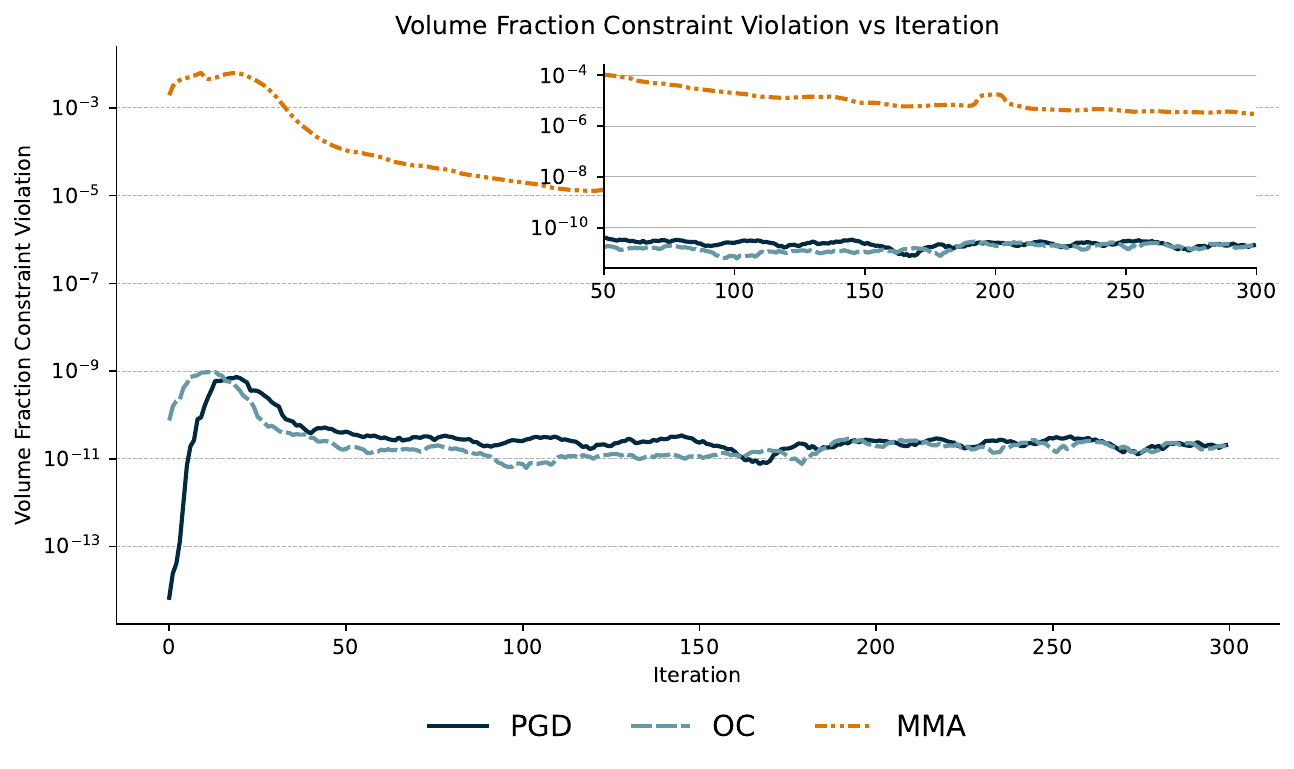}
    \caption{The constraint violation per iteration for all solvers: volume-constrained minimum compliance problem, fine $512 \times 256$ mesh.}
    \label{fig:mincomp_fine_violation}
\end{figure}

\clearpage
\subsubsection{Compliance-Constrained Minimum Volume}
The following subsections give the full results for each mesh resolution.

\paragraph{Coarse Mesh Results}
Here we provide figures for the results of running each optimizer for the coarse $128 \times 64$ mesh.

\begin{figure}[H]
    \centering
    \includegraphics[width=\linewidth]{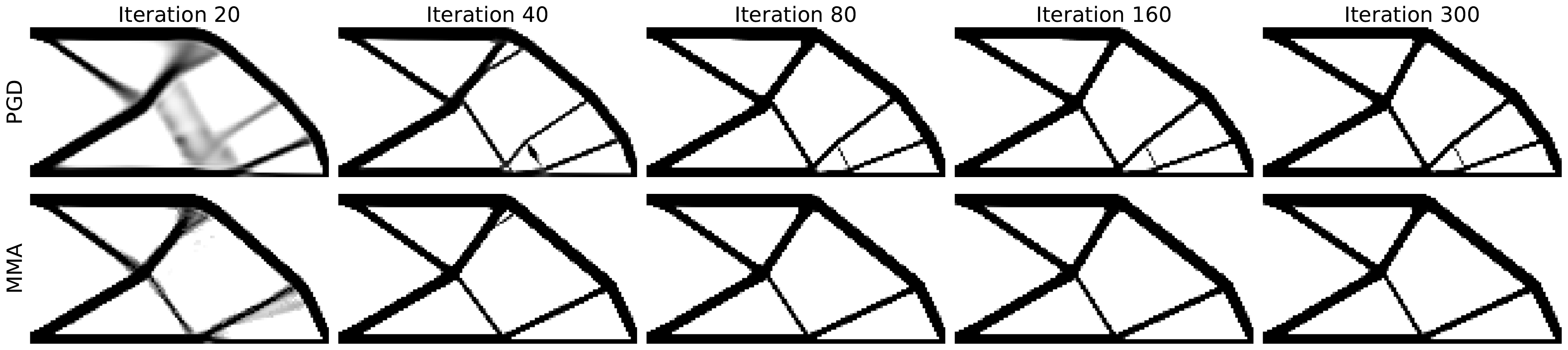}
    \caption{Designs produced by each optimizer for the compliance-constrained minimum volume problem on the cantilever beam with a compliance limit of $C_{\text{max}}=150$, coarse $128 \times 64$ mesh.}
    \label{fig:comp_coarse_designs}
\end{figure}

\begin{figure}[H]
    \centering
    \includegraphics[width=\linewidth]{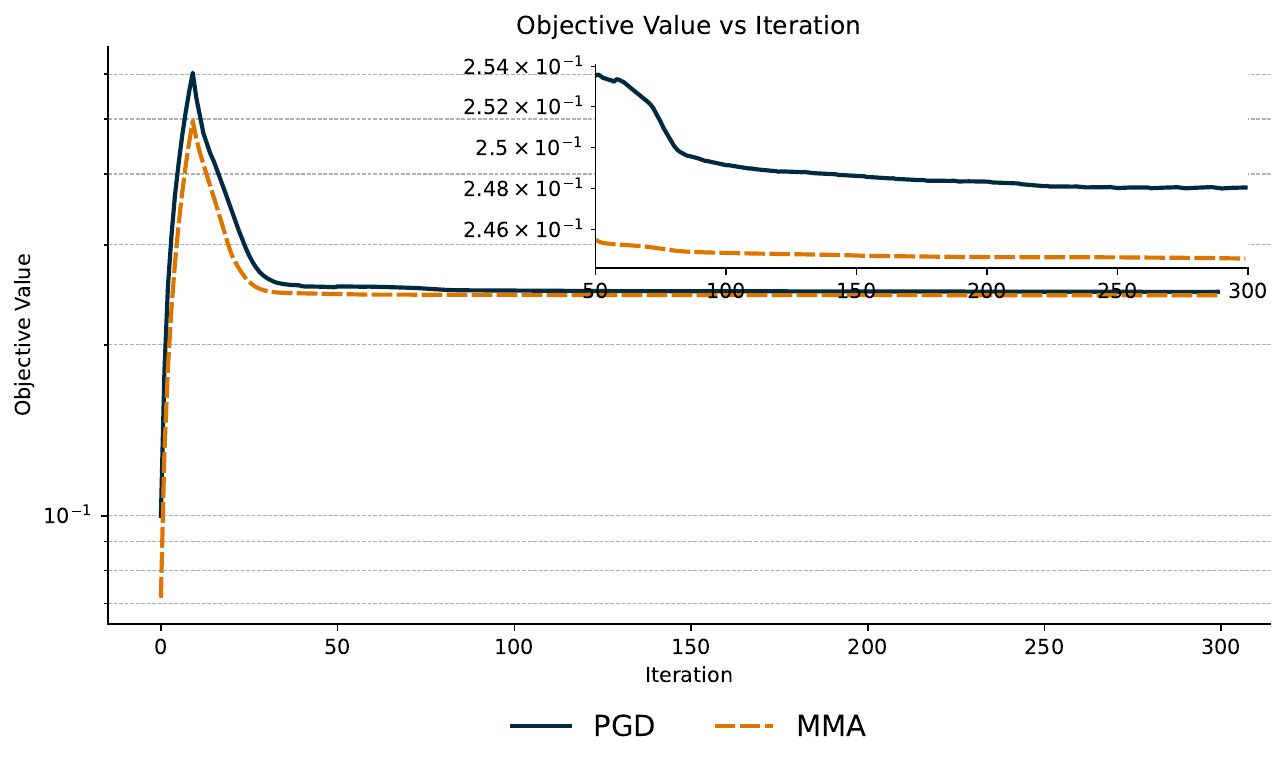}
    \caption{The value of the objective function per iteration for all solvers: compliance-constrained minimum volume problem, coarse $128 \times 64$ mesh.}
    \label{fig:comp_coarse_obj}
\end{figure}

\begin{figure}[H]
    \centering
    \includegraphics[width=\linewidth]{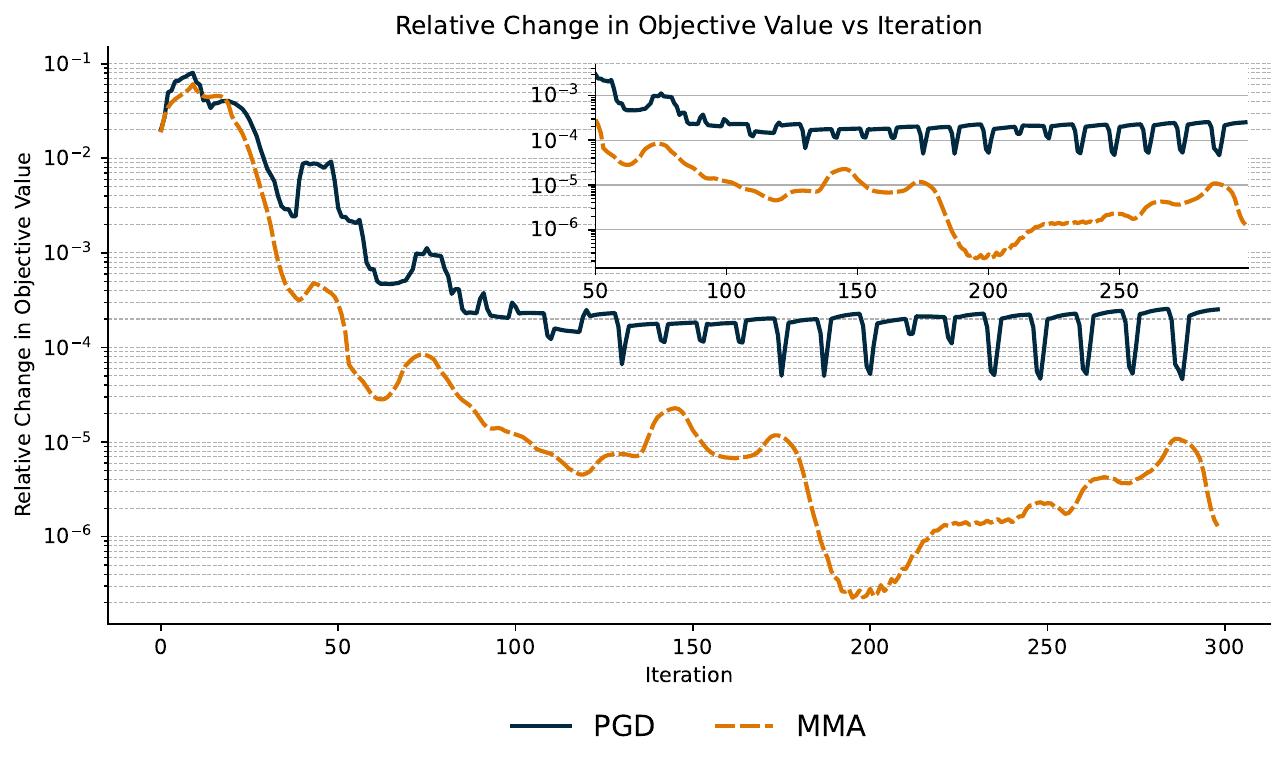}
    \caption{The relative change in the objective function per iteration for all solvers: compliance-constrained minimum volume problem, coarse $128 \times 64$ mesh.}
    \label{fig:comp_coarse_rel_obj}
\end{figure}

\begin{figure}[H]
    \centering
    \includegraphics[width=\linewidth]{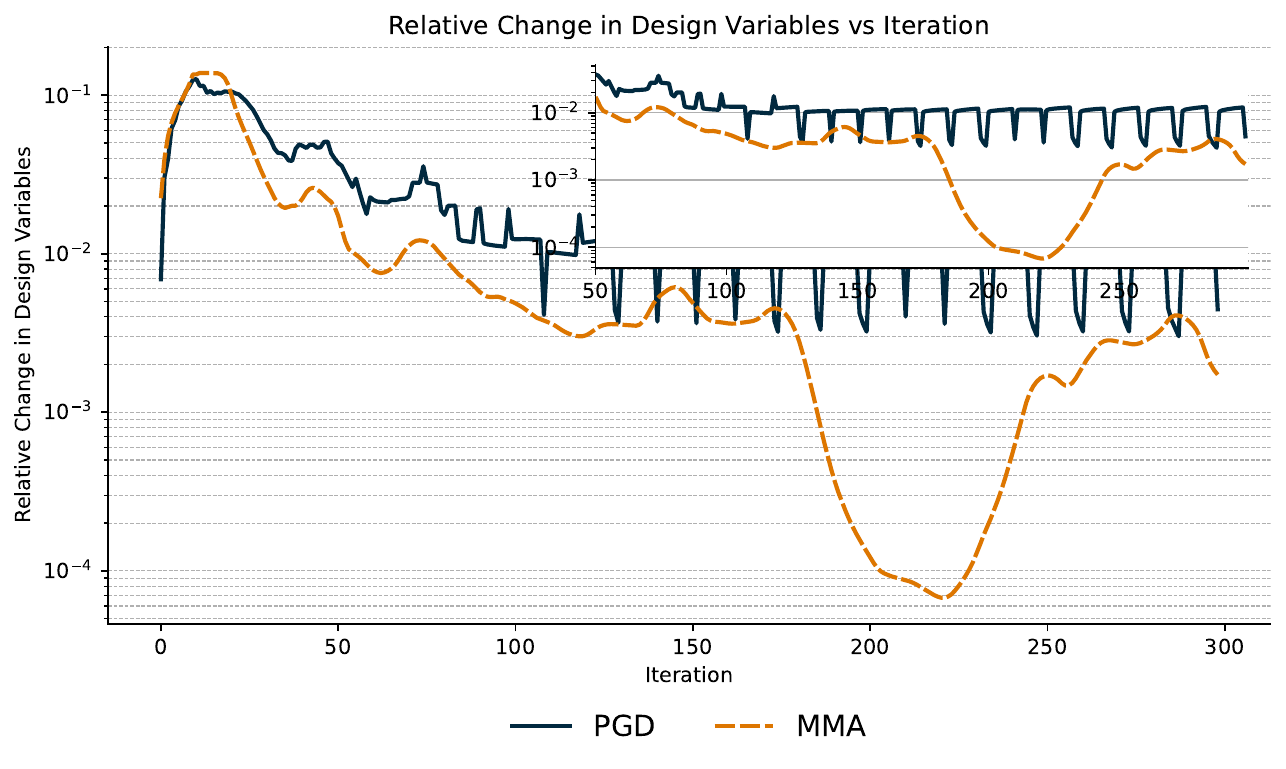}
    \caption{The relative change in the design variable norm per iteration for all solvers: compliance-constrained minimum volume problem, coarse $128 \times 64$ mesh.}
    \label{fig:comp_coarse_rel_change}
\end{figure}

\begin{figure}[H]
    \centering
    \includegraphics[width=\linewidth]{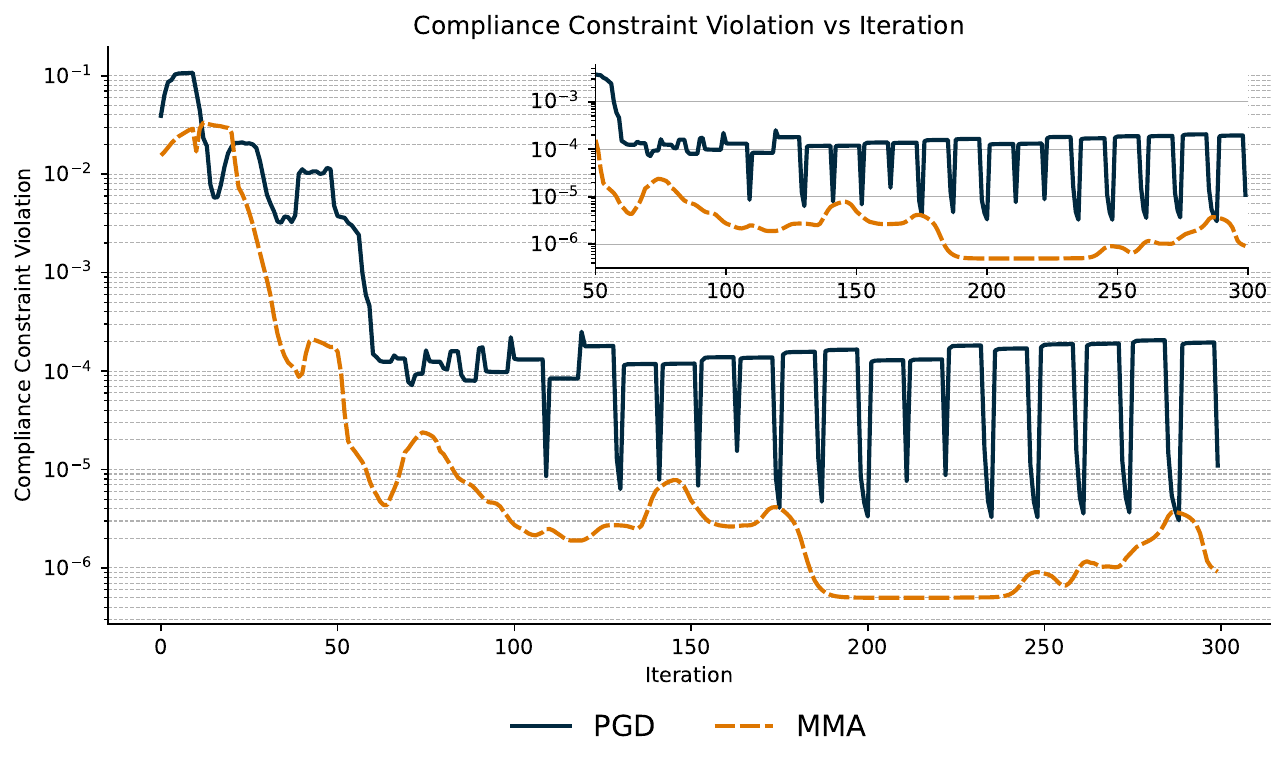}
    \caption{The constraint violation per iteration for all solvers: compliance-constrained minimum volume problem, coarse $128 \times 64$ mesh.}
    \label{fig:comp_coarse_violation}
\end{figure}

\paragraph{Medium Mesh Results}
Here we provide figures for the results of running each optimizer for the medium $256 \times 128$ mesh.

\begin{figure}[H]
    \centering
    \includegraphics[width=\linewidth]{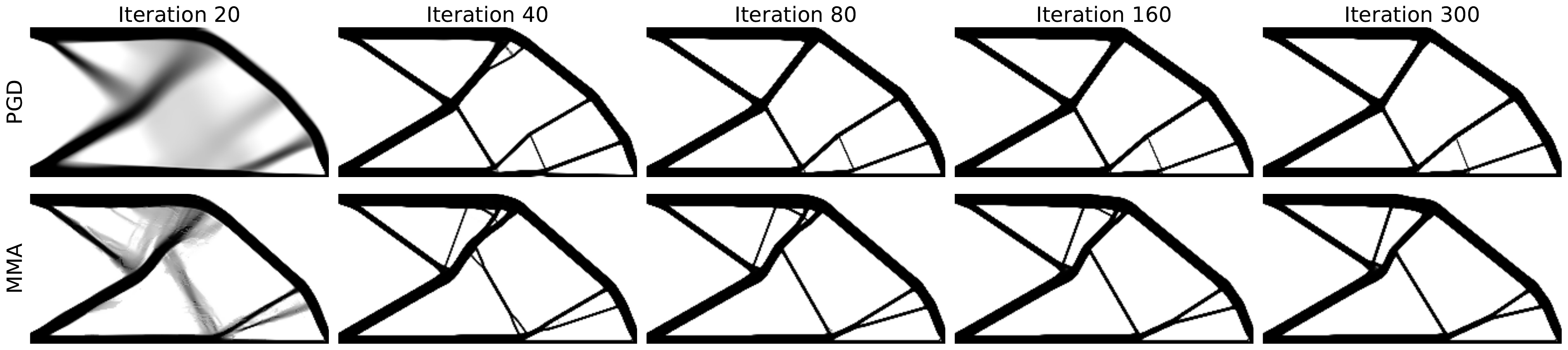}
    \caption{Designs produced by each optimizer for the compliance-constrained minimum volume problem on the cantilever beam with a compliance limit of $C_{\text{max}}=150$, medium $256 \times 128$ mesh.}
    \label{fig:comp_med_designs}
\end{figure}

\begin{figure}[H]
    \centering
    \includegraphics[width=\linewidth]{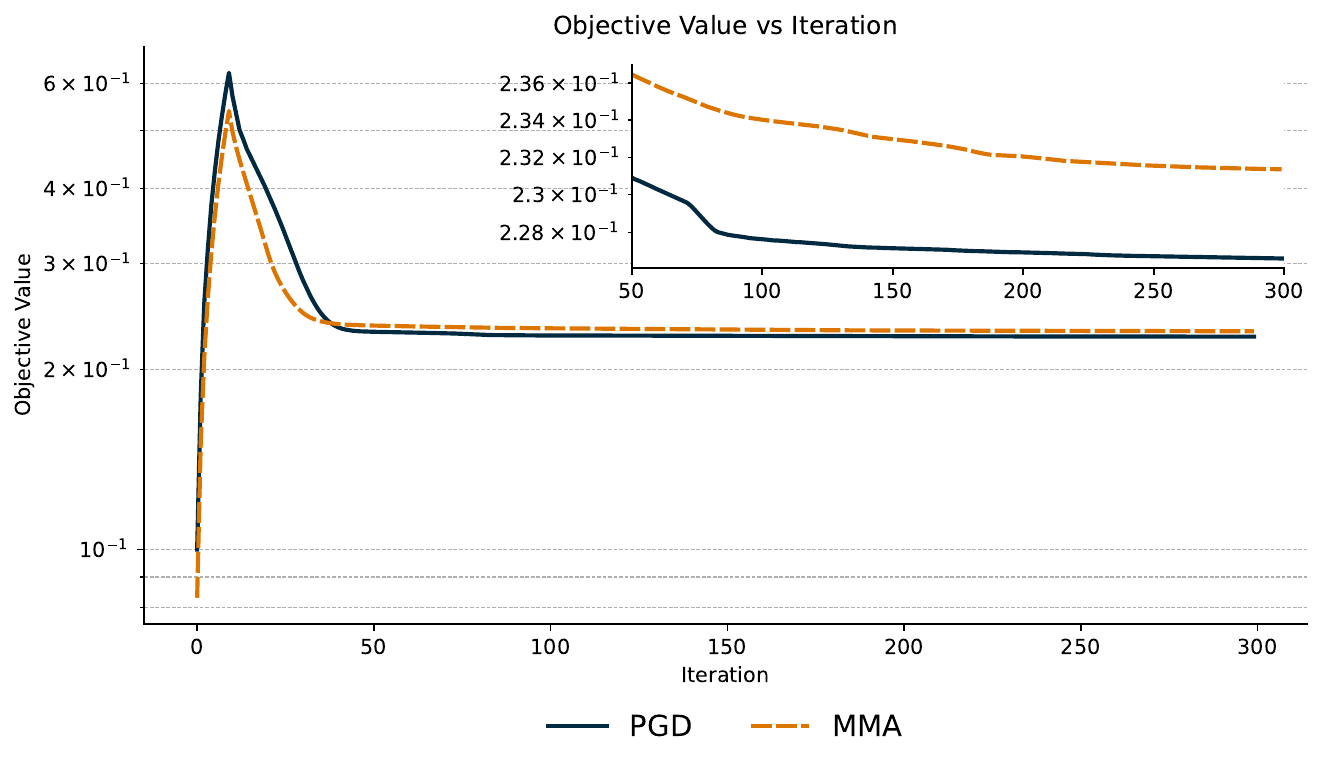}
    \caption{The value of the objective function per iteration for all solvers: compliance-constrained minimum volume problem, medium $256 \times 128$ mesh.}
    \label{fig:comp_med_obj}
\end{figure}

\begin{figure}[H]
    \centering
    \includegraphics[width=\linewidth]{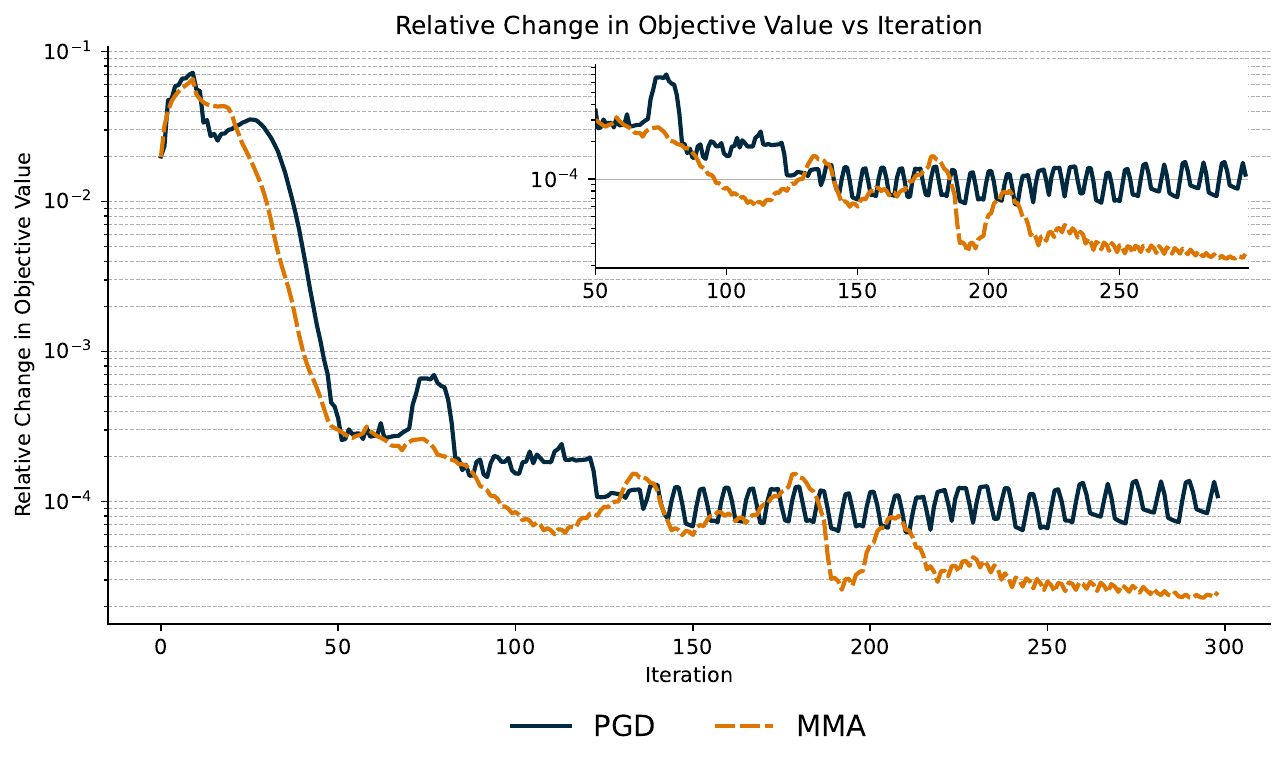}
    \caption{The relative change in the objective function per iteration for all solvers: compliance-constrained minimum volume problem, medium $256 \times 128$ mesh.}
    \label{fig:comp_med_rel_obj}
\end{figure}

\begin{figure}[H]
    \centering
    \includegraphics[width=\linewidth]{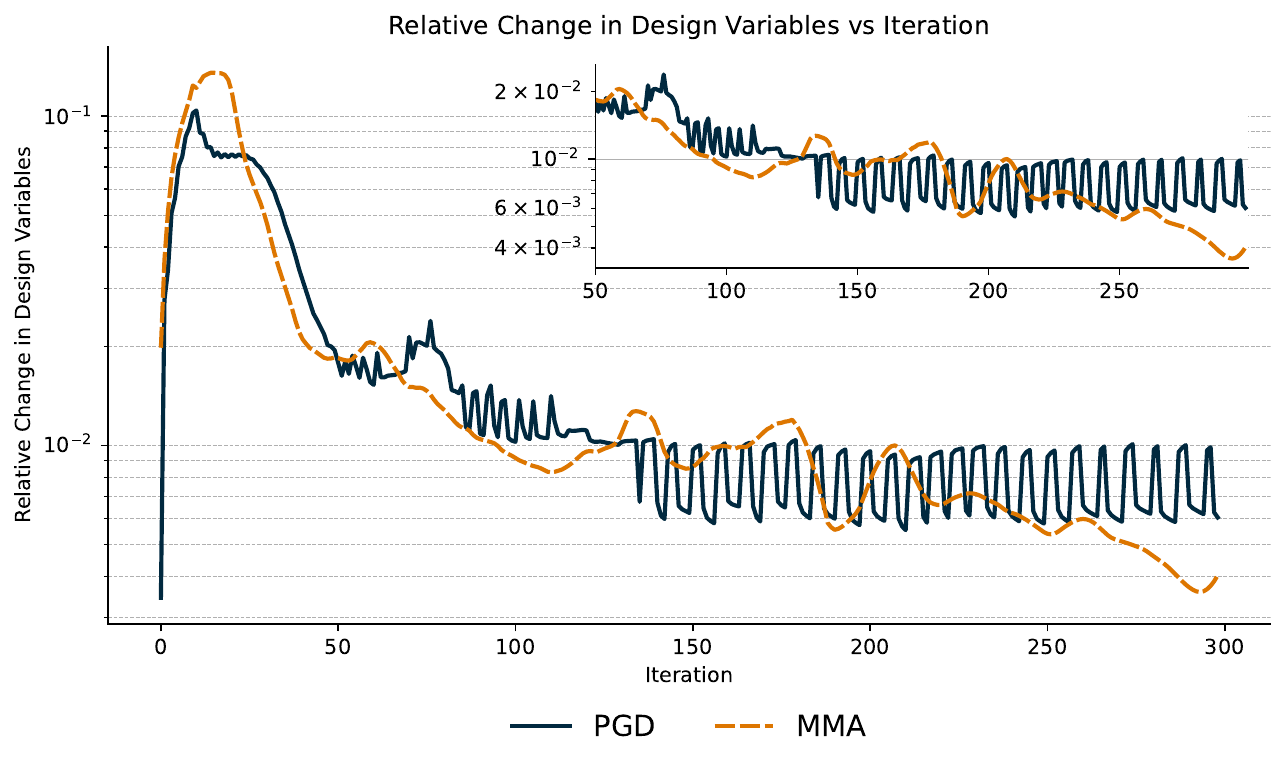}
    \caption{The relative change in the design variable norm per iteration for all solvers: compliance-constrained minimum volume problem, medium $256 \times 128$ mesh.}
    \label{fig:comp_med_rel_change}
\end{figure}

\begin{figure}[H]
    \centering
    \includegraphics[width=\linewidth]{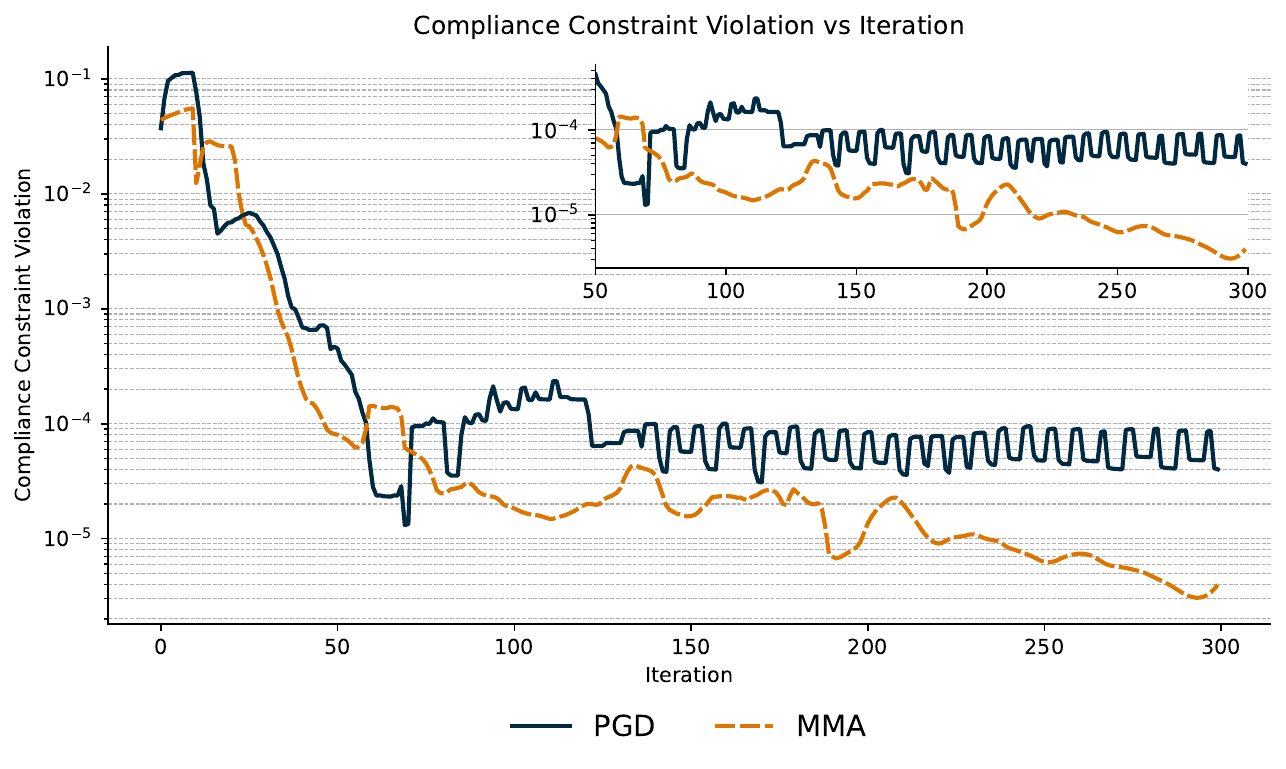}
    \caption{The constraint violation per iteration for all solvers: compliance-constrained minimum volume problem, medium $256 \times 128$ mesh.}
    \label{fig:comp_med_violation}
\end{figure}

\paragraph{Fine Mesh Results}
Here we provide figures for the results of running each optimizer for the fine $512 \times 256$ mesh.

\begin{figure}[H]
    \centering
    \includegraphics[width=\linewidth]{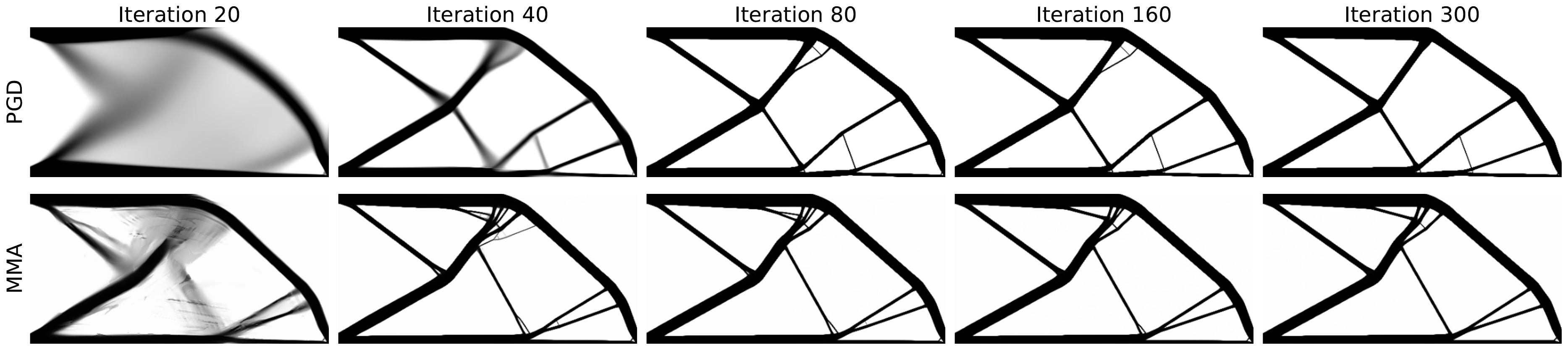}
    \caption{Designs produced by each optimizer for the compliance-constrained minimum volume problem on the cantilever beam with a compliance limit of $C_{\text{max}}=150$, fine $512 \times 256$ mesh.}
    \label{fig:comp_fine_designs}
\end{figure}

\begin{figure}[H]
    \centering
    \includegraphics[width=\linewidth]{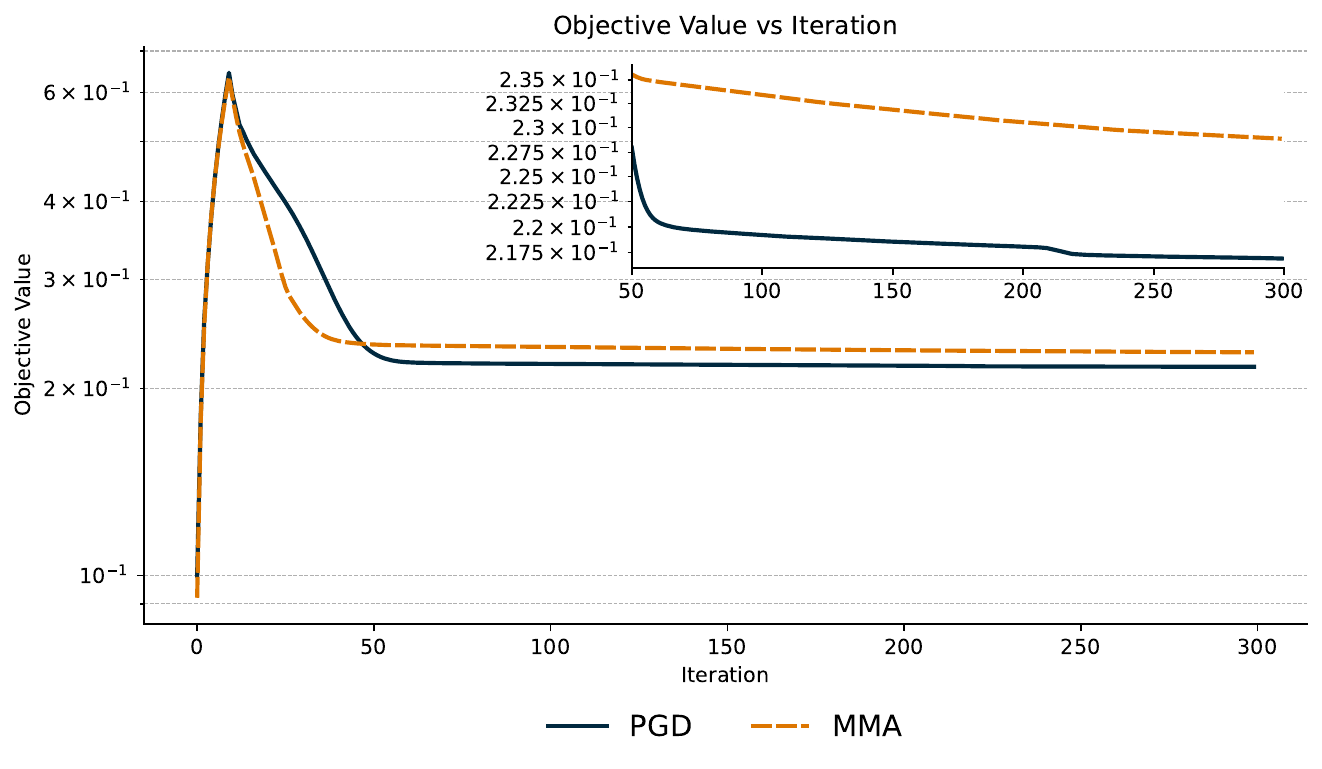}
    \caption{The value of the objective function per iteration for all solvers: compliance-constrained minimum volume problem, fine $512 \times 256$ mesh.}
    \label{fig:comp_fine_obj}
\end{figure}

\begin{figure}[H]
    \centering
    \includegraphics[width=\linewidth]{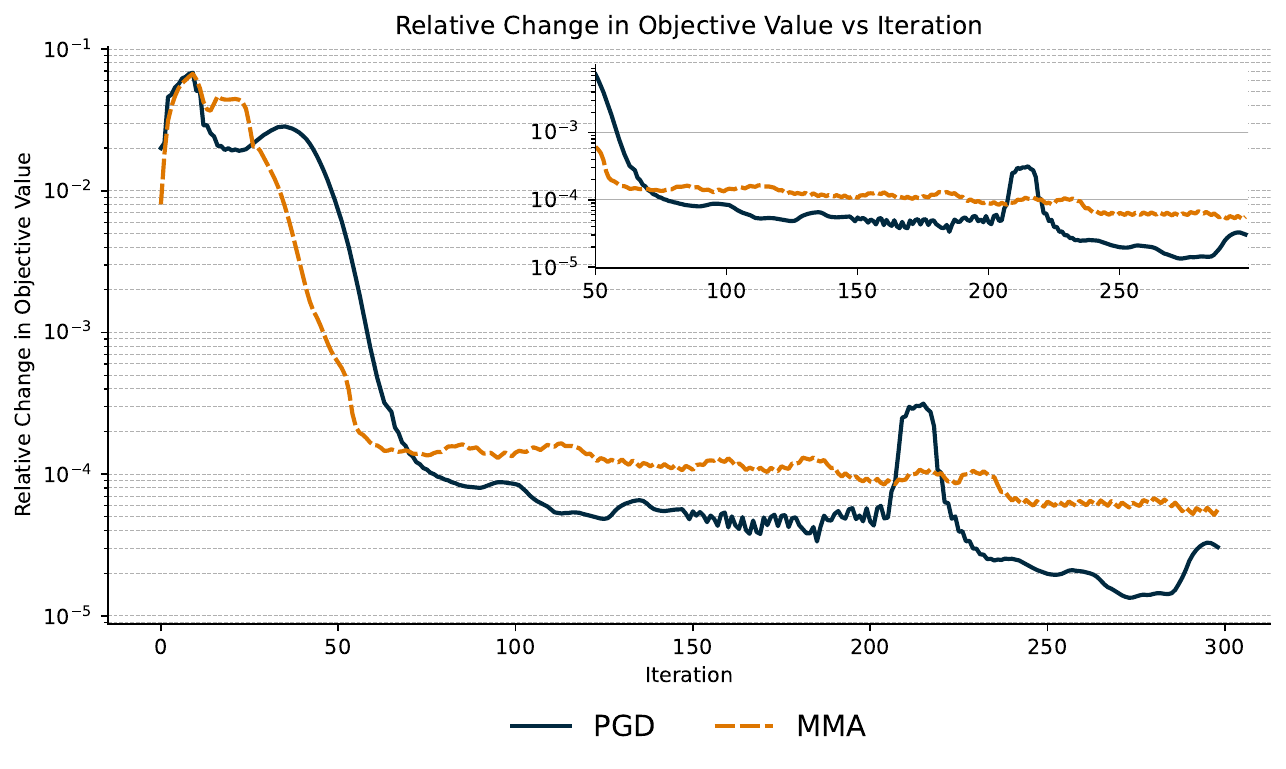}
    \caption{The relative change in the objective function per iteration for all solvers: compliance-constrained minimum volume problem, fine $512 \times 256$ mesh.}
    \label{fig:mincomp_fine_rel_obj}
\end{figure}

\begin{figure}[H]
    \centering
    \includegraphics[width=\linewidth]{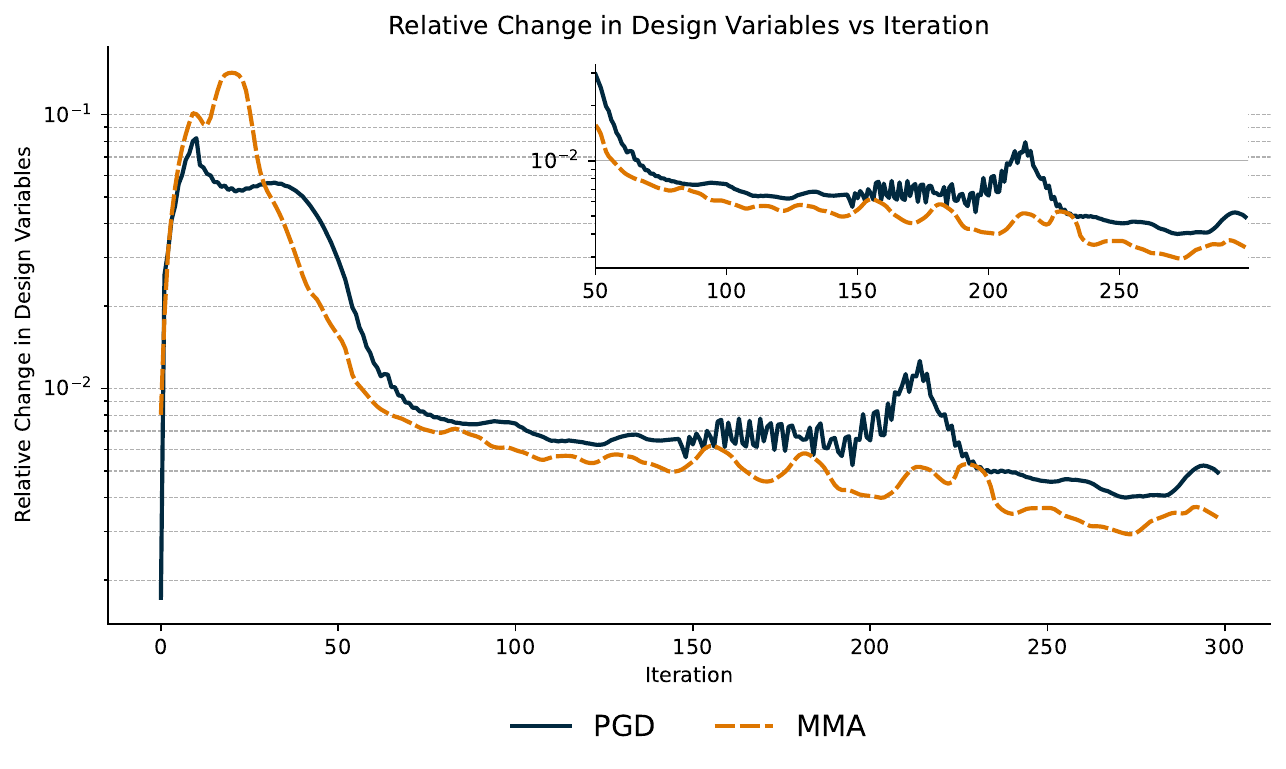}
    \caption{The relative change in the design variable norm per iteration for all solvers: compliance-constrained minimum volume problem, fine $512 \times 256$ mesh.}
    \label{fig:comp_fine_rel_change}
\end{figure}

\begin{figure}[H]
    \centering
    \includegraphics[width=\linewidth]{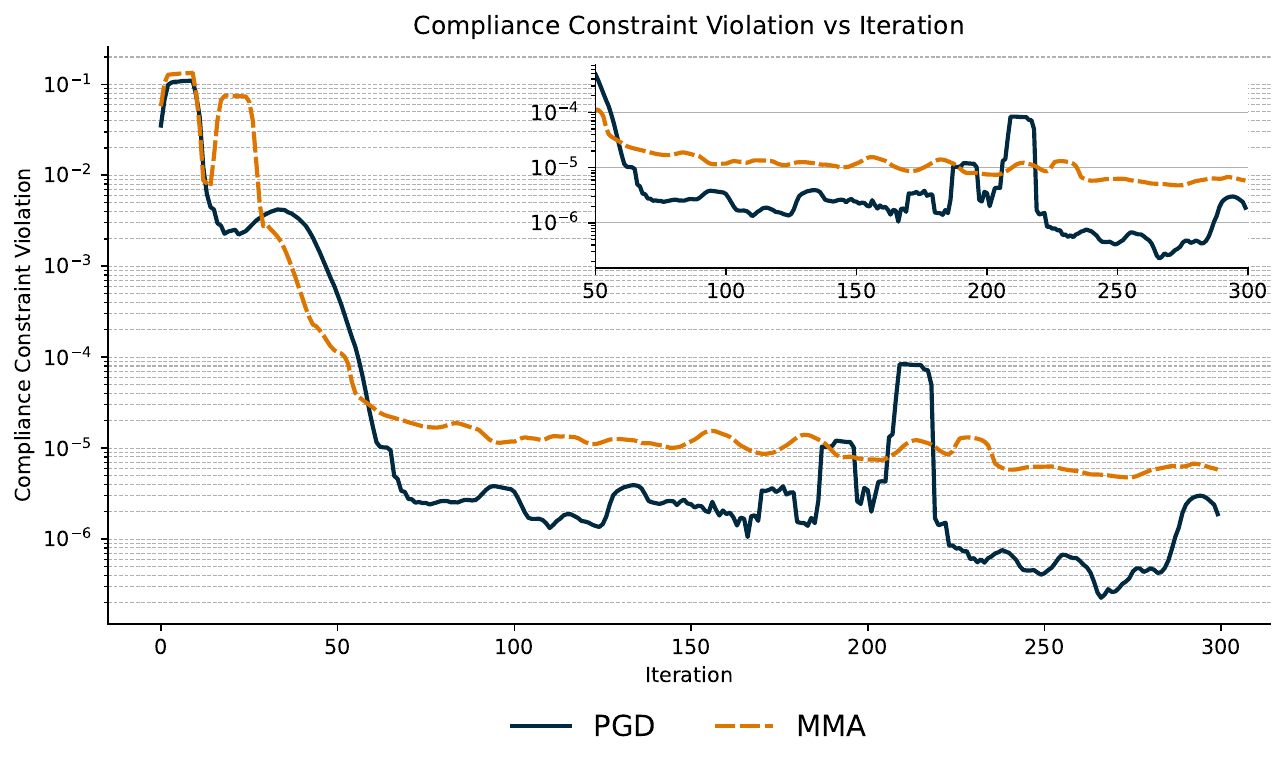}
    \caption{The constraint violation per iteration for all solvers: compliance-constrained minimum volume problem, fine $512 \times 256$ mesh.}
    \label{fig:comp_fine_violation}
\end{figure}

\clearpage
\subsubsection{Multi-Material Minimum Compliance}
The following subsections give the full results for each mesh resolution.

\paragraph{Coarse Mesh Results}
Here we provide figures for the results of running each optimizer for the coarse $128 \times 64$ mesh.

\begin{figure}[H]
    \centering
    \includegraphics[width=\linewidth]{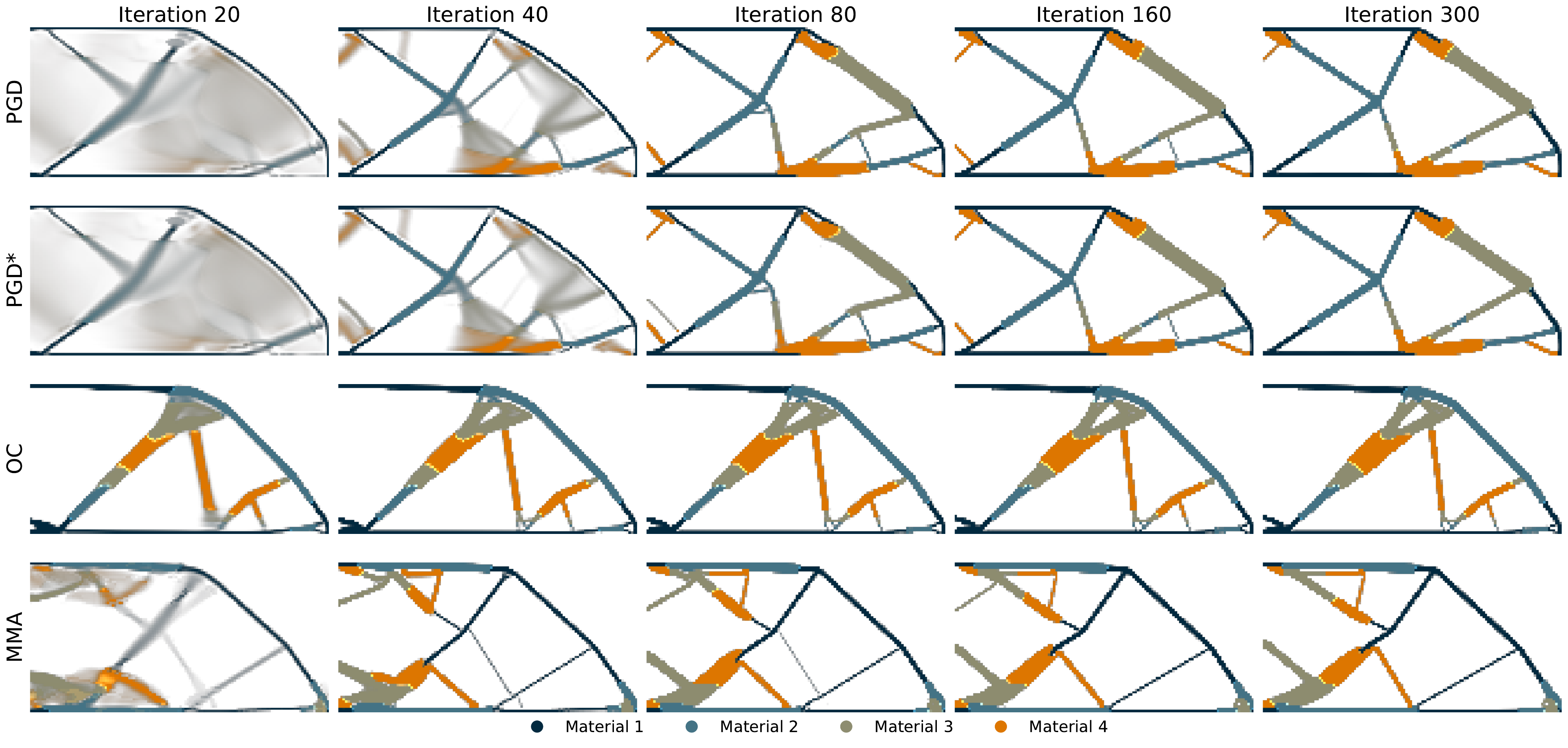}
    \caption{Designs produced by each optimizer for the multi-material minimum compliance problem on the cantilever beam with a volume fraction target of $0.05$ per material, coarse $128 \times 64$ mesh.}
    \label{fig:mincomp_coarse_designs}
\end{figure}

\begin{figure}[H]
    \centering
    \includegraphics[width=\linewidth]{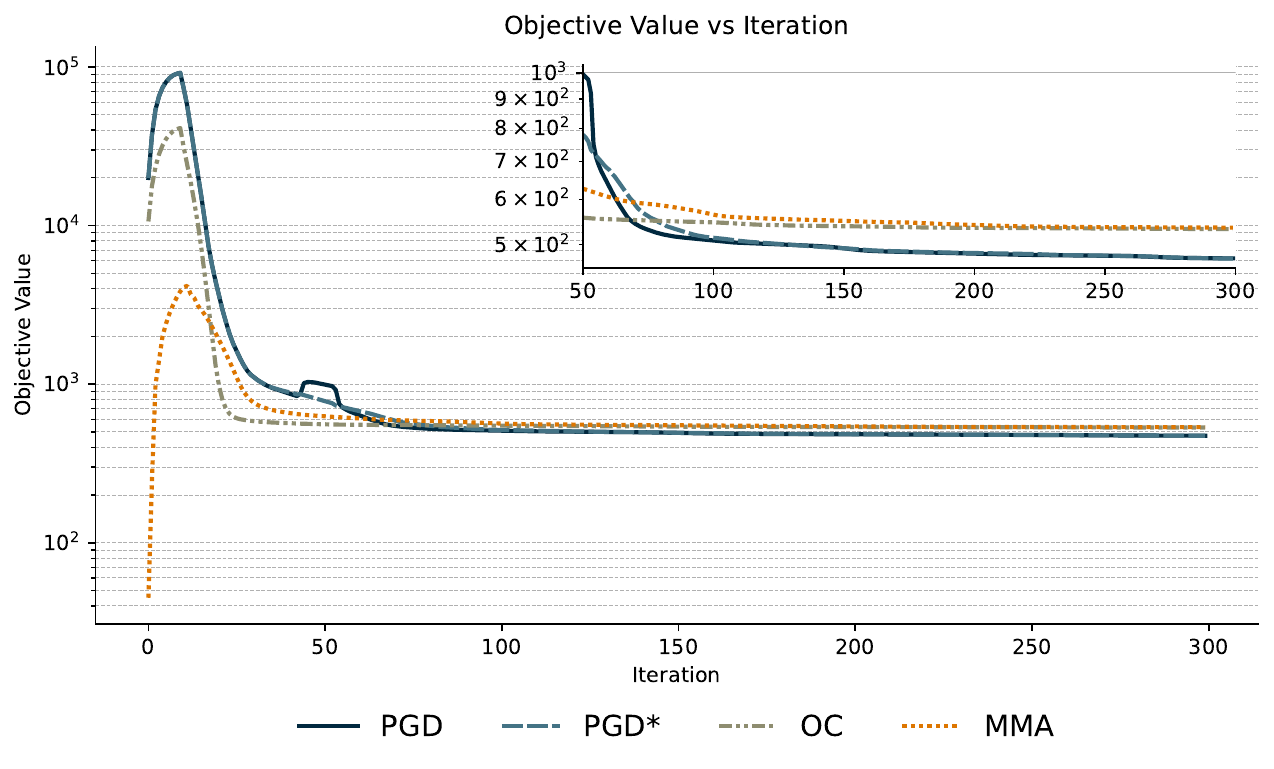}
    \caption{The value of the objective function per iteration for all solvers: multi-material minimum compliance problem, coarse $128 \times 64$ mesh.}
    \label{fig:mincomp_coarse_obj}
\end{figure}

\begin{figure}[H]
    \centering
    \includegraphics[width=\linewidth]{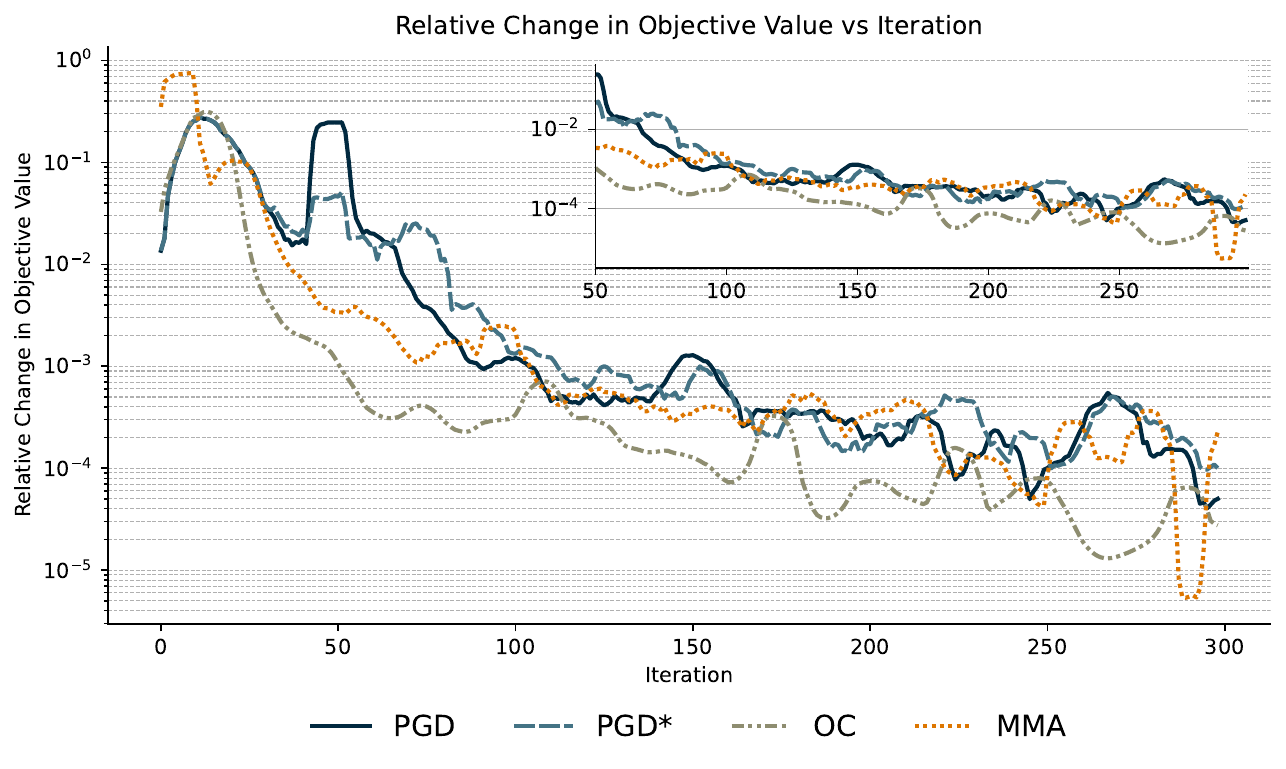}
    \caption{The relative change in the objective function per iteration for all solvers: multi-material minimum compliance problem, coarse $128 \times 64$ mesh.}
    \label{fig:mincomp_coarse_rel_obj}
\end{figure}

\begin{figure}[H]
    \centering
    \includegraphics[width=\linewidth]{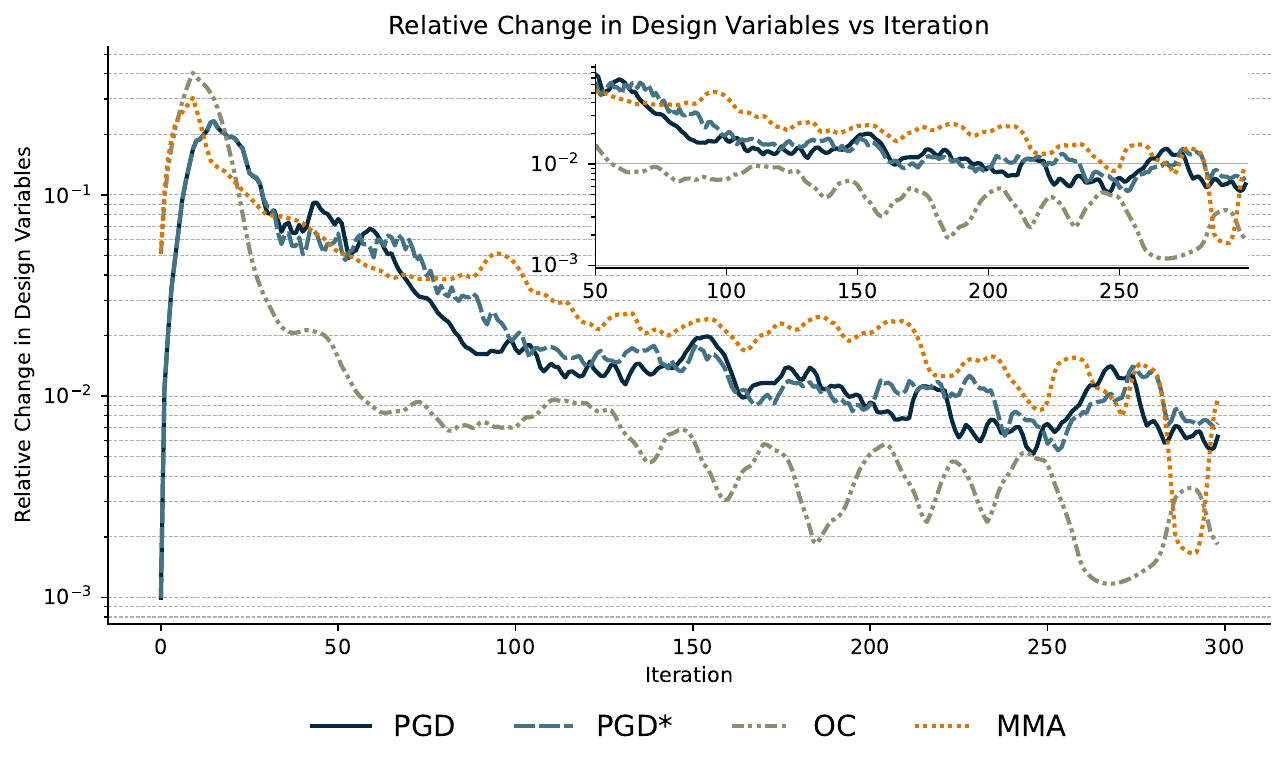}
    \caption{The relative change in the design variable norm per iteration for all solvers: multi-material minimum compliance problem, coarse $128 \times 64$ mesh.}
    \label{fig:mincomp_coarse_rel_change}
\end{figure}

\begin{figure}[H]
    \centering
    \includegraphics[width=\linewidth]{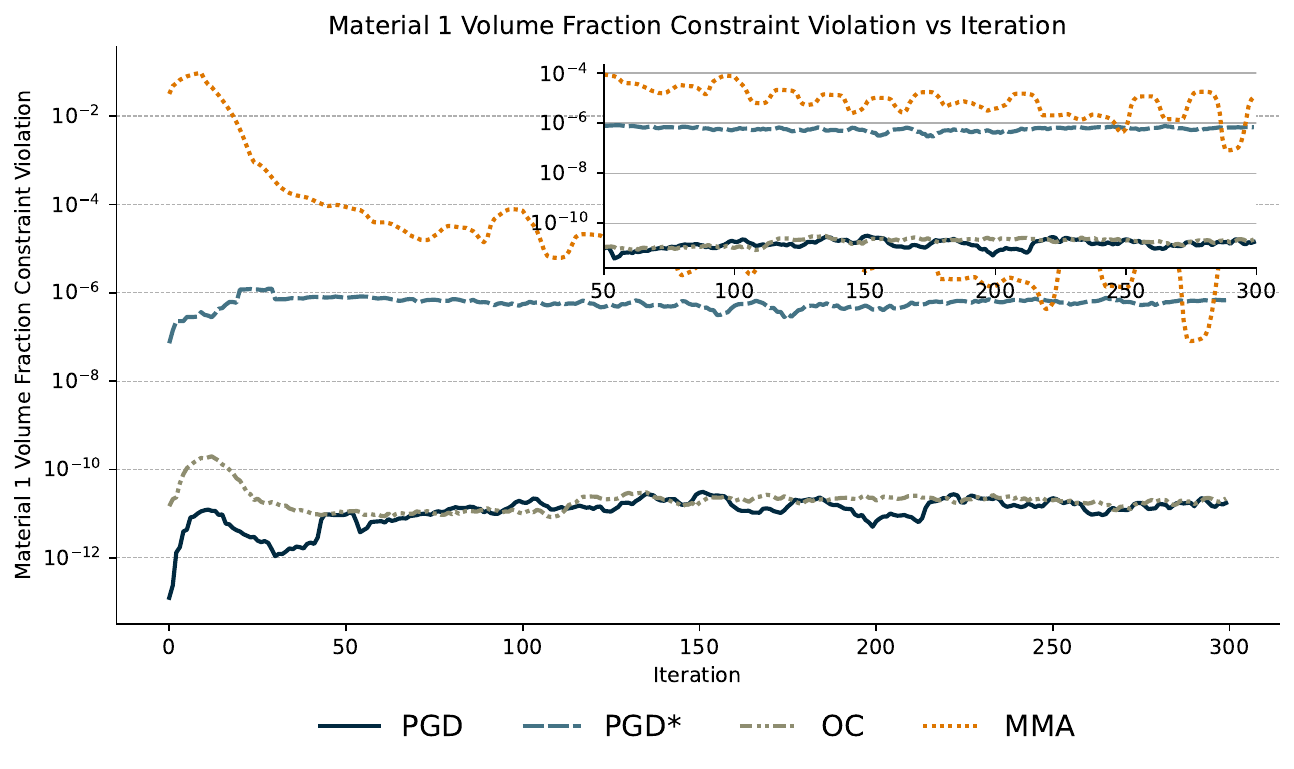}
    \caption{The constraint violation (material 1) per iteration for all solvers: multi-material minimum compliance problem, coarse $128 \times 64$ mesh.}
    \label{fig:mincomp_coarse_violation_1}
\end{figure}

\begin{figure}[H]
    \centering
    \includegraphics[width=\linewidth]{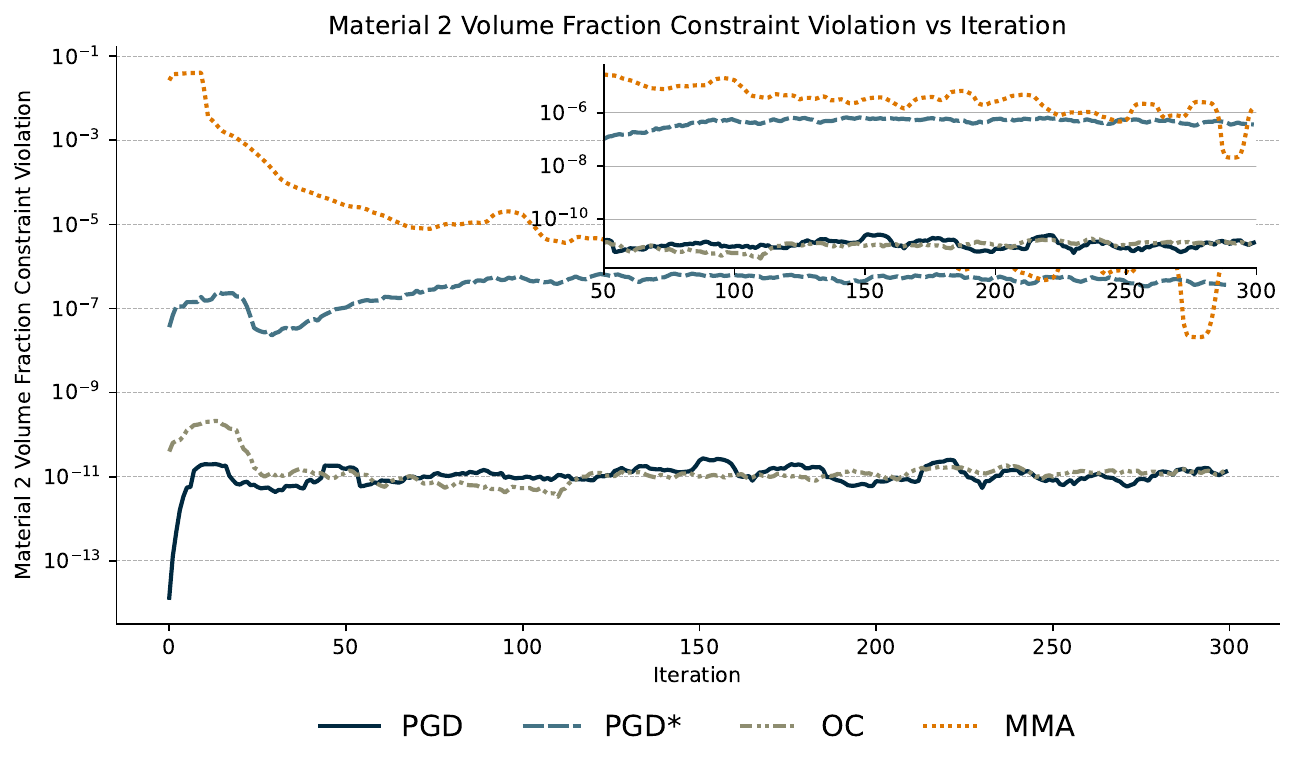}
    \caption{The constraint violation (material 2) per iteration for all solvers: multi-material minimum compliance problem, coarse $128 \times 64$ mesh.}
    \label{fig:mincomp_coarse_violation_2}
\end{figure}

\begin{figure}[H]
    \centering
    \includegraphics[width=\linewidth]{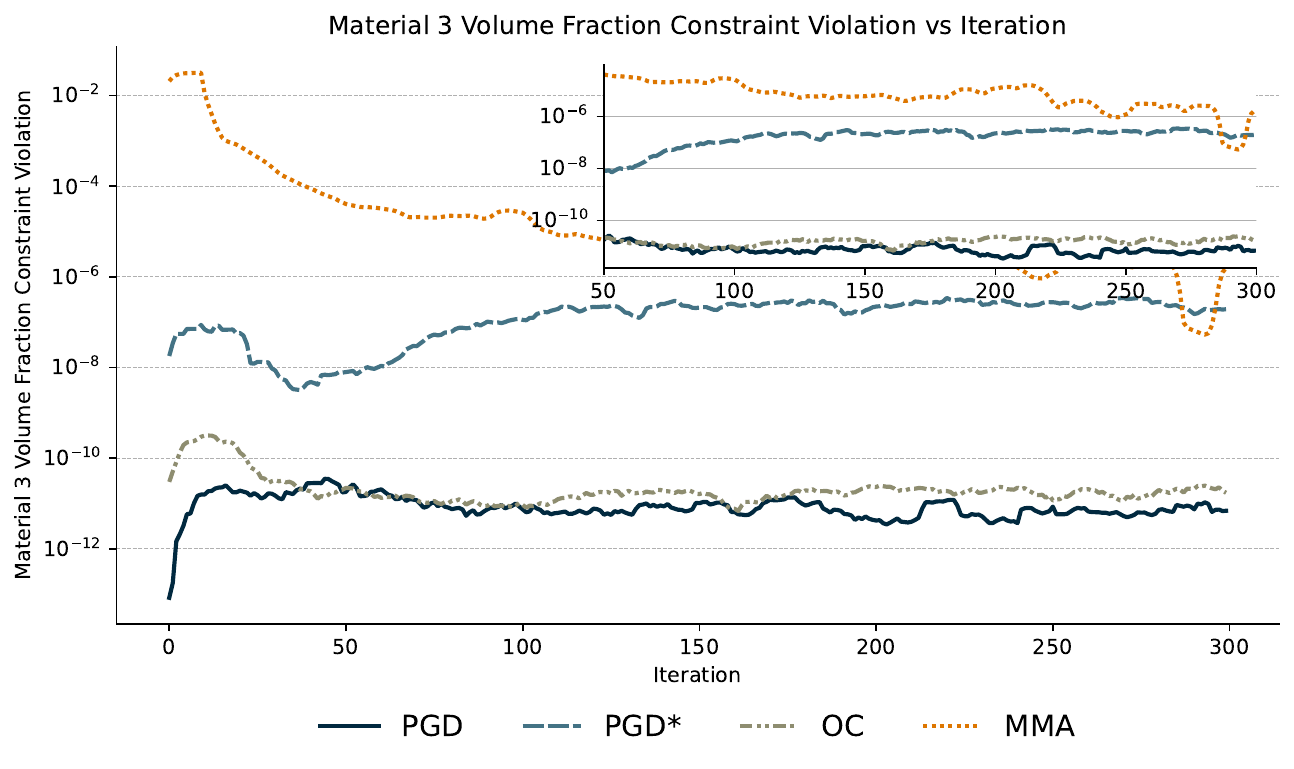}
    \caption{The constraint violation (material 3) per iteration for all solvers: multi-material minimum compliance problem, coarse $128 \times 64$ mesh.}
    \label{fig:mincomp_coarse_violation_3}
\end{figure}

\begin{figure}[H]
    \centering
    \includegraphics[width=\linewidth]{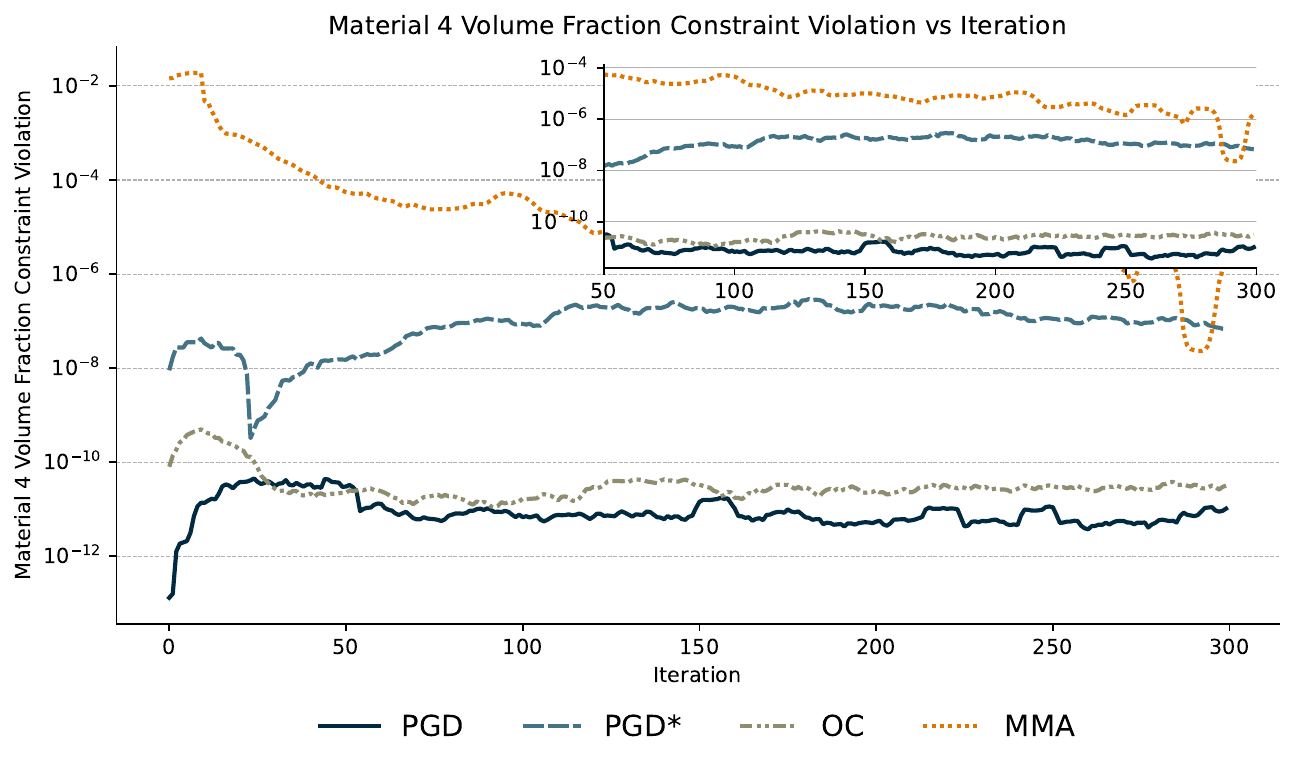}
    \caption{The constraint violation (material 4) per iteration for all solvers: multi-material minimum compliance problem, coarse $128 \times 64$ mesh.}
    \label{fig:mincomp_coarse_violation_4}
\end{figure}

\paragraph{Medium Mesh Results}
Here we provide figures for the results of running each optimizer for the medium $256 \times 128$ mesh.

\begin{figure}[H]
    \centering
    \includegraphics[width=\linewidth]{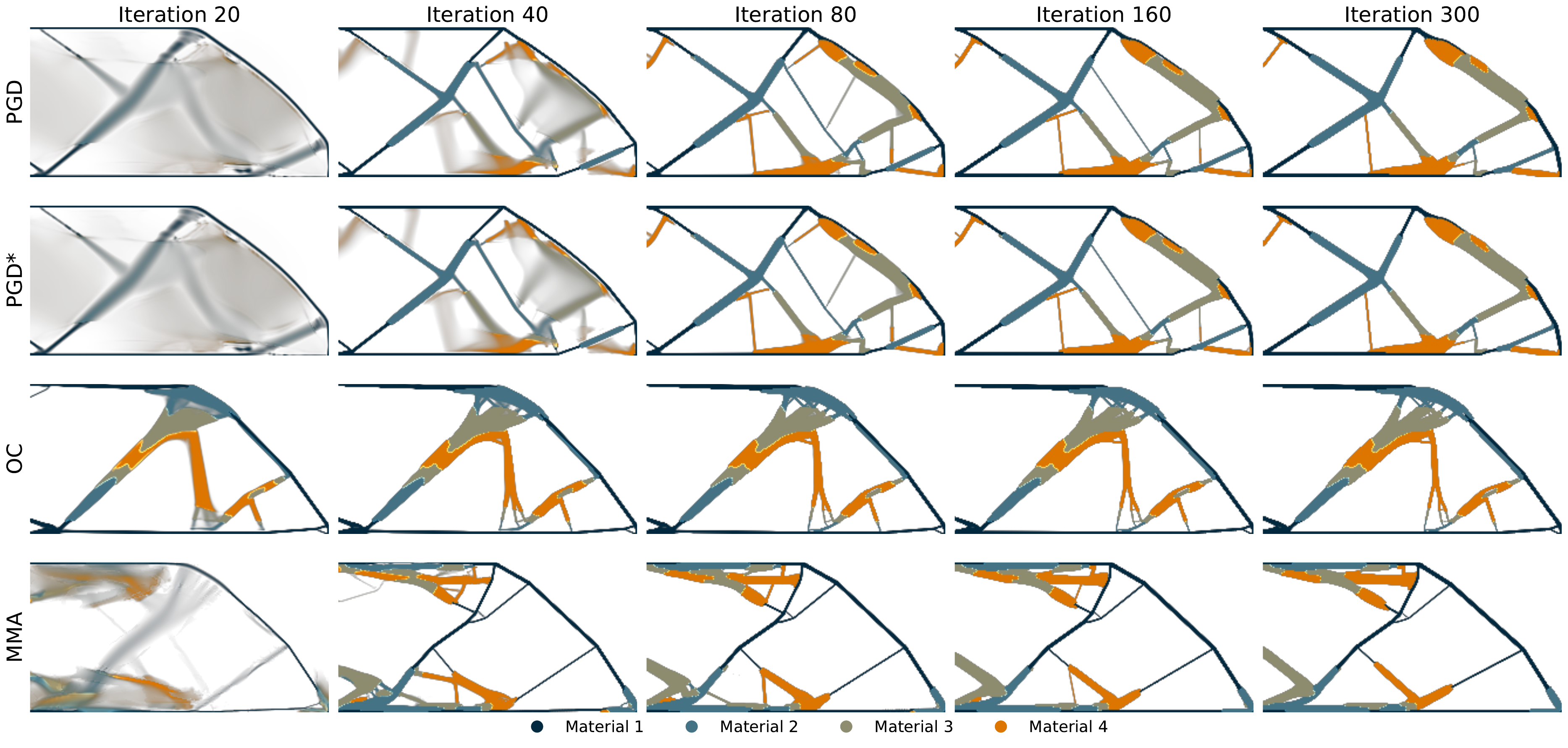}
    \caption{Designs produced by each optimizer for the multi-material minimum compliance problem on the cantilever beam with a volume fraction target of $0.05$ per material, medium $256 \times 128$ mesh.}
    \label{fig:mincomp_med_designs}
\end{figure}

\begin{figure}[H]
    \centering
    \includegraphics[width=\linewidth]{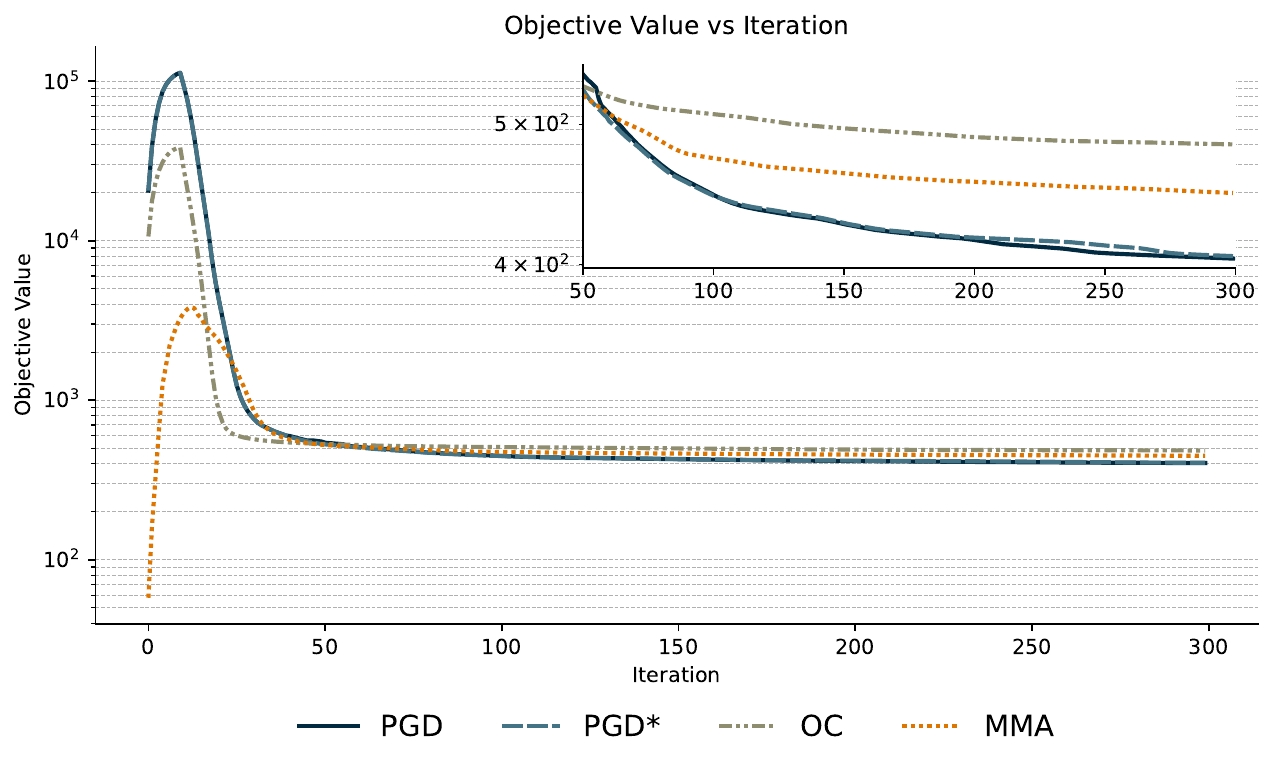}
    \caption{The value of the objective function per iteration for all solvers: multi-material minimum compliance problem, medium $256 \times 128$ mesh.}
    \label{fig:mincomp_med_obj}
\end{figure}

\begin{figure}[H]
    \centering
    \includegraphics[width=\linewidth]{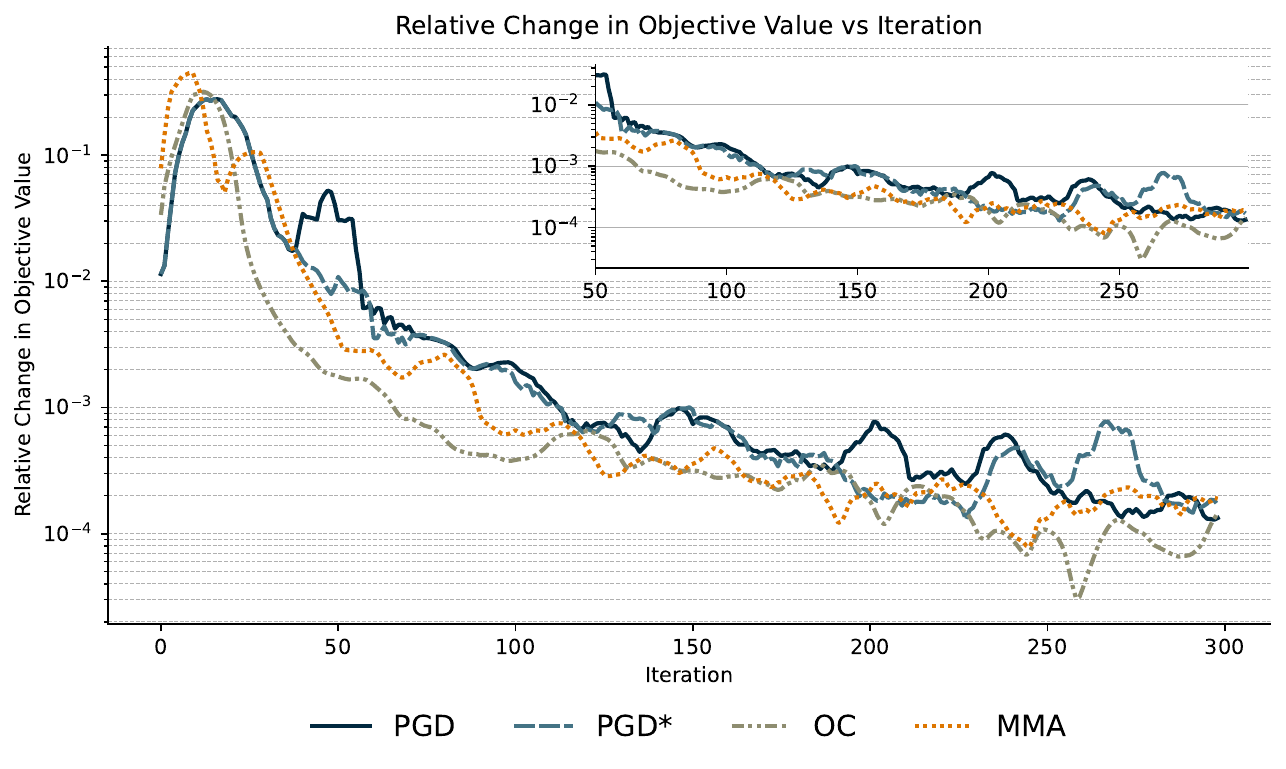}
    \caption{The relative change in the objective function per iteration for all solvers: multi-material minimum compliance problem, medium $256 \times 128$ mesh.}
    \label{fig:mincomp_med_rel_obj}
\end{figure}

\begin{figure}[H]
    \centering
    \includegraphics[width=\linewidth]{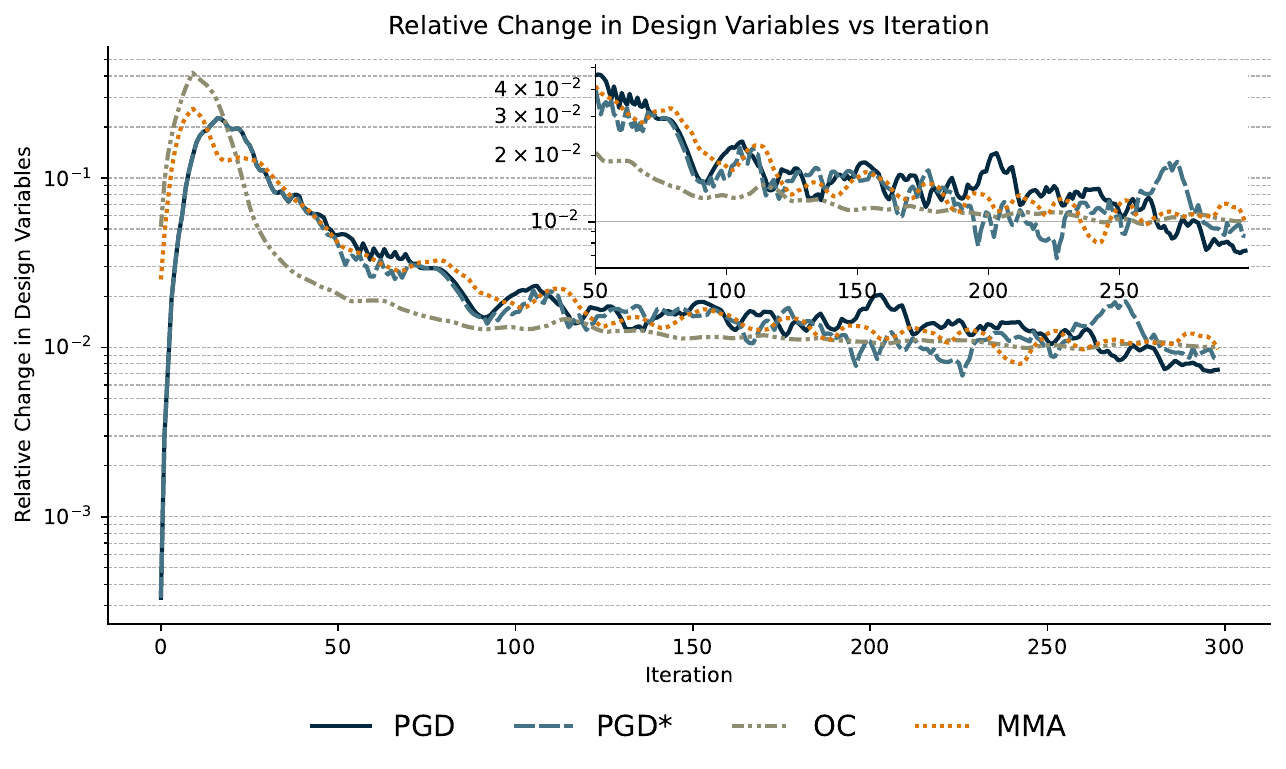}
    \caption{The relative change in the design variable norm per iteration for all solvers: multi-material minimum compliance problem, medium $256 \times 128$ mesh.}
    \label{fig:mincomp_med_rel_change}
\end{figure}

\begin{figure}[H]
    \centering
    \includegraphics[width=\linewidth]{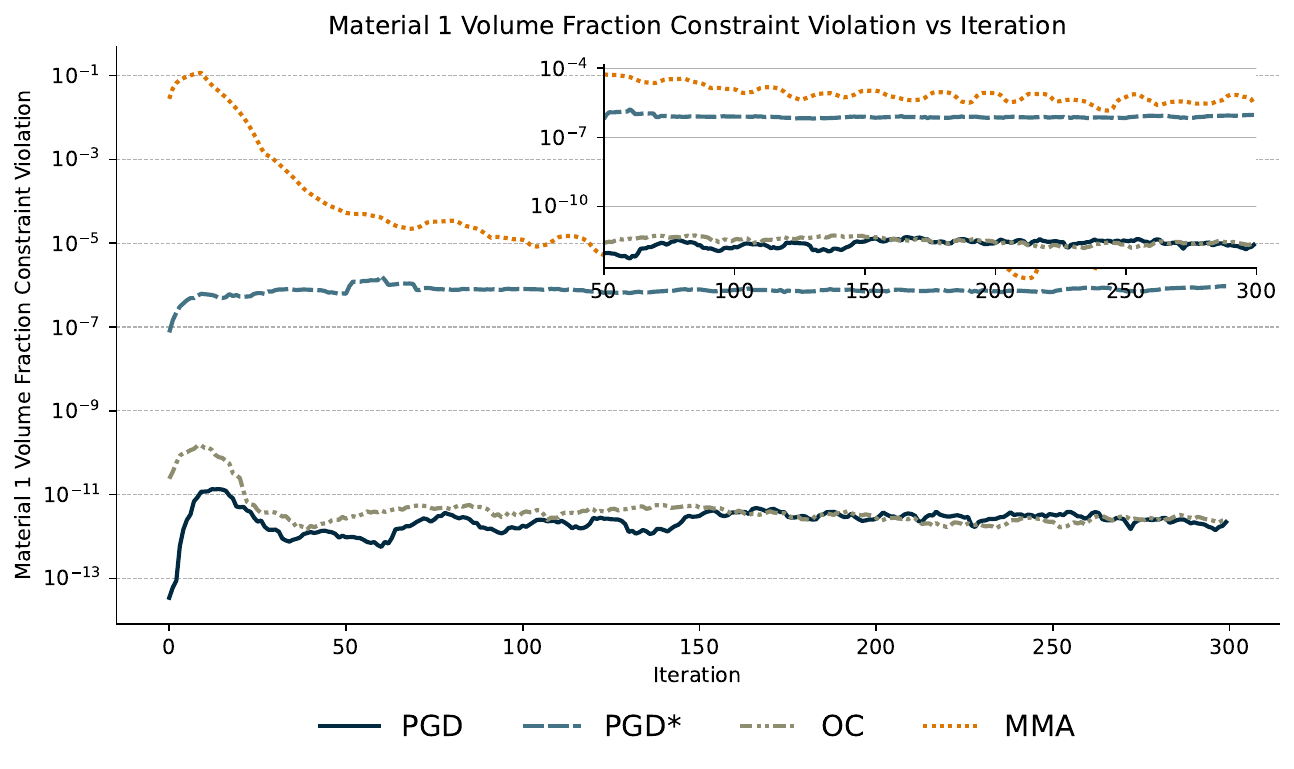}
    \caption{The constraint violation (material 1) per iteration for all solvers: multi-material minimum compliance problem, medium $256 \times 128$ mesh.}
    \label{fig:mincomp_med_violation_1}
\end{figure}

\begin{figure}[H]
    \centering
    \includegraphics[width=\linewidth]{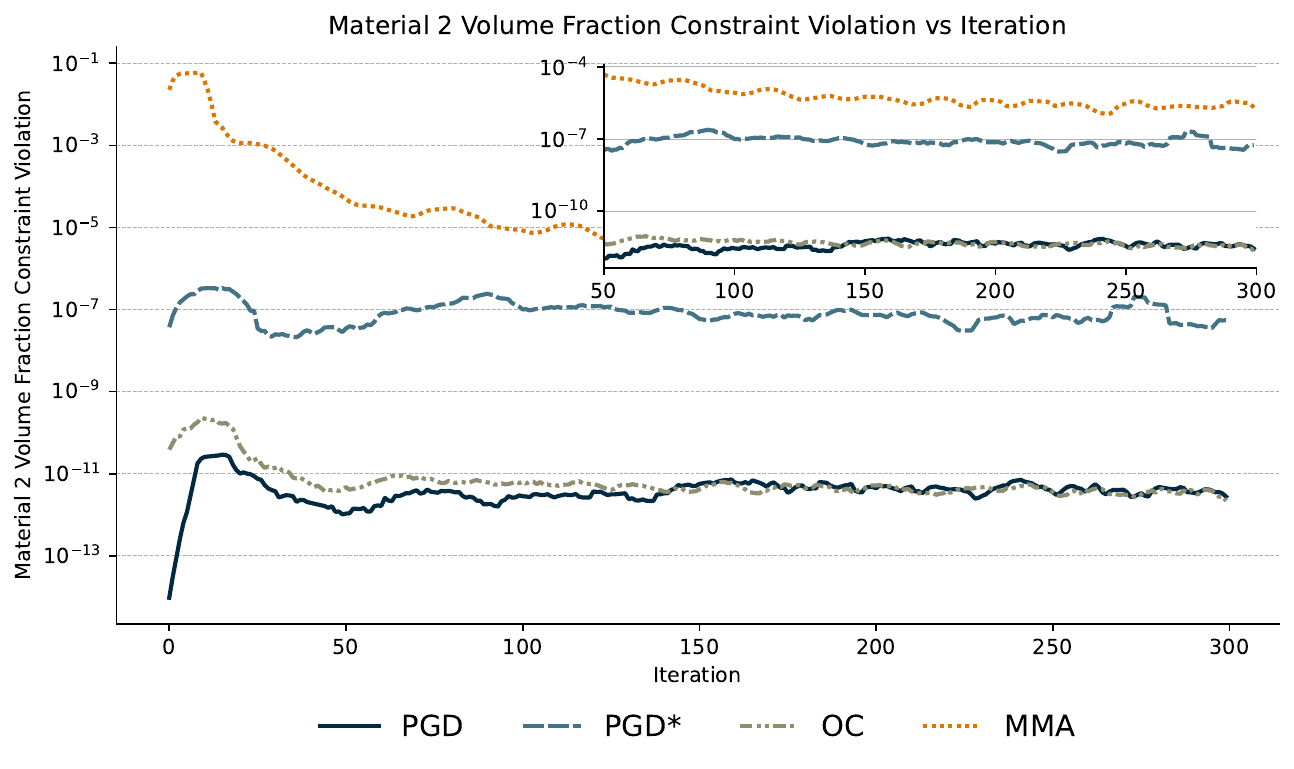}
    \caption{The constraint violation (material 2) per iteration for all solvers: multi-material minimum compliance problem, medium $256 \times 128$ mesh.}
    \label{fig:mincomp_med_violation_2}
\end{figure}

\begin{figure}[H]
    \centering
    \includegraphics[width=\linewidth]{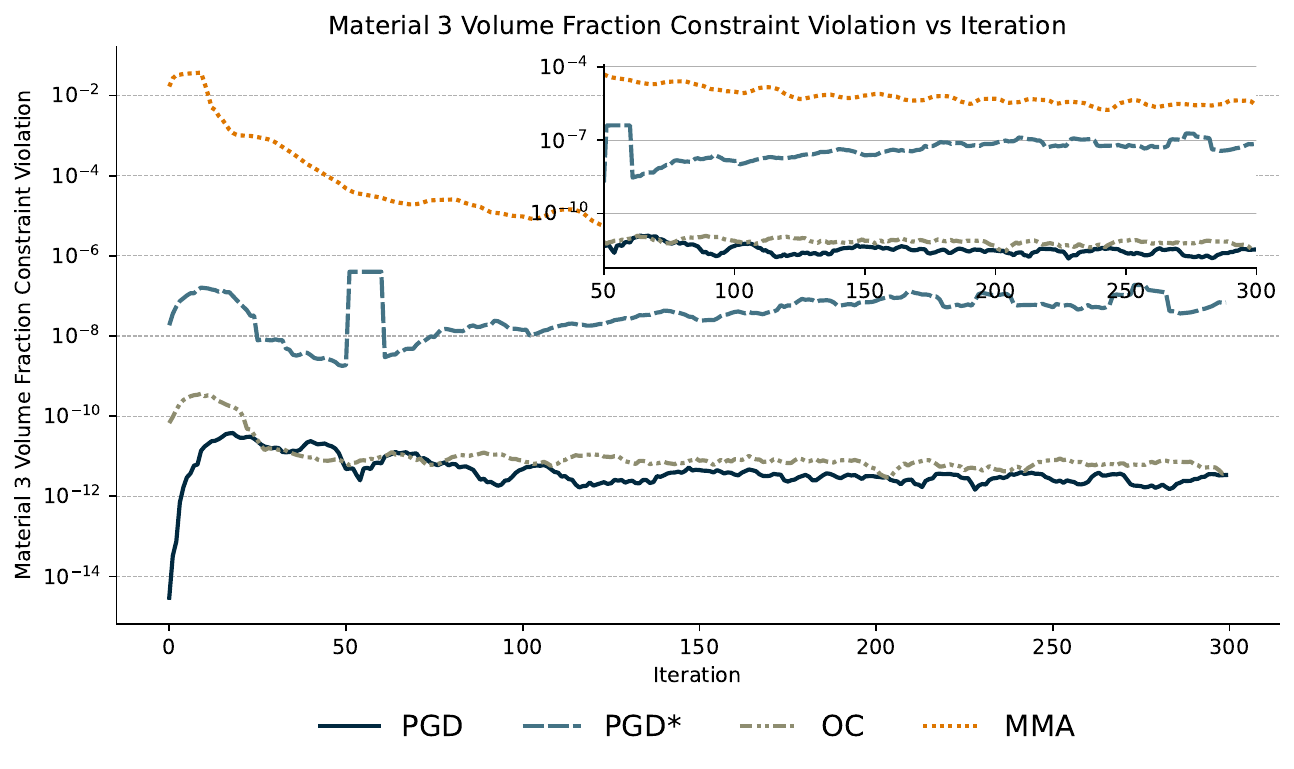}
    \caption{The constraint violation (material 3) per iteration for all solvers: multi-material minimum compliance problem, medium $256 \times 128$ mesh.}
    \label{fig:mincomp_med_violation_3}
\end{figure}

\begin{figure}[H]
    \centering
    \includegraphics[width=\linewidth]{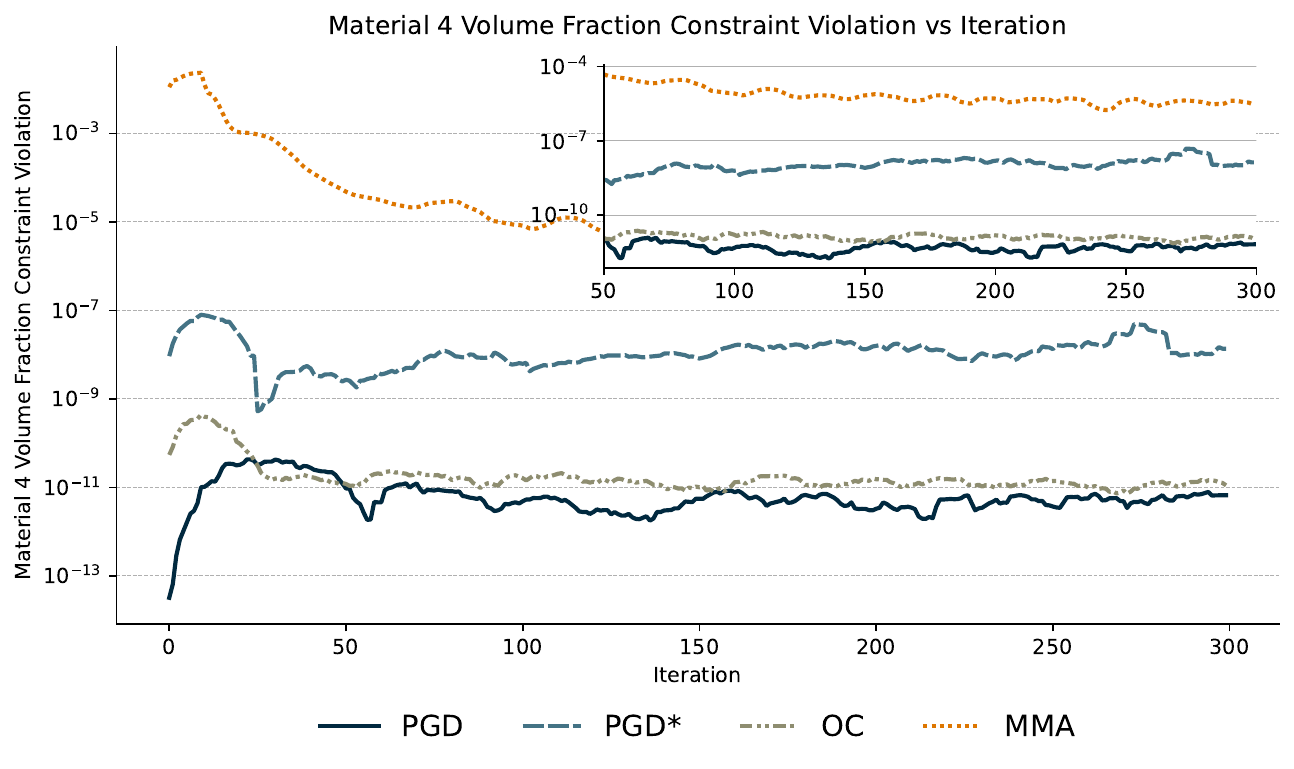}
    \caption{The constraint violation (material 4) per iteration for all solvers: multi-material minimum compliance problem, medium $256 \times 128$ mesh.}
    \label{fig:mincomp_med_violation_4}
\end{figure}

\paragraph{Fine Mesh Results}
Here we provide figures for the results of running each optimizer for the fine $512 \times 256$ mesh.

\begin{figure}[H]
    \centering
    \includegraphics[width=\linewidth]{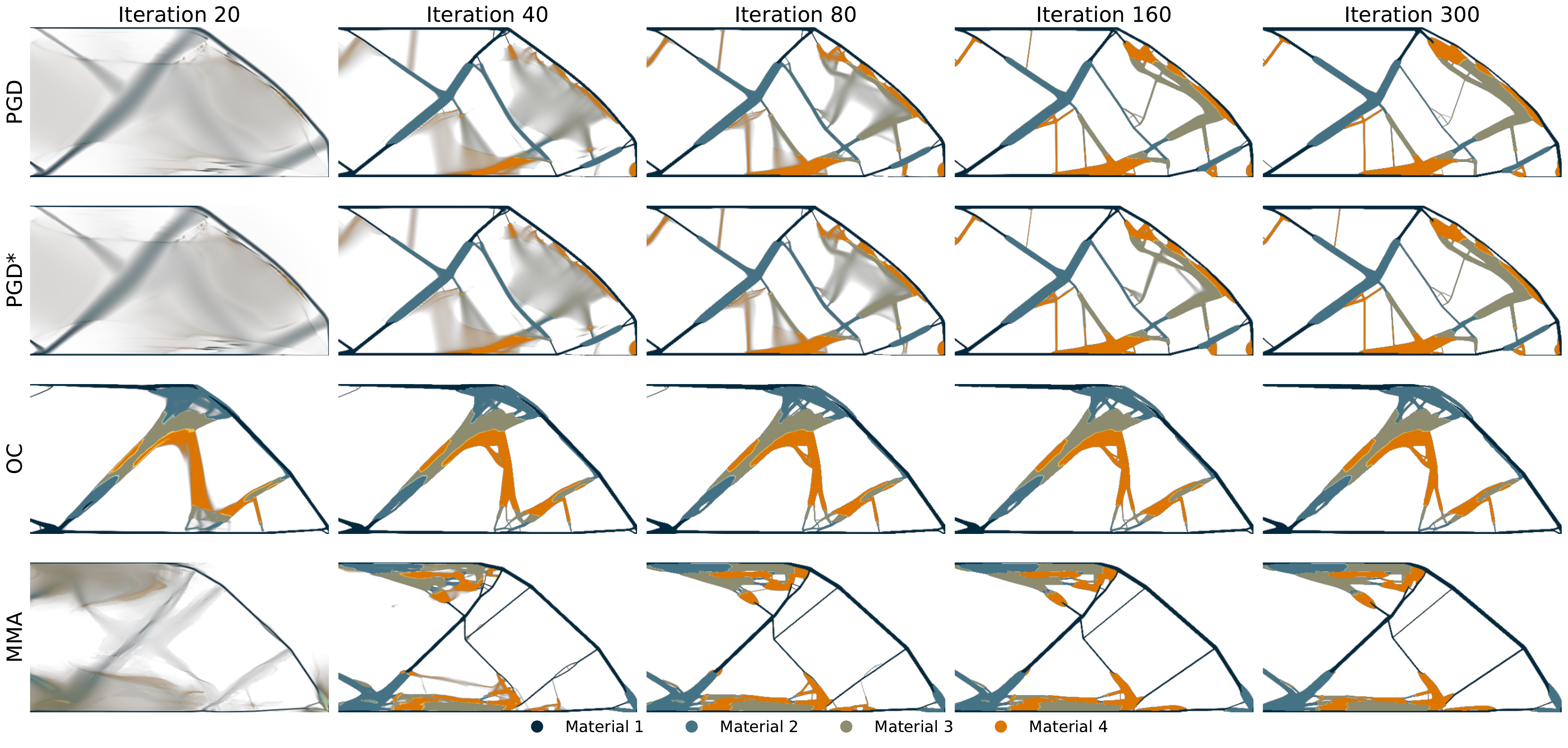}
    \caption{Designs produced by each optimizer for the multi-material minimum compliance problem on the cantilever beam with a volume fraction target of $0.05$ per material, fine $512 \times 256$ mesh.}
    \label{fig:mincomp_fine_designs}
\end{figure}

\begin{figure}[H]
    \centering
    \includegraphics[width=\linewidth]{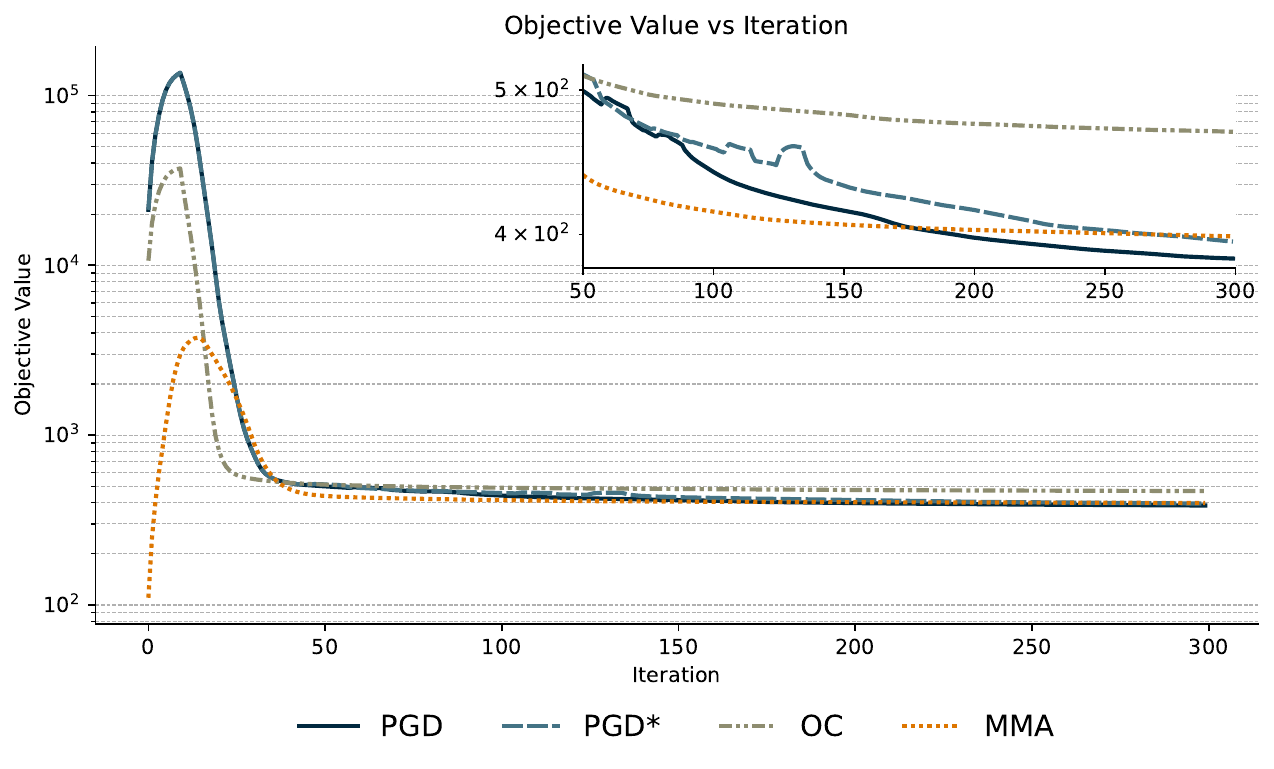}
    \caption{The value of the objective function per iteration for all solvers: multi-material minimum compliance problem, fine $512 \times 256$ mesh.}
    \label{fig:mincomp_fine_obj}
\end{figure}

\begin{figure}[H]
    \centering
    \includegraphics[width=\linewidth]{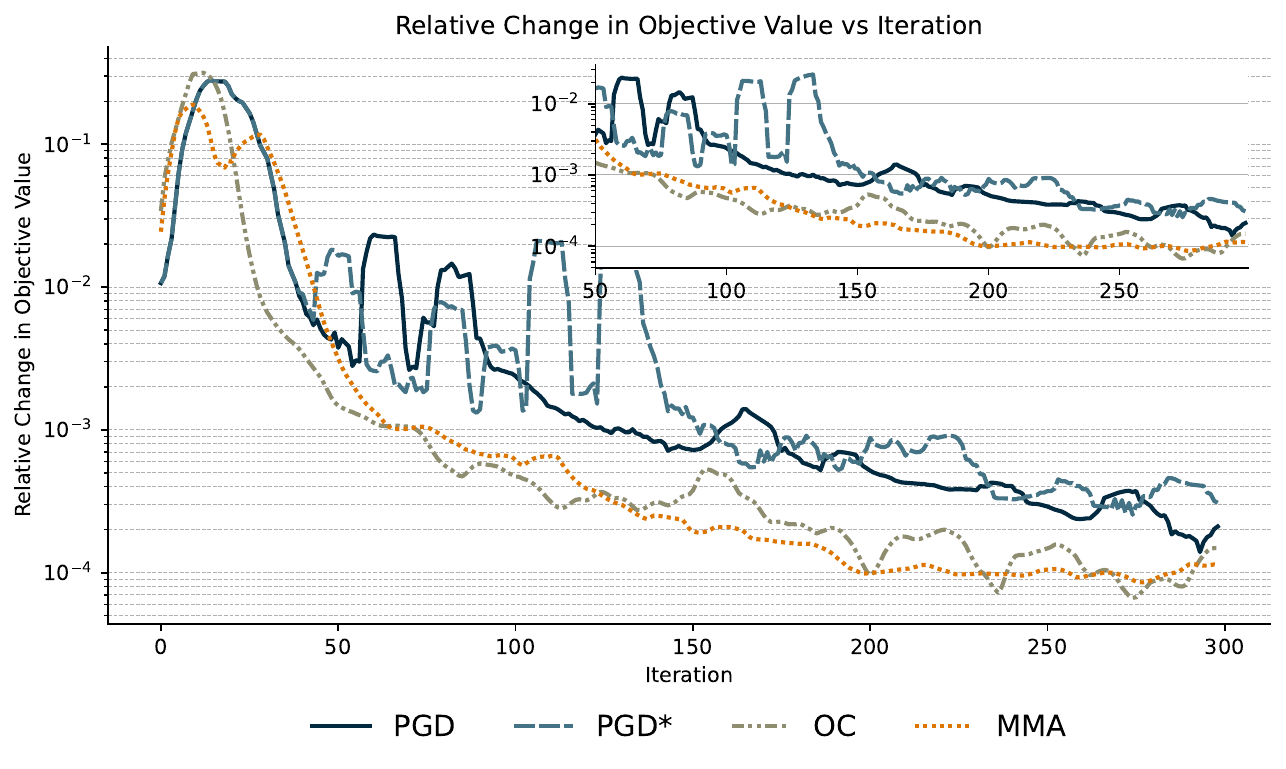}
    \caption{The relative change in the objective function per iteration for all solvers: multi-material minimum compliance problem, fine $512 \times 256$ mesh.}
    \label{fig:mincomp_fine_rel_obj}
\end{figure}

\begin{figure}[H]
    \centering
    \includegraphics[width=\linewidth]{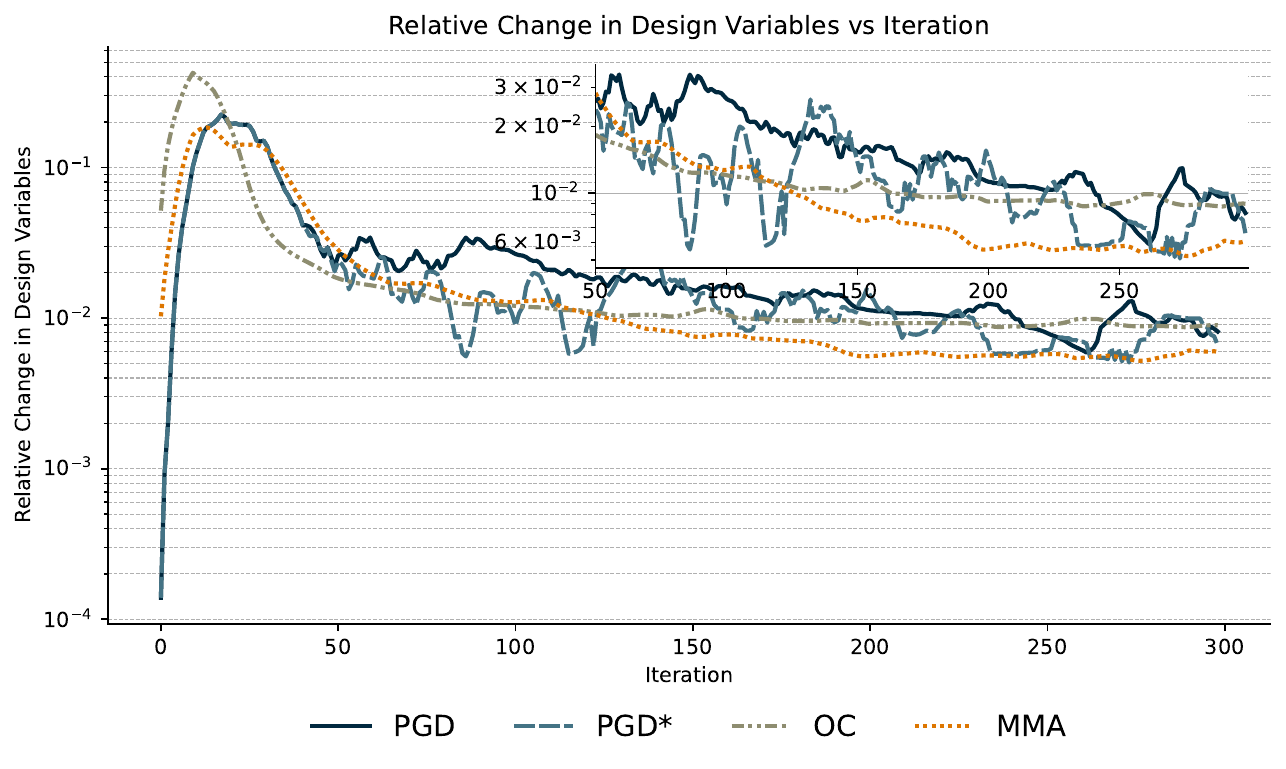}
    \caption{The relative change in the design variable norm per iteration for all solvers: multi-material minimum compliance problem, fine $512 \times 256$ mesh.}
    \label{fig:mincomp_fine_rel_change}
\end{figure}

\begin{figure}[H]
    \centering
    \includegraphics[width=\linewidth]{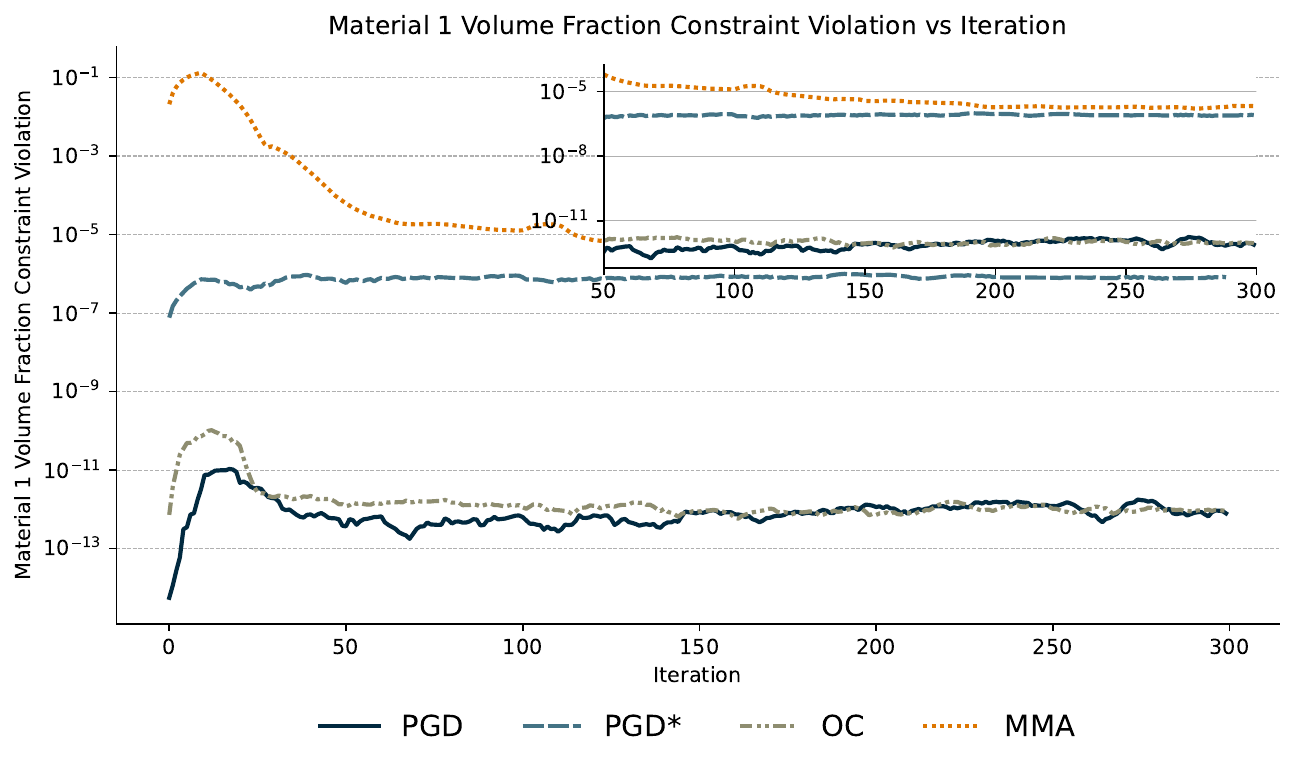}
    \caption{The constraint violation (material 1) per iteration for all solvers: multi-material minimum compliance problem, fine $512 \times 256$ mesh.}
    \label{fig:mincomp_fine_violation_1}
\end{figure}

\begin{figure}[H]
    \centering
    \includegraphics[width=\linewidth]{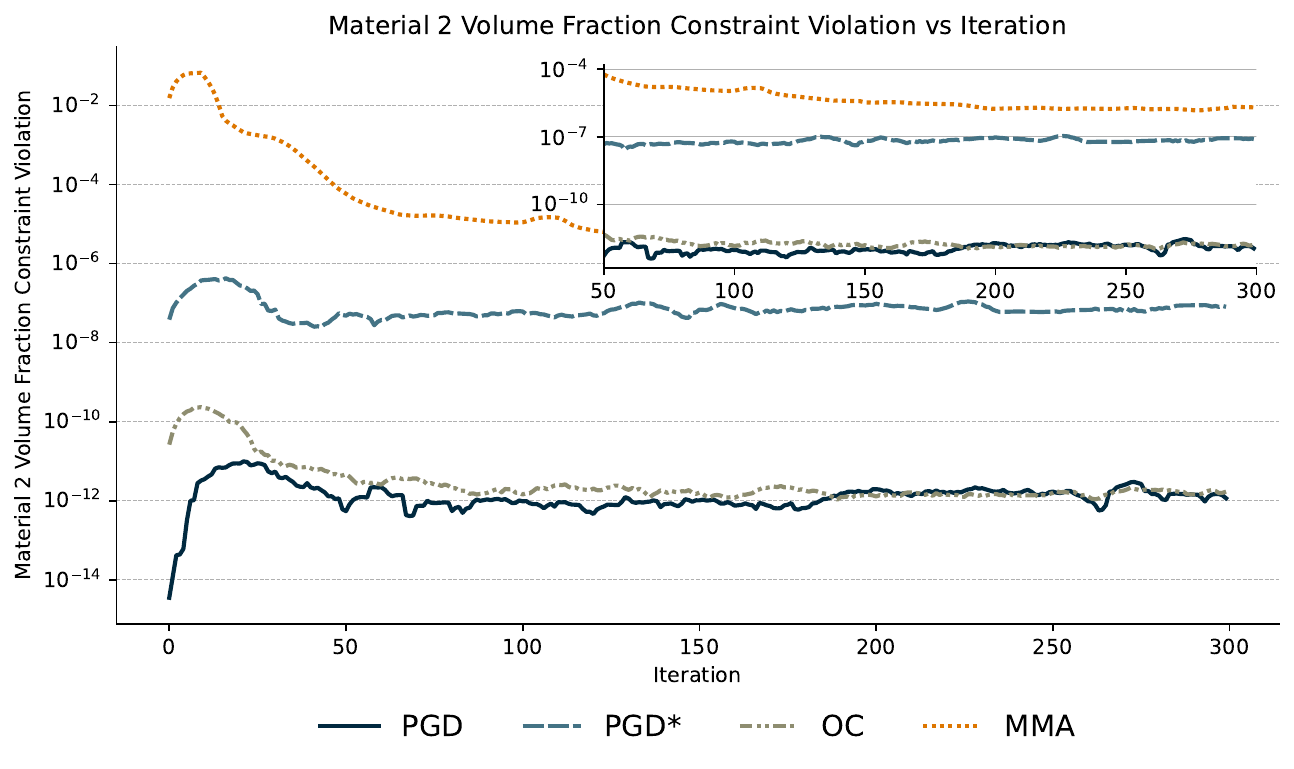}
    \caption{The constraint violation (material 2) per iteration for all solvers: multi-material minimum compliance problem, fine $512 \times 256$ mesh.}
    \label{fig:mincomp_fine_violation_2}
\end{figure}

\begin{figure}[H]
    \centering
    \includegraphics[width=\linewidth]{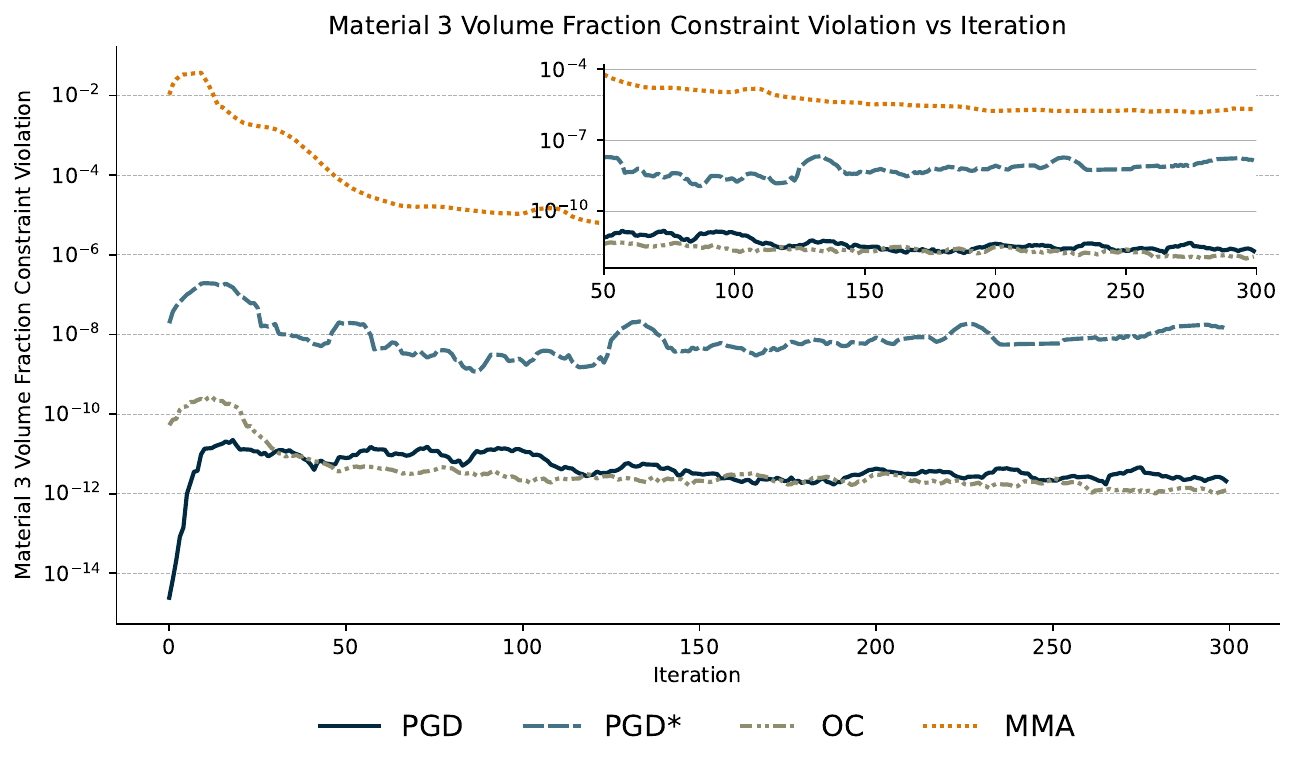}
    \caption{The constraint violation (material 3) per iteration for all solvers: multi-material minimum compliance problem, fine $512 \times 256$ mesh.}
    \label{fig:mincomp_fine_violation_3}
\end{figure}

\begin{figure}[H]
    \centering
    \includegraphics[width=\linewidth]{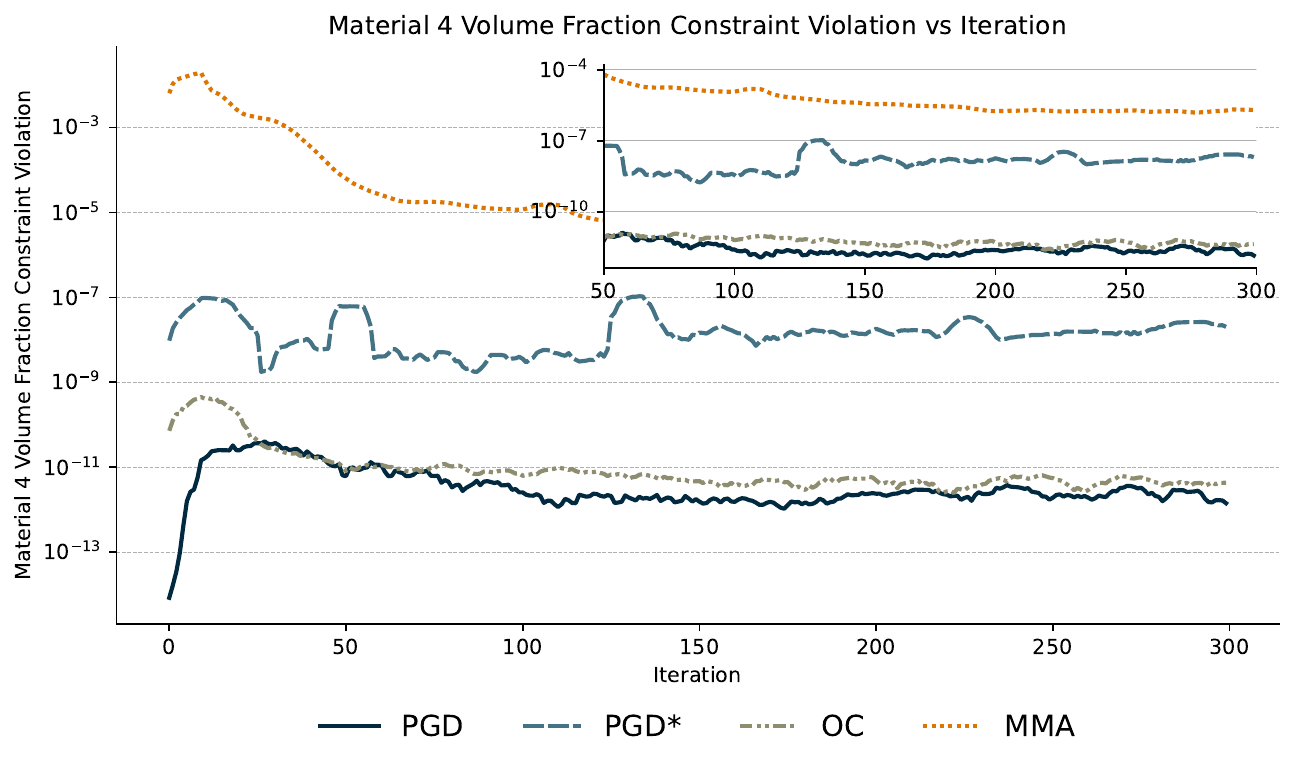}
    \caption{The constraint violation (material 4) per iteration for all solvers: multi-material minimum compliance problem, fine $512 \times 256$ mesh.}
    \label{fig:mincomp_fine_violation_4}
\end{figure}

\clearpage
\subsubsection{Volume and Weight Distribution Constrained Minimum Compliance}
The following subsections give the full results for each mesh resolution.

\paragraph{Coarse Mesh Results}
Here we provide figures for the results of running each optimizer for the coarse $128 \times 64$ mesh.

\begin{figure}[H]
    \centering
    \includegraphics[width=\linewidth]{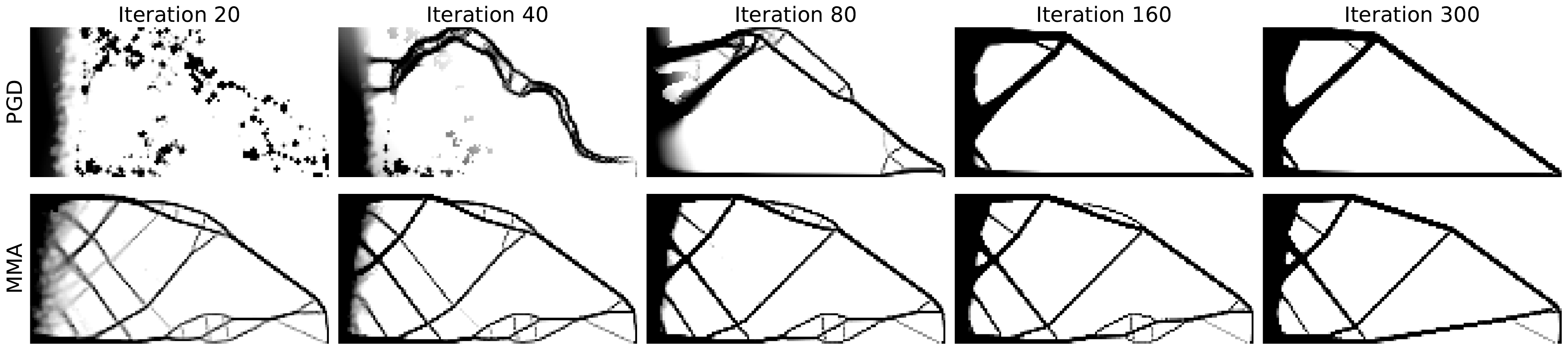}
    \caption{Designs produced by each optimizer for the volume and weight distribution constrained minimum compliance problem on the cantilever beam with a volume fraction target of $0.2$, coarse $128 \times 64$ mesh.}
    \label{fig:mincomp_coarse_designs}
\end{figure}

\begin{figure}[H]
    \centering
    \includegraphics[width=\linewidth]{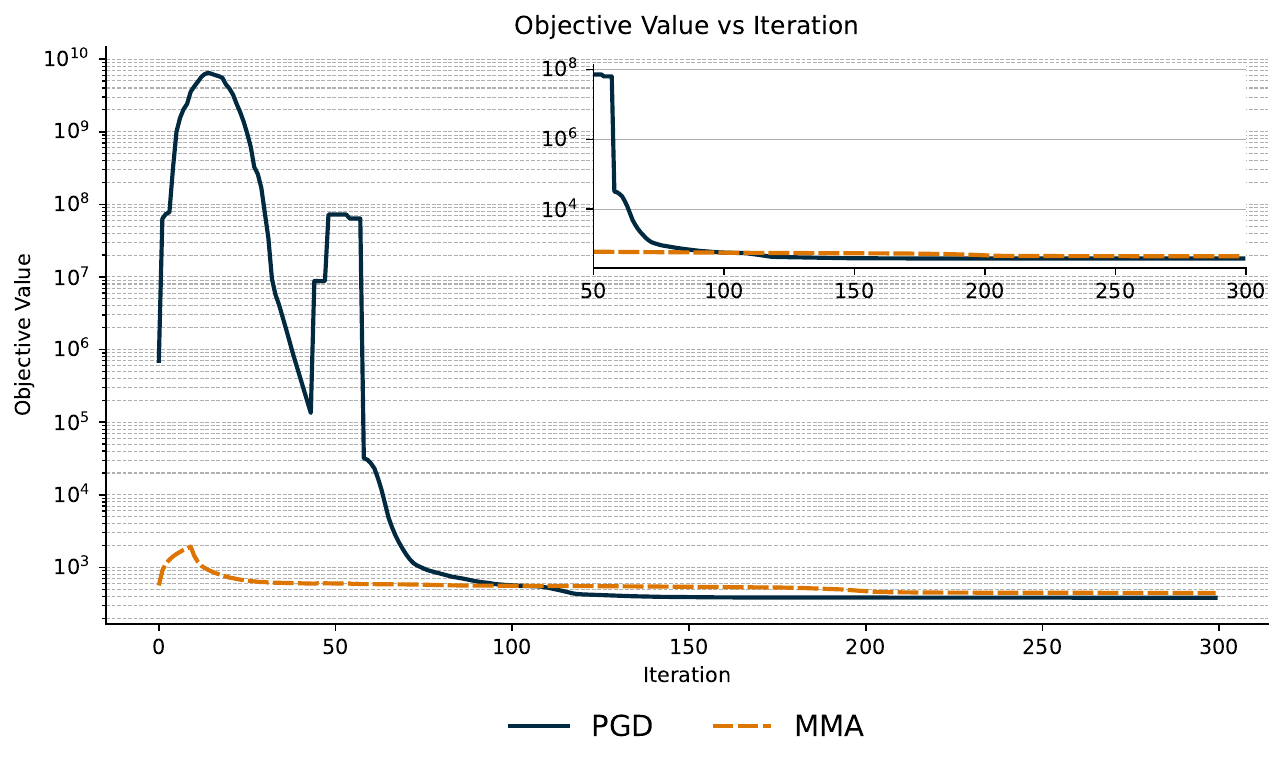}
    \caption{The value of the objective function per iteration for all solvers: volume and weight distribution constrained minimum compliance problem, coarse $128 \times 64$ mesh.}
    \label{fig:mincomp_coarse_obj}
\end{figure}

\begin{figure}[H]
    \centering
    \includegraphics[width=\linewidth]{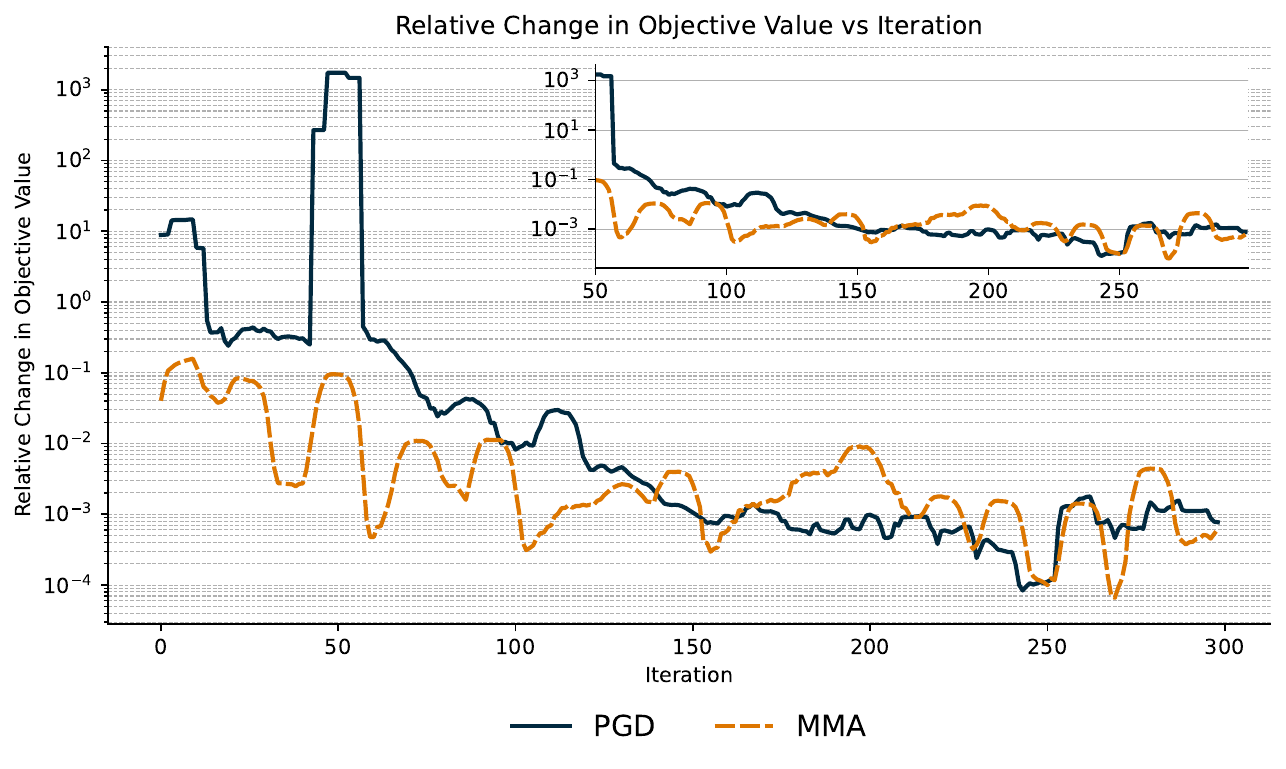}
    \caption{The relative change in the objective function per iteration for all solvers: volume and weight distribution constrained minimum compliance problem, coarse $128 \times 64$ mesh.}
    \label{fig:mincomp_coarse_rel_obj}
\end{figure}

\begin{figure}[H]
    \centering
    \includegraphics[width=\linewidth]{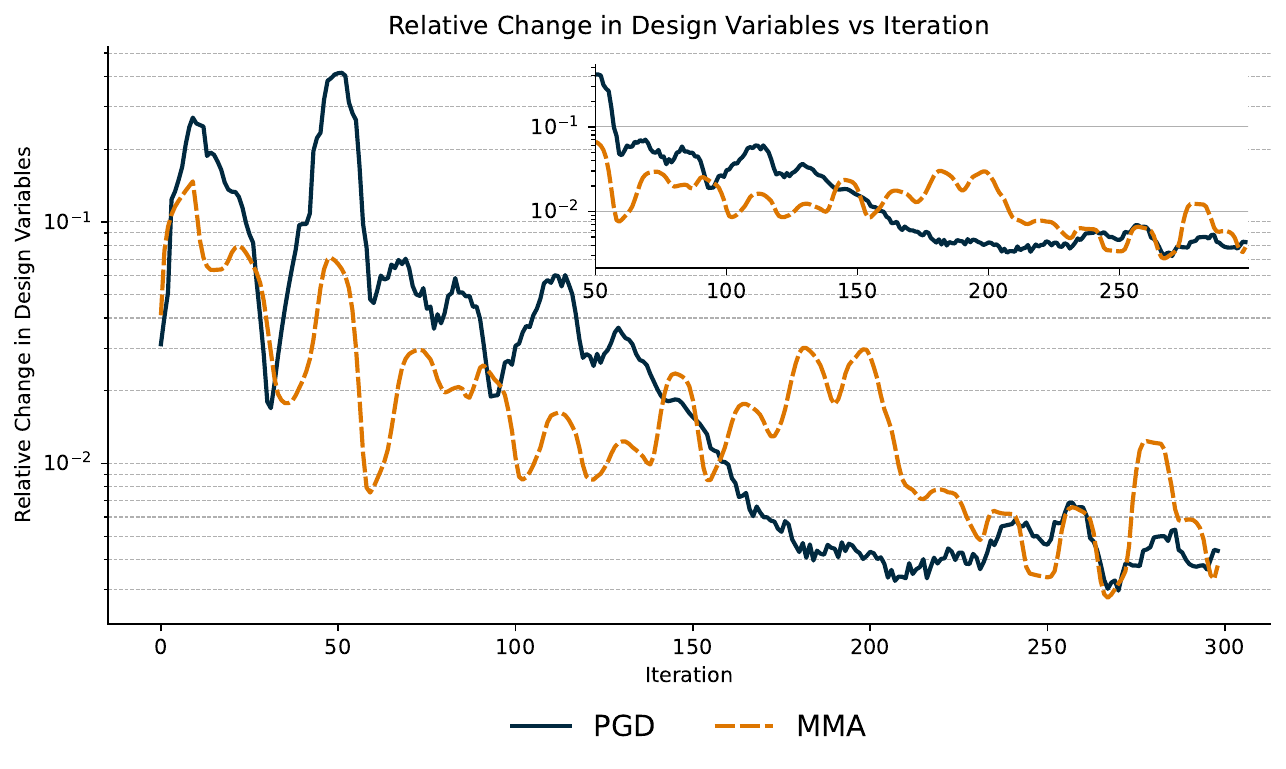}
    \caption{The relative change in the design variable norm per iteration for all solvers: volume and weight distribution constrained minimum compliance problem, coarse $128 \times 64$ mesh.}
    \label{fig:mincomp_coarse_rel_change}
\end{figure}

\begin{figure}[H]
    \centering
    \includegraphics[width=\linewidth]{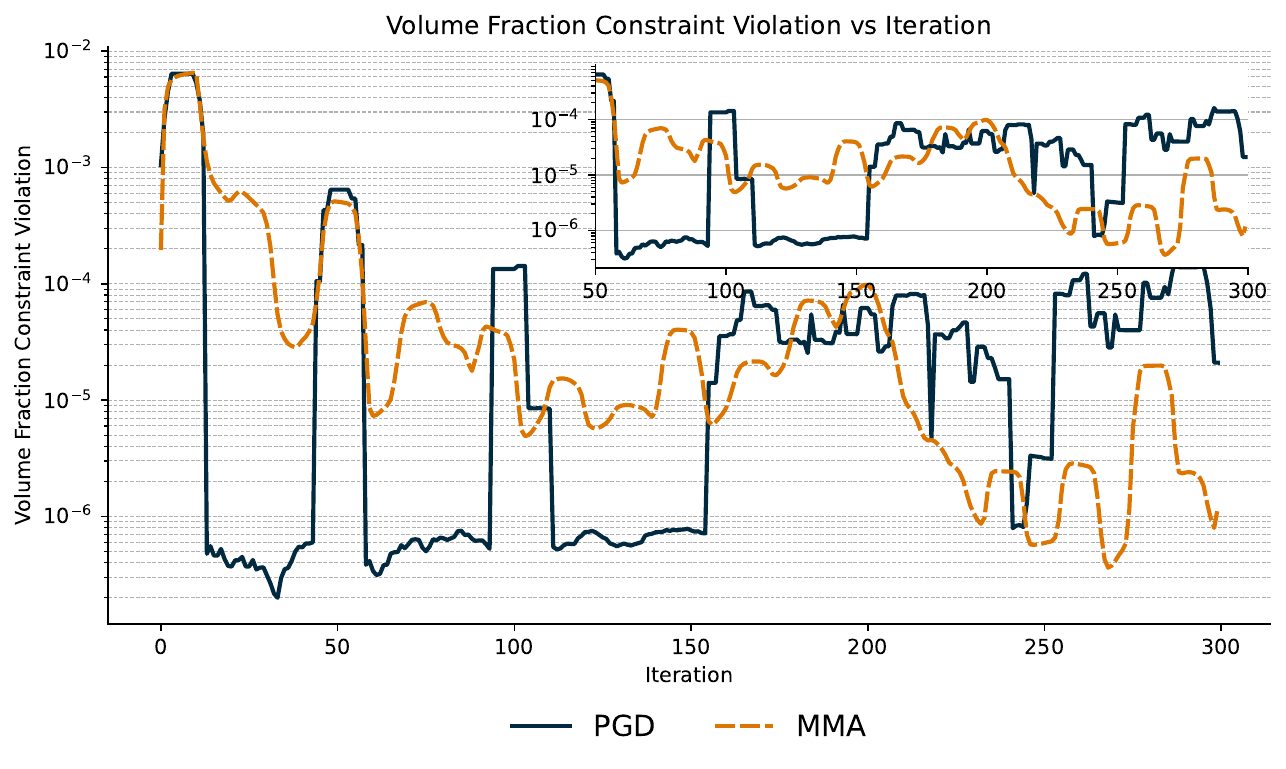}
    \caption{The constraint violation (volume constraint) per iteration for all solvers: volume and weight distribution constrained minimum compliance problem, coarse $128 \times 64$ mesh.}
    \label{fig:mincomp_coarse_violation_1}
\end{figure}

\begin{figure}[H]
    \centering
    \includegraphics[width=\linewidth]{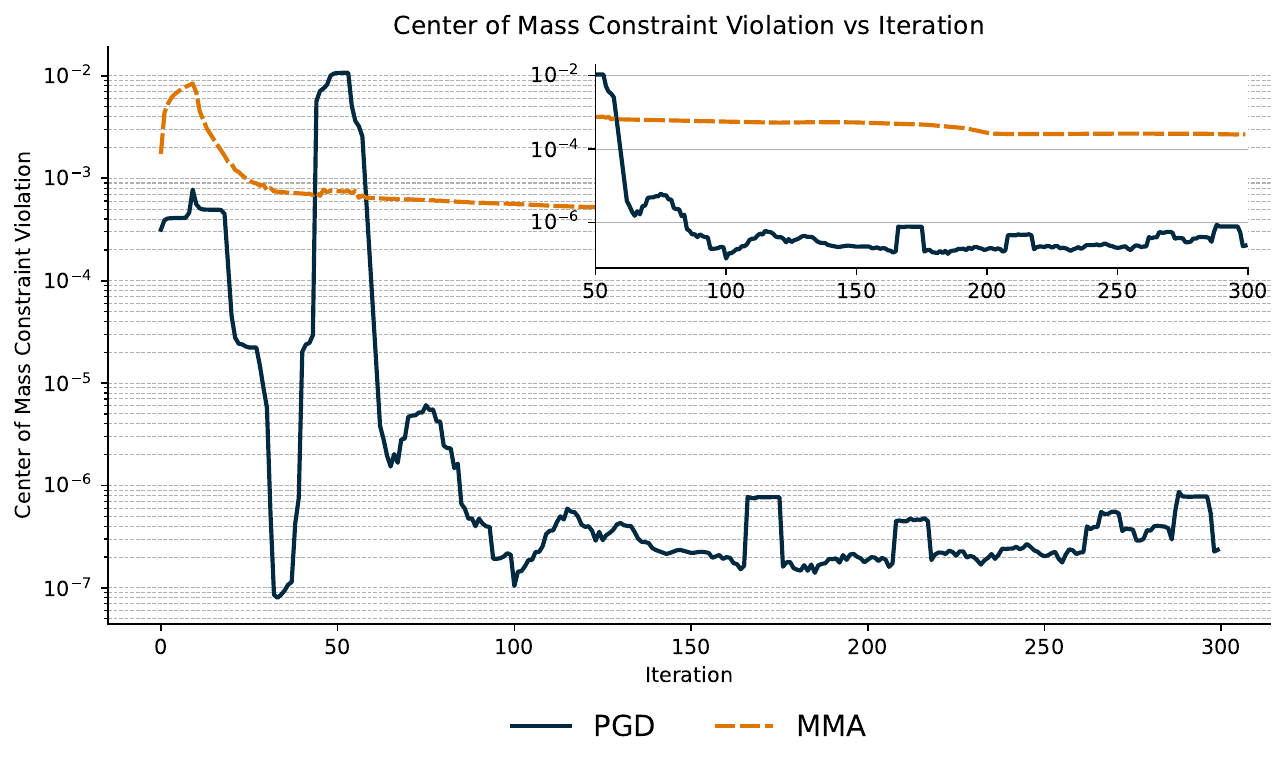}
    \caption{The constraint violation (center-of-mass deviation) per iteration for all solvers: volume and weight distribution constrained minimum compliance problem, coarse $128 \times 64$ mesh.}
    \label{fig:mincomp_coarse_violation_2}
\end{figure}

\paragraph{Medium Mesh Results}
Here we provide figures for the results of running each optimizer for the medium $256 \times 128$ mesh.

\begin{figure}[H]
    \centering
    \includegraphics[width=\linewidth]{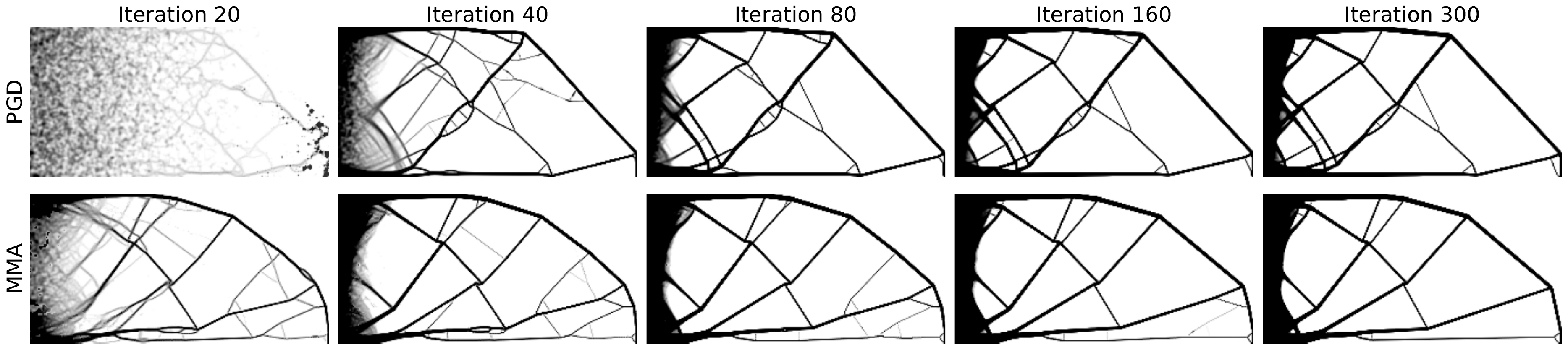}
    \caption{Designs produced by each optimizer for the volume and weight distribution constrained minimum compliance problem on the cantilever beam with a volume fraction target of $0.2$, medium $256 \times 128$ mesh.}
    \label{fig:mincomp_med_designs}
\end{figure}

\begin{figure}[H]
    \centering
    \includegraphics[width=\linewidth]{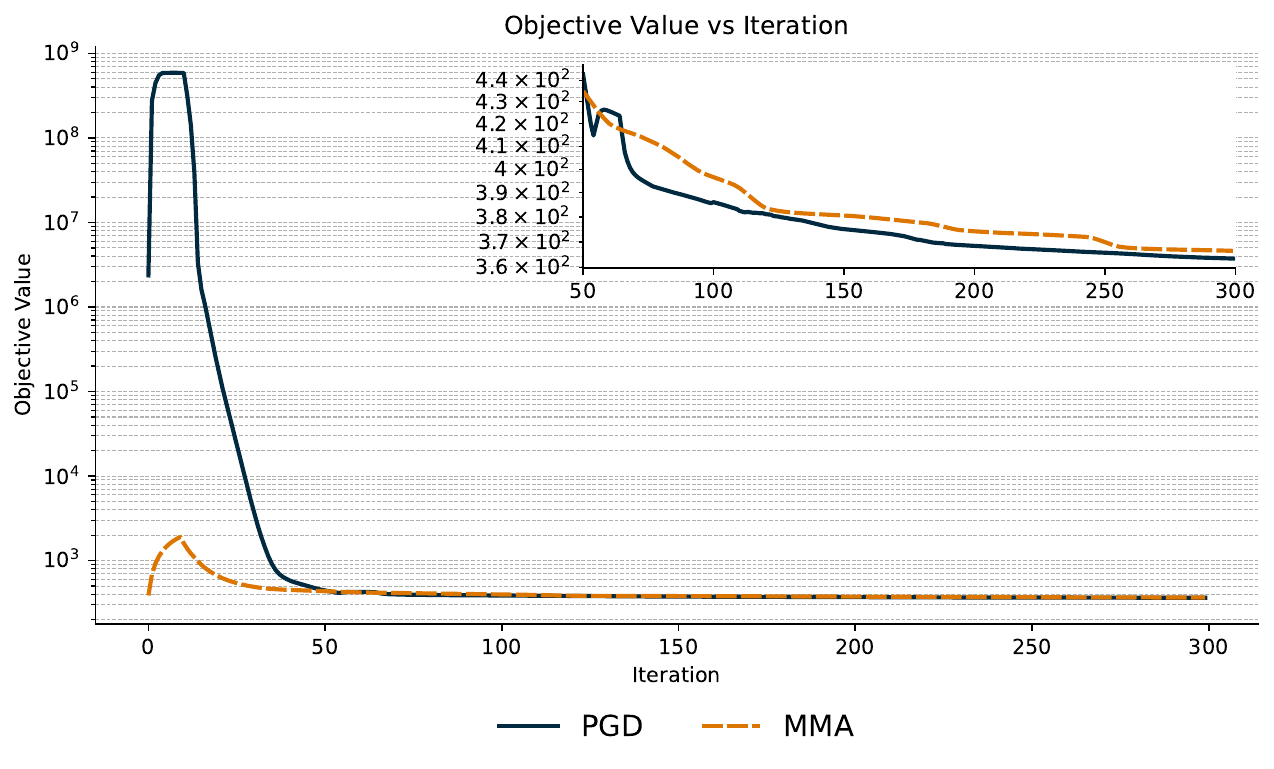}
    \caption{The value of the objective function per iteration for all solvers: volume and weight distribution constrained minimum compliance problem, medium $256 \times 128$ mesh.}
    \label{fig:mincomp_med_obj}
\end{figure}

\begin{figure}[H]
    \centering
    \includegraphics[width=\linewidth]{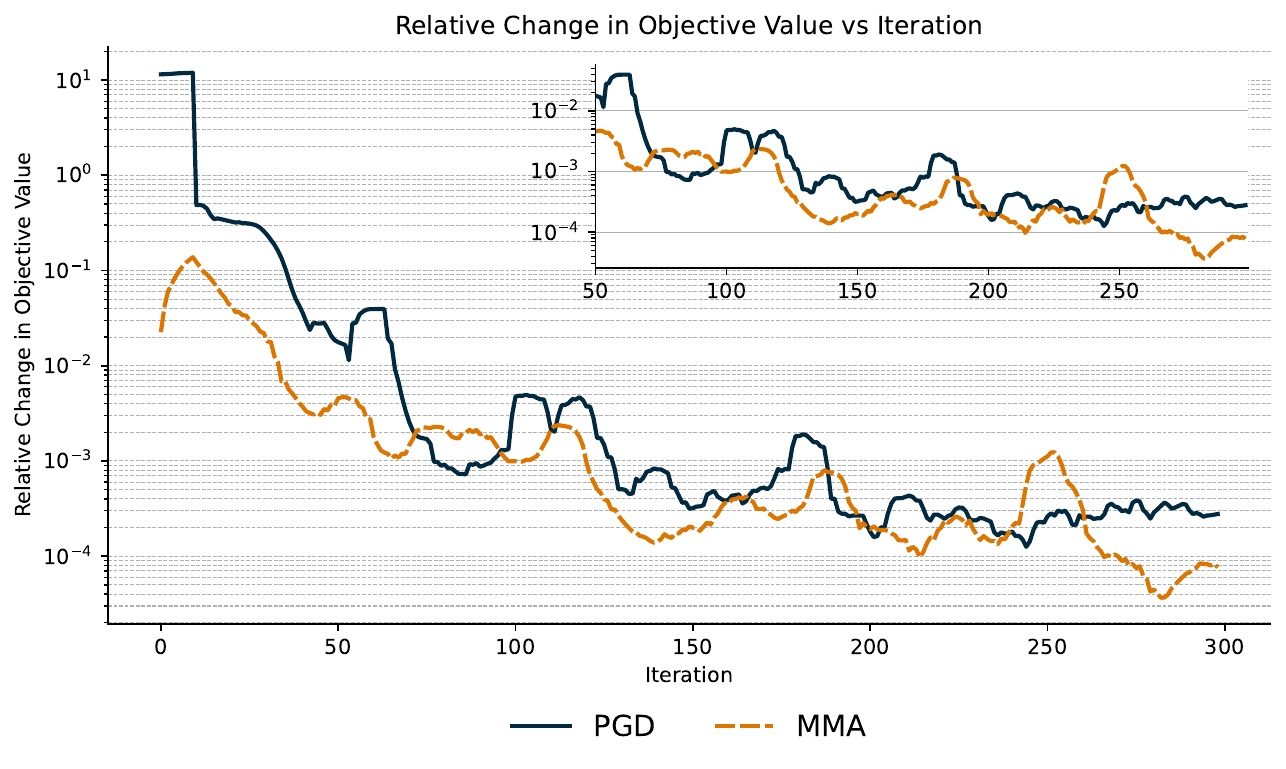}
    \caption{The relative change in the objective function per iteration for all solvers: volume and weight distribution constrained minimum compliance problem, medium $256 \times 128$ mesh.}
    \label{fig:mincomp_med_rel_obj}
\end{figure}

\begin{figure}[H]
    \centering
    \includegraphics[width=\linewidth]{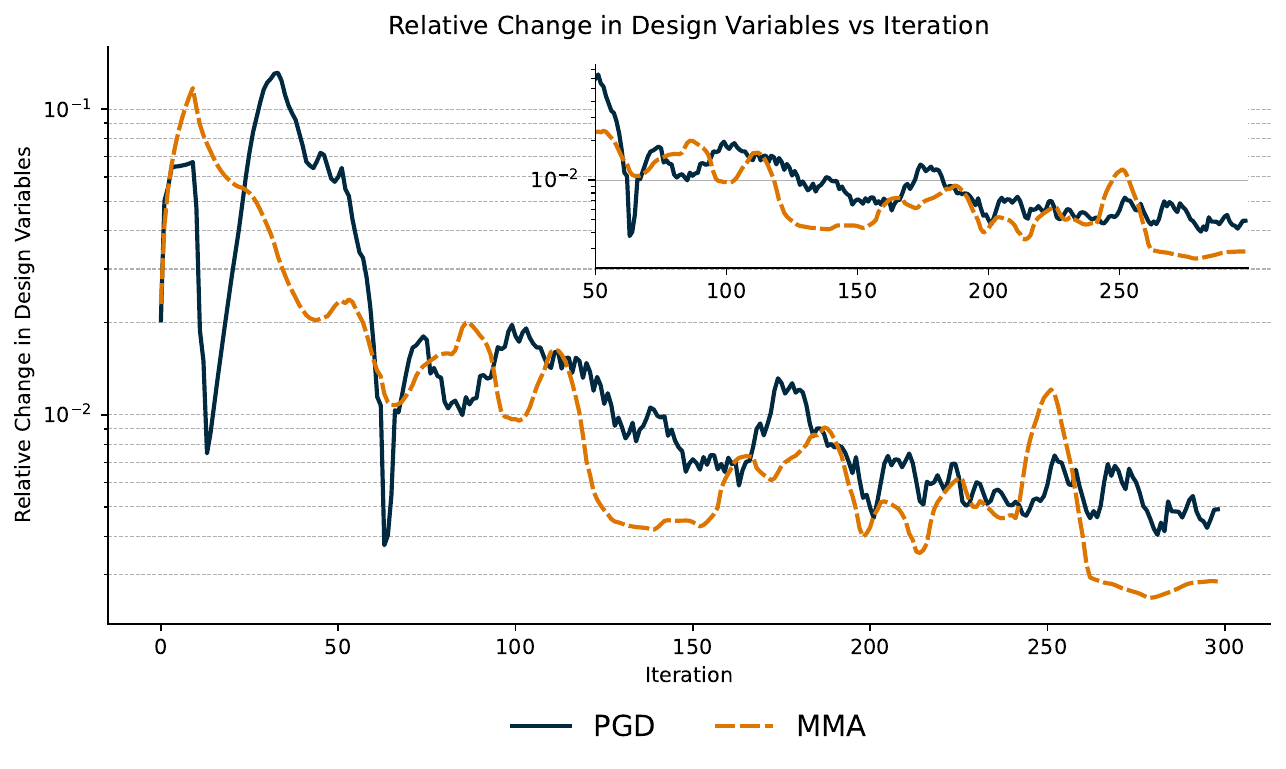}
    \caption{The relative change in the design variable norm per iteration for all solvers: volume and weight distribution constrained minimum compliance problem, medium $256 \times 128$ mesh.}
    \label{fig:mincomp_med_rel_change}
\end{figure}

\begin{figure}[H]
    \centering
    \includegraphics[width=\linewidth]{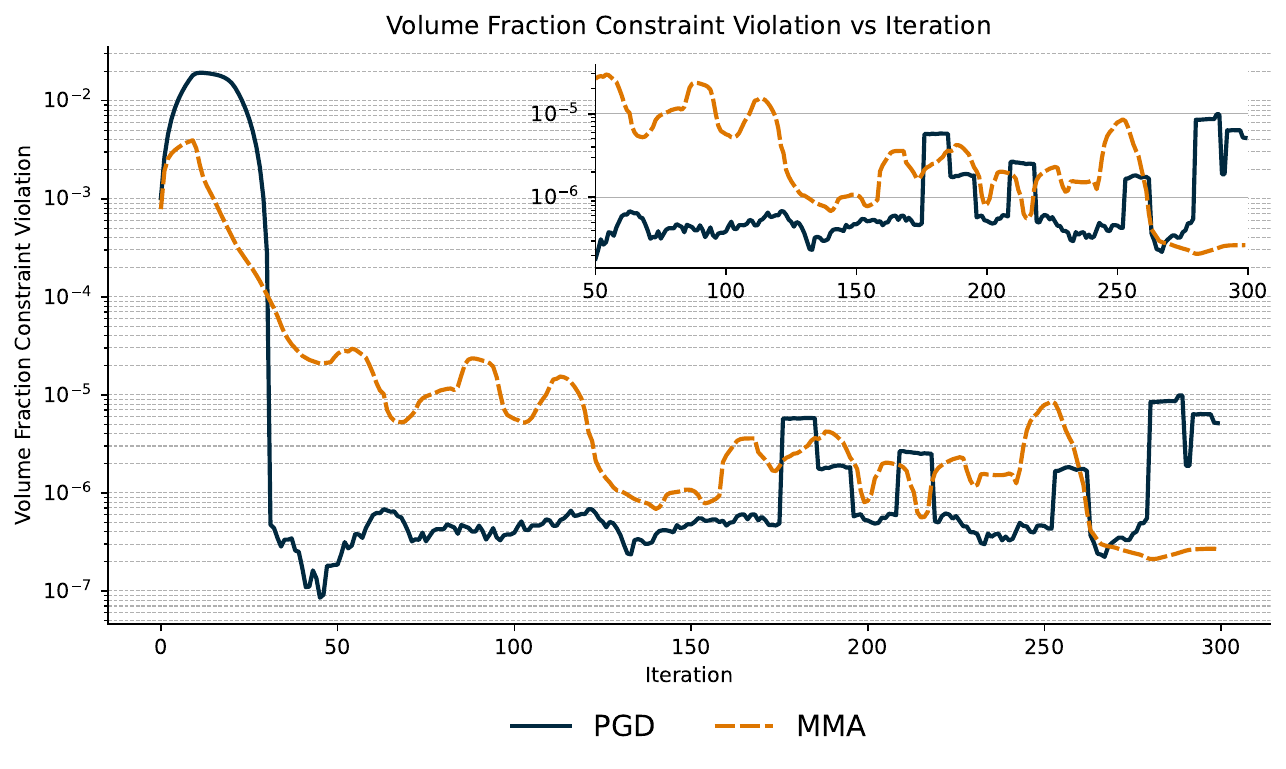}
    \caption{The constraint violation (volume constraint) per iteration for all solvers: volume and weight distribution constrained minimum compliance problem, medium $256 \times 128$ mesh.}
    \label{fig:mincomp_med_violation_1}
\end{figure}

\begin{figure}[H]
    \centering
    \includegraphics[width=\linewidth]{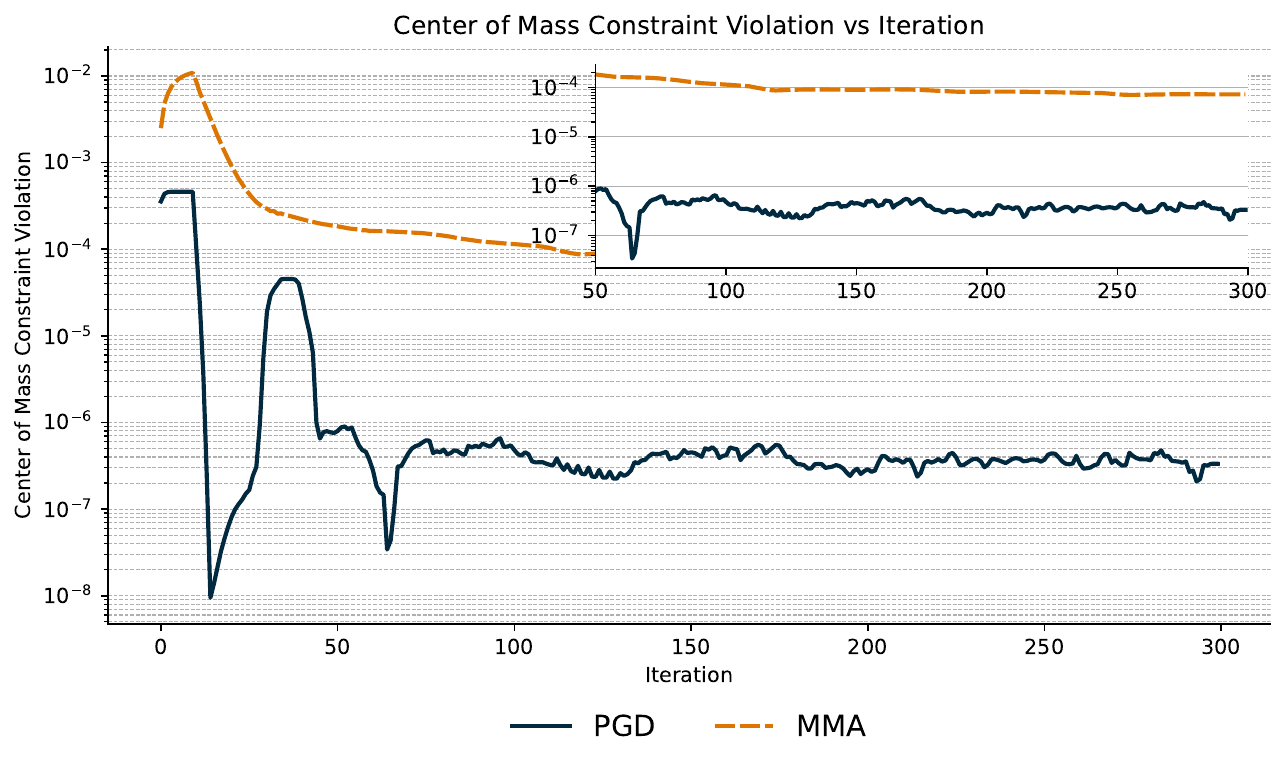}
    \caption{The constraint violation (center-of-mass deviation) per iteration for all solvers: volume and weight distribution constrained minimum compliance problem, medium $256 \times 128$ mesh.}
    \label{fig:mincomp_med_violation_2}
\end{figure}

\paragraph{Fine Mesh Results}
Here we provide figures for the results of running each optimizer for the fine $512 \times 256$ mesh.

\begin{figure}[H]
    \centering
    \includegraphics[width=\linewidth]{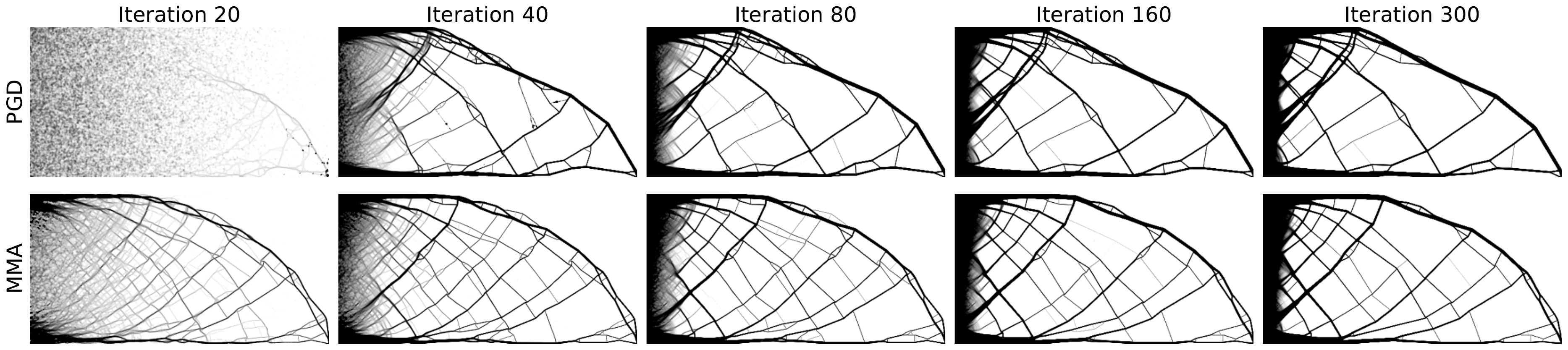}
    \caption{Designs produced by each optimizer for the volume and weight distribution constrained minimum compliance problem on the cantilever beam with a volume fraction target of $0.2$, fine $512 \times 256$ mesh.}
    \label{fig:mincomp_fine_designs}
\end{figure}

\begin{figure}[H]
    \centering
    \includegraphics[width=\linewidth]{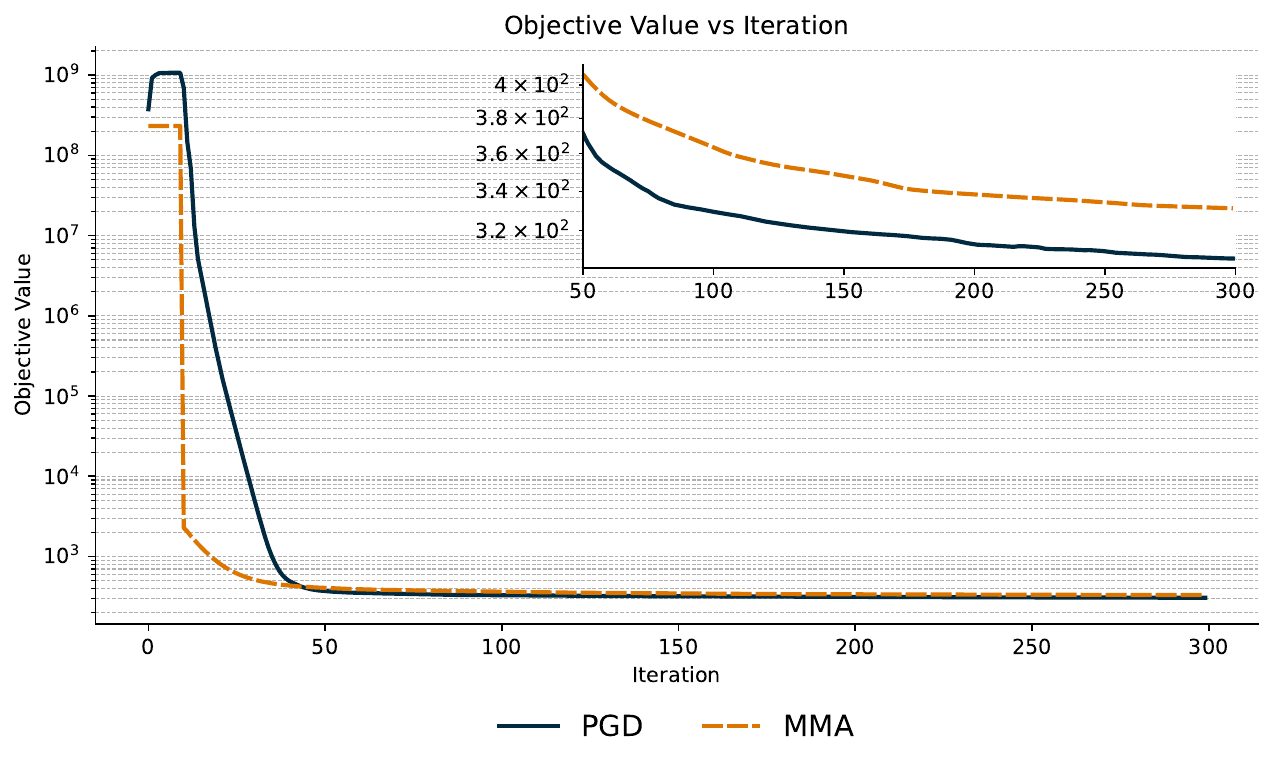}
    \caption{The value of the objective function per iteration for all solvers: volume and weight distribution constrained minimum compliance problem, fine $512 \times 256$ mesh.}
    \label{fig:mincomp_fine_obj}
\end{figure}

\begin{figure}[H]
    \centering
    \includegraphics[width=\linewidth]{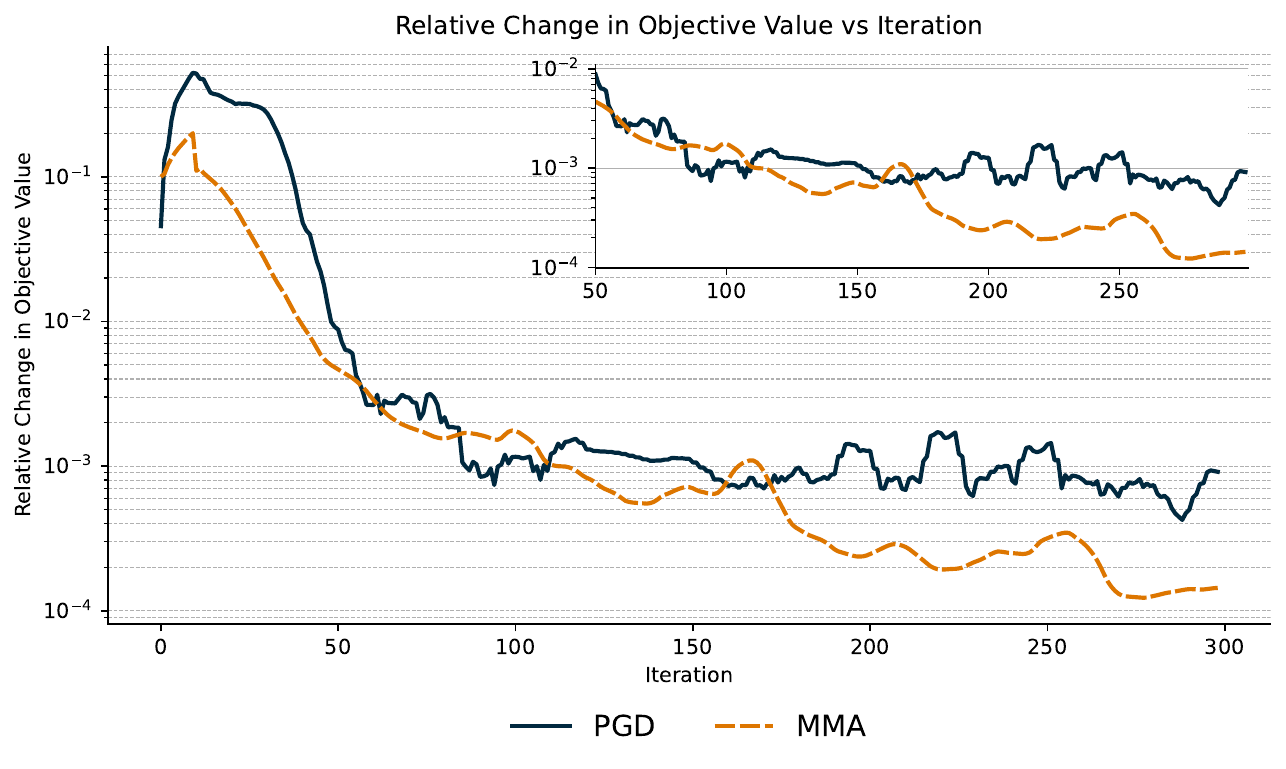}
    \caption{The relative change in the objective function per iteration for all solvers: volume and weight distribution constrained minimum compliance problem, fine $512 \times 256$ mesh.}
    \label{fig:mincomp_fine_rel_obj}
\end{figure}

\begin{figure}[H]
    \centering
    \includegraphics[width=\linewidth]{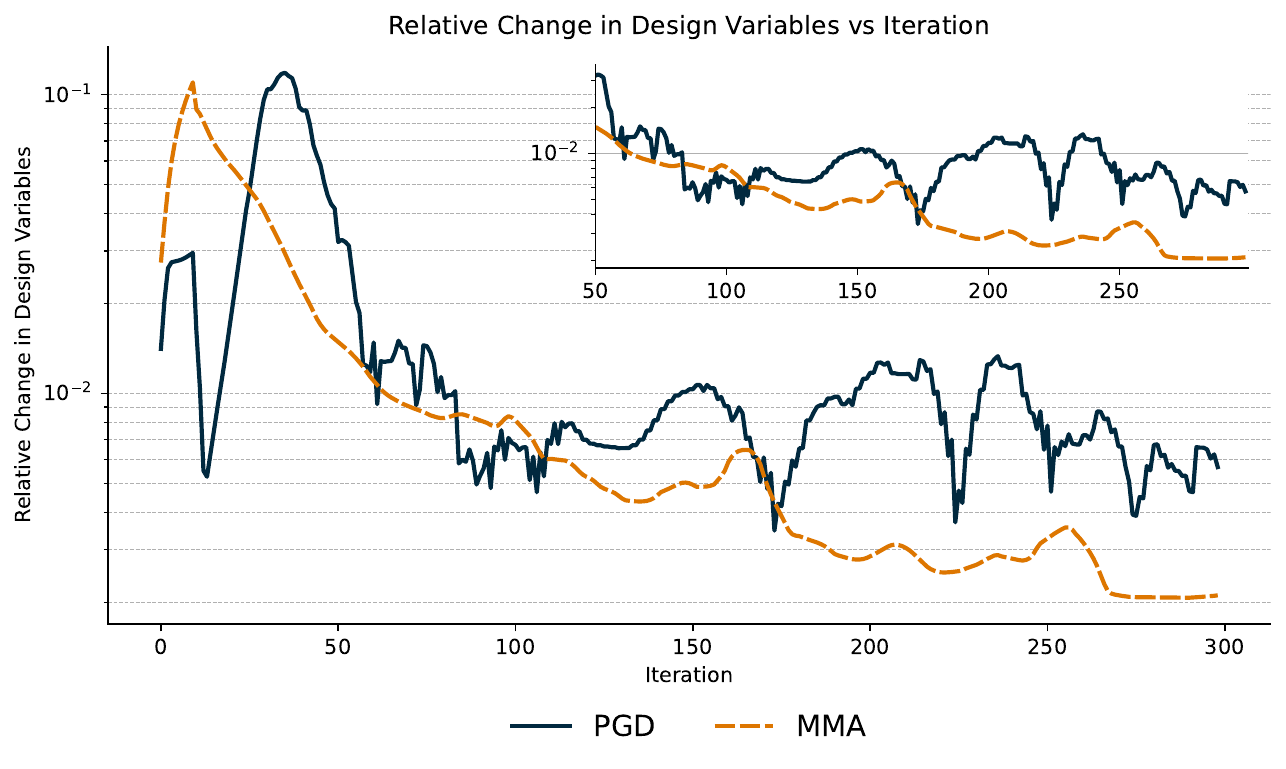}
    \caption{The relative change in the design variable norm per iteration for all solvers: volume and weight distribution constrained minimum compliance problem, fine $512 \times 256$ mesh.}
    \label{fig:mincomp_fine_rel_change}
\end{figure}

\begin{figure}[H]
    \centering
    \includegraphics[width=\linewidth]{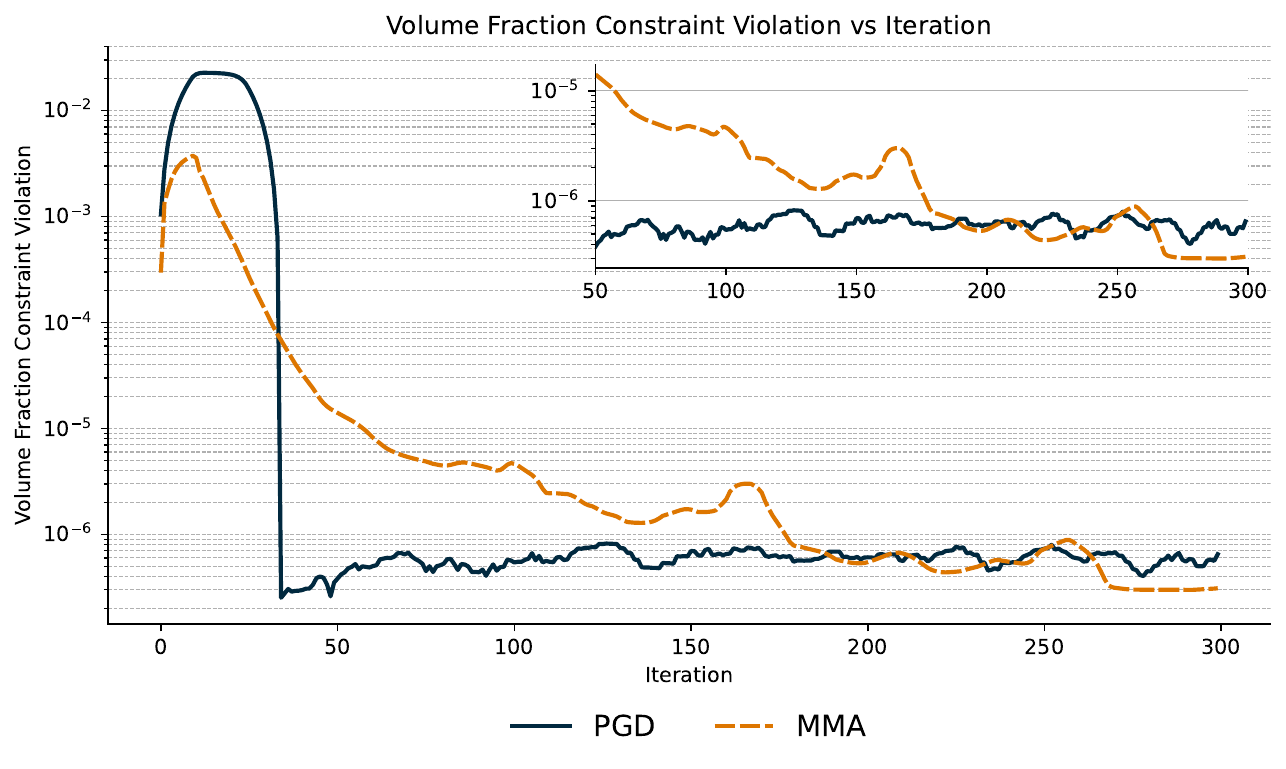}
    \caption{The constraint violation (volume constraint) per iteration for all solvers: volume and weight distribution constrained minimum compliance problem, fine $512 \times 256$ mesh.}
    \label{fig:mincomp_fine_violation_1}
\end{figure}

\begin{figure}[H]
    \centering
    \includegraphics[width=\linewidth]{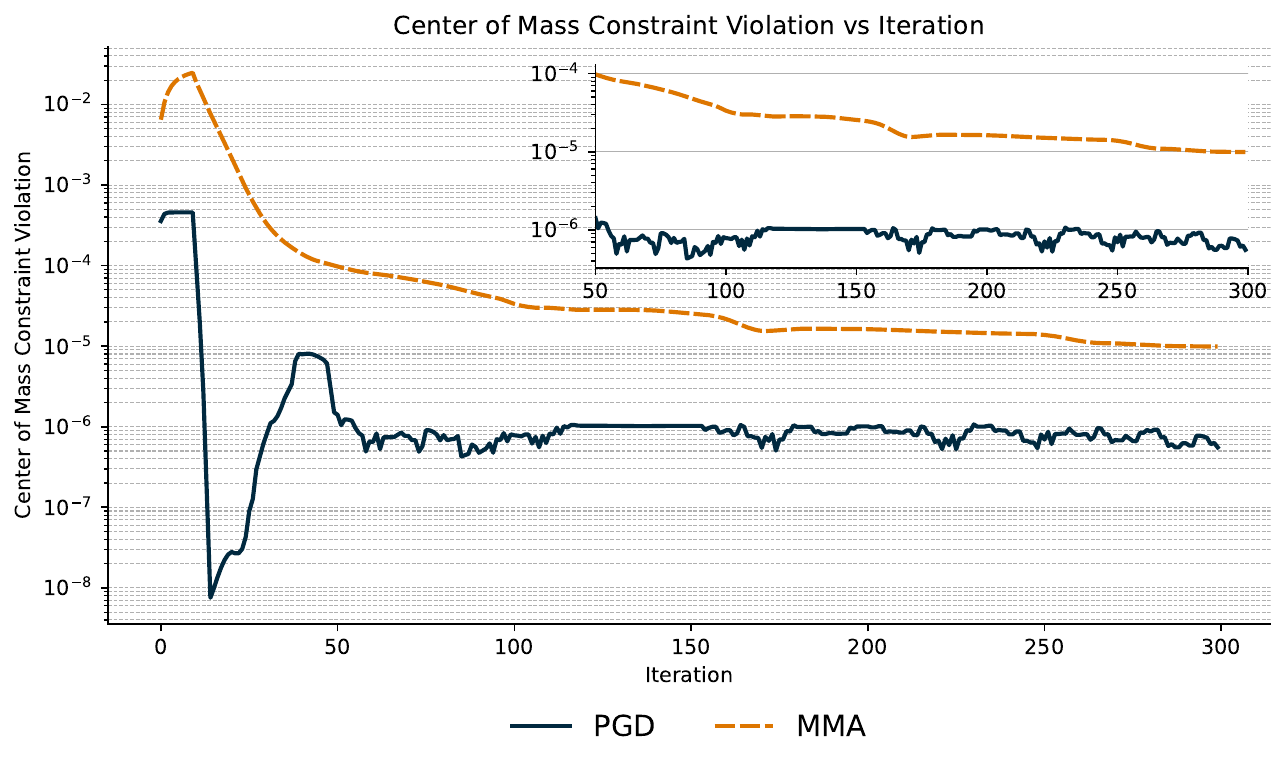}
    \caption{The constraint violation (center-of-mass deviation) per iteration for all solvers: volume and weight distribution constrained minimum compliance problem, fine $512 \times 256$ mesh.}
    \label{fig:mincomp_fine_violation_2}
\end{figure}

\end{appendices}
\end{document}